\documentclass[a4paper,11pt]{article}
\usepackage{jheppub}
\usepackage{tcolorbox}
\usepackage{hyperref}
\usepackage{tensor}
\usepackage{enumerate}
\usepackage{braket}
\usepackage{bbm}

\makeatletter
\gdef\@fpheader{}
\makeatother

\tcbuselibrary{theorems}

\title{Compact holography via polarized IKKT}

\author[]{Arsam Najafian,}
\author[]{Mark Van Raamsdonk,}
\author[]{Alejandro Vilar L\'opez}
\affiliation{Department of Physics and Astronomy, University of British Columbia, \\
6224 Agricultural Road, Vancouver, B.C.\ V6T 1Z1, Canada}

\emailAdd{arsamn@student.ubc.ca}
\emailAdd{mav@phas.ubc.ca}
\emailAdd{alejandro.vilarlopez@ubc.ca}

\abstract{The polarized IKKT matrix model is a $U(N)$ supersymmetric matrix integral proposed to be dual to an $N$ D-instanton sector of Euclidean type IIB string theory. 
We show a direct path from matrix model saddles to supergravity saddles: matrix model resolvent $\to$ holomorphic function $\to$ 2D Coulomb potential $\to$ 4D Coulomb potential $\to$ supergravity fields, where the last step was previously described. Here, the 2D electrostatics picture is an exact finite $N$ reformulation of a localization deformation of the matrix model that gives a constrained gas of dipoles in a background field. We show a matching between supergravity and matrix model on-shell actions and protected observables for all saddles. We show that the supergravity saddles have a compact and non-singular metric in string frame with the Einstein frame asymptotic region replaced by a pointlike D-instanton source. 
We provide evidence that the physics of particular gravity saddles can be obtained through restricting the domain of the matrix model integral to a basin associated with the corresponding matrix model saddle. We describe certain large $N$ decoupling limits where families of the saddles decouple from the rest to give the gravity duals of certain higher-dimensional field theories.}

\newcommand{\cF}{\mathcal{F}}

\newcommand{\cI}{\mathcal{I}}
\newcommand{\cK}{\mathcal{K}}

\newcommand{\cO}{\mathcal{O}}

\newcommand{\rd}{\mathrm{d}}
\newcommand{\rs}{\rho_{\star}}
\newcommand{\zs}{z_{\star}}
\newcommand{\Qs}{Q_{\star}}
\newcommand{\hV}{\hat{V}}
\newcommand{\hpsi}{\hat{\psi}}
\newcommand{\hz}{\hat{z}}
\newcommand{\hzs}{\hat{z}_{\star}}
\newcommand{\hr}{\hat{\rho}}
\newcommand{\Tr}{\mathrm{Tr}}
\newcommand{\dd}{\mathrm{d}}

\usepackage{xcolor}

\begin{document}

\maketitle
\flushbottom

\section{Introduction}
\label{sec:Intro}
In this paper, we study various aspects of the proposed duality between the $U(N)$ ``polarized IKKT model'' and a sector of Euclidean type IIB string theory. This represents perhaps the simplest full-fledged example of holographic duality where the gravitational theory includes Einstein gravity in $D \ge 4$ spacetime dimensions. The model was written down originally in \cite{Bonelli:2002mb} as a deformation of the original IKKT model \cite{Ishibashi:1996xs}. Decades later, its holographic interpretation was developed by Hartnoll and Liu in \cite{Hartnoll:2024csr} and shortly afterwards in the papers \cite{Komatsu:2024bop,Komatsu:2024ydh} by Komatsu, Martina, Penedones, Vuignier, and Zhao. Additional results that we will mention below were obtained in \cite{Chou:2025rwy,Hartnoll:2025ecj,DHoker:2026eox}.

The model is defined by a matrix integral\footnote{Our mass parameter is related to the one in previous works by $\omega=\frac38\Omega$. }
\begin{equation}
Z_{\omega}^{U(N)}=
\int \prod_{I=1}^{10}\dd X_I\prod_{\alpha=1}^{16}\dd\psi_\alpha\,
\exp(-S_\omega) ,
\label{eq:backgroundZ}
\end{equation}
where the $X_I$ and $\psi_\alpha$ are bosonic and Grassmann-valued Hermitian matrices transforming in the adjoint of $U(N)$ and the action is 
\begin{align}
S_\omega=\Tr\Bigg[&-\frac14[X_I,X_J]^2
-\frac{i}{2}\bar\psi\Gamma^I[X_I,\psi]
+\frac{\omega^2}{3}X_iX_i
+\frac{\omega^2}{9}X_p X_p
\nonumber\\
&\hspace{2.2cm}
+i\frac{8\omega}{9}\epsilon_{ijk}X_iX_jX_k
-\frac{\omega}{3}\bar\psi\Gamma^{123}\psi\Bigg] .
\label{eq:polarizedaction}
\end{align}
This is a mass deformation of the original IKKT action which is obtained by setting $\omega = 0$.\footnote{Within the class of supersymmetry-preserving mass deformations classified in \cite{Martina2026MassDeformations}, polarized IKKT is the unique deformation of the ten-dimensional SYM matrix model.} In addition to the $U(N)$ symmetry,\footnote{The $U(N)$ can be considered either as a global symmetry or a gauge symmetry. For the latter, we divide the partition function by the volume of $U(N)$ and consider only $U(N)$-invariant observables. However, the expectation values for these gauge-invariant observables are the same in the global model.  In section \ref{subsec:rank-fugacity-axion} below, we will see that considering the $U(N)$ symmetry as gauged and choosing a specific measure for the bosonic integrals leads to a very simple expression for the generating function for partition functions at different ranks.} the model preserves sixteen supercharges and an $SO(3)\times SO(7)$ global symmetry. 
Here, the bosonic matrices are divided into $X_i$ transforming as a vector under $SO(3)$ and the remaining seven matrices $X_p$ transforming as a vector under $SO(7)$. Conventions for fermions and gamma matrices are given in \cite{Komatsu:2024ydh}.

The matrix integral has various saddles described by bosonic matrix configurations that solve the classical equations of motion. The supersymmetric saddles are in one-to-one correspondence with partitions of $N$, labeling $N$-dimensional representations of $SU(2)$. Explicitly, the saddles are fuzzy-sphere configurations described by 
\begin{equation}
    X_i=\omega L_i\,,
\end{equation}
where $\omega$ is the mass deformation parameter of the model  and $L_i$ is a generator of $SU(2)$ in an arbitrary representation $\mathcal{R}$ defined by $n_s$ copies of the $N_s$-dimensional irreducible representation, with $\sum_s n_s N_s = N$ \cite{Hartnoll:2024csr,Komatsu:2024bop,Komatsu:2024ydh}.

In \cite{Komatsu:2024bop}, the authors constructed solutions of type IIB supergravity with isometries matching the symmetries of the matrix model and having a specific asymptotic behavior associated with D-instantons. They found that these solutions are also labelled by $N$-dimensional representations $\mathcal{R}$ of $SU(2)$, so there is apparently a one-to-one correspondence between matrix model saddles and supergravity saddles with the same symmetry.  Each supergravity solution can be understood as a warped product with $S^2$ and $S^6$ fibered over a two-dimensional half-plane with coordinates $(\rho,z) \in \mathbb{R}^+ \times \mathbb{R}^+$. All metric components and other supergravity fields can be expressed directly in terms of a single function $V(\rho,z)$ that may be interpreted as an axially symmetric potential solving an auxiliary 4D electrostatics problem. This problem involves a set of parallel conducting 3D balls (the 4D version of parallel conducting disks) sitting above an infinite conducting plane with a certain background potential. The sizes and multiplicities of the irreducible representations in $\mathcal{R}$ defining the saddle determine the locations, sizes, and charges on the balls.\footnote{A similar electrostatics description \cite{LinMaldacena2006Fivebranes} in terms of charged disks in three dimensions determines the solutions dual to vacua of the BMN matrix model \cite{Berenstein_2002}, a supersymmetric mass deformation of the D0-brane quantum mechanics}

The connection between the matrix model and the supergravity theory is made most transparent via a supersymmetric localization deformation of the matrix model through which a class of protected observables built from $\phi=X_3-iX_{10}$ can be calculated directly. Via localization, the full matrix integral can be reduced to a sum of distinct contributions associated with the matrix model saddles. For the saddle labeled by representation $\mathcal{R}$, the contribution is an integral over moduli, one for each copy of an irreducible representation in $\mathcal{R}$. In a limit where the multiplicity of each representation is large, the set of moduli can be approximated by a continuous density, and the saddle point equations for this density turn out to be precisely equivalent to the equations determining the electrostatics potential in the supergravity problem \cite{Komatsu:2024ydh}. An analogous identification between localized matrix-model densities and supergravity electrostatics was established for the BMN matrix model in \cite{AsanoEtAl2014Bubbling}

In this paper, we extend these foundational results in various ways that we summarize below. After section 2, the various sections can mostly be read independently depending on the interests of the reader.

\paragraph{2. Background.}
Section \ref{sec:Background} provides a review of various features of the model, the electrostatics description, and the dual supergravity solutions. We review various regimes of parameters where analytic approximations are possible, including both small and large $\omega$ approximations.

\paragraph{3. Supergravity solutions in string frame.}
In section \ref{sec:SupergravityStringFrame}, we study some aspects of the supergravity solutions in string frame. We show that the asymptotic part of the spacetime maps to the vicinity of a single point at which the metric is smooth. Thus, the string frame gravity duals represent compact geometries. The asymptotic point carries $N$ units of D-instanton charge, while the string coupling field vanishes there. We also discuss when the supergravity description is valid, finding a scaling of the parameters $\{(n_s, N_s)\}$ and $\omega$ that controls string loop and higher-curvature effects everywhere and analyzing in detail the validity of the supergravity description near the asymptotic point.

\paragraph{4. Matching of actions.}
In section \ref{sec:OnShellMatching}, we complete a calculation from \cite{Komatsu:2024bop,Komatsu:2024ydh}, showing that with boundary counterterms in the supergravity action chosen to cancel divergences and make the full action stationary under the appropriate variations, the on-shell supergravity action for each saddle matches with the on-shell matrix model action for the corresponding matrix model saddle. The matrix model results are computed via localization in the electrostatics approximation.

\begin{figure}
    \centering\includegraphics[width=\linewidth]{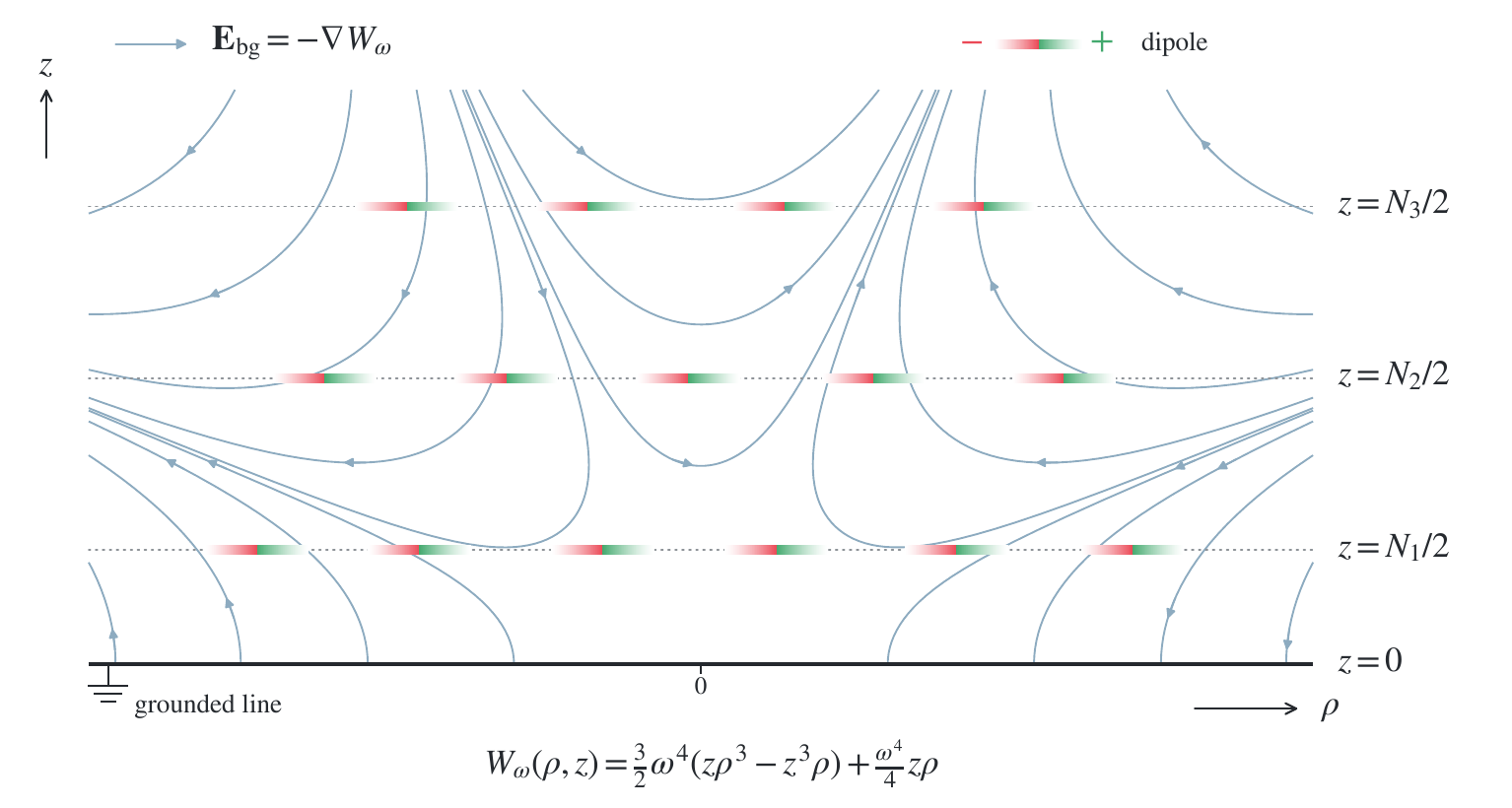}
    \caption{The contribution $Z_\mathcal{R}$ to the partition function from a representation sector $\mathcal{R}$ with $n_s$ copies of the $N_s$-dimensional irreducible representation is exactly equivalent to a statistical model of dipoles with a specific charge profile, with $n_s$ dipoles at height $N_s/2$ above a grounded line, and with a particular background electric field. The variables in the model are the horizontal positions of the dipoles.}
\label{fig:dipoles}
\end{figure}

\paragraph{5. 2D electrostatics reformulation.}
In section \ref{sec:scalar-localization}, we describe an exact off-shell reformulation of the localization-deformed model (i.e. the protected sector) as a statistical model describing dipoles in 2D interacting with a 2D Coulomb field. For the contribution to the localization integral corresponding to a representation $\mathcal{R}$ we have one dipole (with a specific rigid charge distribution) for each irreducible representation in $\mathcal{R}$. The height of the dipole above a conducting plane is proportional to the representation size, as shown in Figure \ref{fig:dipoles}. The variables in the model are the horizontal positions of the dipoles.

For saddles corresponding to representations with a large multiplicity for each irreducible representation, the density of dipoles corresponding to each representation type can be approximated by a smooth function. We show that the saddle-point equations for this function define a definite distribution for the dipoles and an associated saddle-point configuration for the 2D Coulomb potential. We show that the Coulomb potential solving the 4D electrostatics problem is simply related to this by $W(\rho,z)=\rho V(\rho,z)$, where $W$ is the 2D potential and $V$ is the 4D potential.

We show that a formal generating function for the partition functions of different ranks (also considered in \cite{Hartnoll:2025ecj}) can be expressed in a very simple form as 
\begin{equation}
\mathcal{Z}(\mu) = \sum_N e^{- \mu N} Z_{U(N)} = \left\langle \exp \left(\int \rd x \, \rd z \, \mathcal{J}(z) \mathcal{V}(x,z) \right) \right\rangle_0 \; ,
\end{equation}
where $\langle\rangle_0$ is an expectation value in the path integral for the free Coulomb field $W$, $\mathcal{J}$ is a source with support $\mathbb{Z}^+/2$, and $V$ is a ``normal-ordered'' vertex operator
\begin{equation}
 \mathcal V(x,z)
\equiv :\!\exp[-\hat{\mathcal D}( W + W_{\rm bg})]\!: \; 
 \label{eq:sl-vertex}
\end{equation}
defined in terms of the Coulomb field $W$, a background field $W_{\rm bg}(\omega,\mu)$, and a horizontal smearing operator $\hat{\mathcal{D}}$.

\paragraph{6. Holomorphic electrostatics potential from matrix resolvent.}
 Away from the dipoles, the 2D Coulomb potential solves the Laplace equation and thus can be represented as the imaginary part of a holomorphic function $\mathcal{W}$. In section \ref{sec:ResolventHolomorphic}, we show using localization that this holomorphic function is directly determined by the matrix resolvent for a complex matrix
 \begin{equation}
     \widehat{\phi}={1 \over \omega} (X_3-iX_{10}) \; .
 \end{equation}
 Explicitly, we find:
 \begin{equation}
 \quad
 \mathcal W(\zeta)-\mathcal W_{\rm bg}(\zeta)
 =\partial_\zeta G_{\mathcal R,\mathrm{cont}}(\zeta)
 \qquad \qquad G_{\mathcal R}(\zeta)
 \equiv \left\langle\operatorname{Tr}
  (\zeta I-\widehat\phi)^{-1}\right\rangle_{\mathcal R} \; ,
 \label{IntroResolvent}
 \end{equation}
 where $\zeta = z + i \rho$, $\mathcal W_{\rm bg}$ is the part of $\mathcal W$ associated with the background potential and ``cont'' indicates the continuum, long-wavelength approximation that leads to the 4D electrostatics picture.

The coefficients of inverse powers of $\zeta$ in the asymptotic expansion of the resolvent formula for large $\zeta$ are (up to overall constants) exactly the expectation values of operators $\Tr(\phi^{2k})$. We show that these are directly proportional to the multipole moments $q_{2k+1}$ appearing in the asymptotic expansion of the 4D electrostatic potential.

We show that adding sources for $\Tr(\phi^{2k})$ in the matrix model (which can be thought of as defining a generating function for the protected observables) corresponds to a modification of the background potential in either the 2D or the the 4D electrostatics problems. The conductor positions and total charges remain fixed by the representation data. 

\paragraph{7. Holomorphic function reformulation.}
In section \ref{sec:ConformalMappintElectrostaticsProblem}, we translate the 4D electrostatics problem into a problem to find the holomorphic function $\mathcal{W}$, showing that this can be solved through conformal mapping techniques. We outline a computational procedure to solve this complex analysis problem in order to find explicit metrics. This holomorphic function description is closely related to a recent reformulation of the supergravity solutions in \cite{DHoker:2026eox}.

\paragraph{8. Single ball solutions.}
To investigate the correspondence in more quantitative detail, in section \ref{sec:OneBallQuantitative} we focus on saddles corresponding to $n_\star$ copies of a single $N_\star$-dimensional irreducible representation. Much of the analysis of this family was already carried out in \cite{Komatsu:2024ydh}. Here we provide an independent and systematic derivation of the results, with particular emphasis on the solution method, and extend the analysis in several directions. For this family of saddles, the electrostatic system has a single charged ball above the infinite conducting plane. While the physical system and specification of the saddle depends on three independent parameters $\omega$, $n_{\star}$, and $N_{\star}$, we show that the relevant electrostatics solution and the corresponding supergravity solution depend non-trivially (i.e. up to overall scaling) only on the combination
\begin{equation}
    \xi\equiv\frac{n_\star}{\omega^4N_\star^5}\,.
    \label{Intro:Xi}
\end{equation}
This can be related to the ratio
$\hzs = z_\star/\rho_\star$ between the height of the ball above the conducting plane and the radius of the ball.

We derive analytic charge densities in the limits $\hzs \gg 1$ and $\hzs \ll 1$, and formulate the general problem as a Fredholm equation of the second kind that can be solved numerically in the intermediate regime. Using the numerics, we explicitly calculate expectation values of the protected observables as a function of $\xi$; dependence on the other parameters can be recovered by scaling.

\paragraph{9. D1- and NS5-brane limits.}
In section \ref{sec:Limits}, following \cite{Maldacena:2002rb,Ling:2006up}, we show that the endpoints of the one-ball family have direct brane interpretations. In the high-ball limit $\hzs \to\infty$, or equivalently $\xi\to0$, the grounded plane decouples from the local electrostatic problem.
Taking
\begin{equation}
    N_\star\to\infty\,,
    \qquad
    n_\star\ \text{fixed}\,,
    \qquad
    \omega^4N_\star\ \text{fixed}\,,
\end{equation}
we obtain a geometry whose asymptotic metric, dilaton, and flux have the form
expected for a spherical D1-brane configuration associated with the polarized
D-instantons; the D1 charge is set by $n_\star$. In this limit, the matrix model becomes a continuum two-dimensional field theory.

In the opposite limit $\hzs \to0$, the charged ball becomes infinitely large relative to its separation from the infinite plane. In a certain double-scaling limit, it is possible to preserve a non-trivial solution between the ball and the plate in the limit. The corresponding supergravity solution has asymptotics consistent with being the dual of a decoupled NS5-brane theory on $S^6$. 

The limiting solutions studied in this section may be related to the spherical brane solutions constructed in \cite{BobevEtAl2018SphericalBranes}.

\paragraph{10. Isolating supergravity saddles from the matrix model.}
In section \ref{sec:Basins}, we ask how the physics of a specific supergravity saddle might be extracted from matrix model calculations. We note that (as already emphasized in \cite{Hartnoll:2025ecj}) in the regime of parameters where any given supergravity solution is valid, the matrix model integral is completely dominated by contributions associated with the irreducible saddle for which there is not a good dual backreacted  supergravity description. Thus, the physics of saddles with a good supergravity description will correspond to exponentially suppressed contributions to matrix model observables. 

We argue that there may be practical ways to isolate these exponentially suppressed contributions. For protected observables, the localization calculation already exactly splits matrix model observables into specific contributions associated with the various saddles. Our results demonstrate that the contribution associated with a given matrix model saddle can be matched with observables in the corresponding supergravity saddle.

For unprotected observables, we can try to split up the integration domain for the matrix model into basins associated with the various saddles. We propose a specific method where a basin is defined by gradient flow for the real part of the effective action obtained by integrating out the fermions. Using direct Monte Carlo calculations for $SU(2)$ and $SU(3)$, we find that the contributions to protected observables calculated for various basins of the original matrix model match well with the contributions calculated via localization. The basin method applies to any observable, giving us a natural candidate for defining quantities in the matrix model that can be compared with observables in a specific supergravity saddle. 

An interesting aspect of our analysis in this section is the observation that the saddles for the effective action obtained by integrating out the fermions are modified relative to the classical saddles; for a given classical saddle, there is a critical value of $\omega$ below which the corresponding saddle of the effective action splits into one or more continuous families of saddles. Specifically, the minima appear to correspond to one or more families of $SO(7)$ orbits of configurations with the transverse matrices $X_p$ non-zero and simultaneously diagonalizable. We discuss this more in Appendix \ref{sec:EffectiveActionSaddles}.

\paragraph{11. Discussion.} 
We end in section \ref{sec:Discussion} with a discussion of various points, including understanding stringy physics from the matrix model, calculating more general protected observables and establishing a holographic dictionary, understanding the gravity interpretation of the dipole gas, and testing the correspondence beyond the protected sector.

\section{Background}
\label{sec:Background}

We begin by reviewing key properties of the polarized IKKT model and its corresponding supergravity description from \cite{Hartnoll:2024csr,Komatsu:2024bop,Komatsu:2024ydh,Hartnoll:2025ecj}.

\subsection{Polarized IKKT saddles and parameter regimes}
\label{subsec:MatrixSetup}

The partition function and action were already given in the introduction. The parameters of the model are the matrix size $N$ and the deformation parameter $\omega$.

The bosonic potential has supersymmetric fuzzy-sphere minima
\begin{equation}
    X_i=\omega L_i\,,
    \qquad
    X_p=0\,,
    \qquad
    [L_i,L_j]=i\epsilon_{ijk}L_k\,,
    \label{MatrixSetup:FuzzySphere}
\end{equation}
where the $L_i$ are $SU(2)$ generators in an arbitrary $N$-dimensional representation $\mathcal{R}$. We can write the generators in a general such representation as
\begin{equation}
    L_i
    =
    \bigoplus_{s=1}^{q}
    \left(
        \mathbf 1_{n_s}\otimes L_i^{N_s}
    \right),
    \qquad
    N=\sum_{s=1}^{q}n_sN_s\,,
    \label{MatrixSetup:Representation}
\end{equation}
that is, we have $n_s$ copies of the $N_s$-dimensional irreducible representation. This configuration of matrices preserves a symmetry subalgebra $\prod_s U(n_s) \times SO(3)_\mathcal{R} \times SO(7)$ where the first factor is the subgroup of $U(N)$ defined by
\begin{equation}
U = \bigoplus_{s=1}^{q}
    \left(u_{n_s} \otimes \mathbf{1}_{N_s}
    \right)
\end{equation}
and the second factor combines an $SO(3)$ rotation $R$ with a compensating unitary
\begin{equation}
U = \bigoplus_{s=1}^{q}
    \left(\mathbf{1}_{n_s} \otimes u^R_{N_s}
    \right) \, , \qquad u^R_{N_s} L_i^{N_s} (u^R_{N_s})^\dagger = R_i {}^j L_j^{N_s} \; .
\end{equation}
The classical action evaluated on such a configuration is
\begin{equation}
    S_{\omega,\mathrm{vac}}[\mathcal R]
    =
    -\frac{\omega^4}{72}
    \sum_s n_s\left(N_s^3-N_s\right).
    \label{MatrixSetup:VacuumAction}
\end{equation}

At fixed $N$ and large enough $\omega$ we have a saddle-point expansion about the fuzzy-sphere minima. In the large $\omega$ limit, the action describing the fluctuations around each saddle becomes Gaussian. We note that the scaling $X \to \omega X$, $\psi \to \omega^{3 \over 2} \psi$ gives an overall factor of $\omega^4$ outside the action, so it is natural to identify
\begin{equation}
\omega^4 \sim {1 \over g^2} \, ,
\end{equation}
where $g$ is the standard coupling parameter in a 't Hooft expansion. Taking $g^2 N \sim N/\omega^4$ small ensures that perturbative calculations of simple observables about the $X=0$ vacuum will be reliable \cite{Hartnoll:2024csr}. The parameter controlling perturbation theory is different for different vacua; for example in the vacuum corresponding to $n_{\star} > 1$ copies of the $N_{\star}$-dimensional irreducible representation, perturbation theory is controlled by $n_{\star}/(\omega^4 N_{\star})$. For the irreducible vacuum with $n_{\star} = 1$ we find that perturbation theory is controlled by $1/(\omega^4 N^2)$.\footnote{These last two statements are based on a direct calculation of the leading and subleading non-vanishing terms of $\langle \Tr X_p X_p \rangle$ that vanishes classically, with the perturbation theory parameter defined up to numerical coefficients and logs by the ratio of these. The $n_\star = 1$ result is different from the result obtained by setting $n_\star = 1$ in the $n_\star > 1$ result since the $n_\star=1$ theory is an Abelian gauge theory on the fuzzy sphere while the $n_\star > 1$ theories are non-abelian.}
From \eqref{MatrixSetup:VacuumAction}, it follows that the saddle corresponding to the irreducible representation has the least action for all values of the parameter $\omega$,
\begin{equation}
S = - {\omega^4 \over 72} (N^3 - N) \; .
\end{equation}
At the level of the classical action, contributions from the $(N - 1, 1)$ saddle with the next lowest action are suppressed by a
factor
\begin{equation}
e^{-\Delta S} = e^{-{1 \over 24} \omega^4 (N^2 - N)} \; .
\end{equation}
For large $N$, \cite{Hartnoll:2025ecj} found evidence for a first-order phase transition at a critical value of $\omega^2 N$. Below the transition, the dominant configurations are approximately commuting, and their joint eigenvalues $\vec{x}_a=(x_a^1,\ldots,x_a^{10})$, $a=1,\ldots,N$, can be interpreted as the positions of $N$ particles in ten dimensions. For $\omega^2 N \ll 1$, an effective action obtained by integrating out off-diagonal modes around the diagonal background can be used to reliably compute observables such as $\Tr(X^I X^I)$, as explained in \cite{Komatsu:2024ydh}.

\subsection{Localization}
The partition function of the matrix model at general $\omega$ and a class of protected observables can be calculated via supersymmetric localization \cite{Hartnoll:2024csr,Komatsu:2024ydh}, where the action is deformed by a term that is exact under a chosen supersymmetry variation $\delta_Q$ in such a way that the partition function is unchanged:
\begin{equation}
Z(t)=\int[\dd\Phi]\,\exp[-S_\omega-t \delta_Q\mathcal V] \, ,
\qquad
\frac{\dd Z(t)}{\dd t}=0 \, .
\label{eq:localizationfamily}
\end{equation}
In the limit $t\to\infty$, the classical saddles are
modified to include an additional non-zero bosonic matrix commuting with the $SU(2)$ generators
\begin{equation}
    X_i=\omega L_i\,,
    \qquad
    X_{10}=\omega M\,,
    \qquad
    [M,L_i]=0\,,
    \qquad
    X_4=\cdots=X_9=0\,.
\label{MatrixSetup:LocalizationLocus}
\end{equation}
Explicitly, $M$ takes the form
\begin{equation}
    M
    =
    \bigoplus_s
    \left(
        M_s\otimes\mathbf 1_{N_s}
    \right) \, ,
    \qquad
    M_s
    =
    \operatorname{diag}
    \left(\rho_{s1},\ldots,\rho_{sn_s}\right) \, .
    \label{MatrixSetup:Modulus}
\end{equation}
The $\rho_{si}$ are moduli that parameterize the localization locus. In the $t \to \infty$ limit, the contributions to the path integral from the various saddles decouple from one another and the integrals over fluctuations around each saddle can be performed exactly, leaving a finite-dimensional integral over the moduli $\rho_{si}$. The partition function thus splits into a sum
\begin{equation}
    Z=\sum_{\mathcal R} Z_{\mathcal R} \; .
    \label{MatrixSetup:SectorSum}
\end{equation}
Here, the saddles are still labeled by representations $\mathcal{R}$.

The localization calculation can be extended to compute any gauge-invariant functions of the complex matrix
\begin{equation}
    \phi\equiv X_3-iX_{10} \, ,
    \qquad
    \delta_Q \phi=0\,.
    \label{MatrixSetup:ProtectedMatrix}
\end{equation}
As for the partition function, expectation values of these observables can be decomposed exactly as
\begin{equation}
Z \langle f(\phi) \rangle = \sum_{\cal R} {Z_{\mathcal R}} \langle f(\phi) \rangle_{\cal R} \; .
\label{Zfsum}
\end{equation}

The explicit result for the contribution to $\langle f(\phi) \rangle_{\cal R}$ from representation $\mathcal{R}$ takes the form \cite{Komatsu:2024ydh}
\begin{align}
    {Z_{\mathcal R}} \bigl\langle f(\phi)\bigr\rangle_{\mathcal R}
    \equiv
    C_{\mathcal R}
    &\exp\left[
        \frac{\omega^4}{72}
        \sum_s n_s\left(N_s^3-N_s\right)
    \right]
    \int
    \prod_s\prod_{i=1}^{n_s}\rd\rho_{si}\,
    Z_{\text{1-loop}}^{(\mathcal R)}(\rho)
    \nonumber\\
    &\times
    f\left[
        \omega\left(L_3-iM\right)
    \right]
    \exp\left[
        -\frac{\omega^4}{6}
        \sum_s N_s\sum_{i=1}^{n_s}\rho_{si}^2
    \right],
    \label{MatrixSetup:FixedSectorObservable}
\end{align}
where $C_{\mathcal R}$ and $Z_{\text{1-loop}}^{(\mathcal R)}(\rho)$ are given explicitly below in section \ref{sec:scalar-localization}.
In this paper, the observables that we will be primarily interested in are single-trace operators
\begin{equation}
    \mathcal O_k
    \equiv
    \Tr\left(\phi^{2k}\right)
    \label{MatrixSetup:ProtectedMoments}
\end{equation}
constructed from even powers of $\phi$; expectation values of odd powers vanish by symmetry.

\subsection{The electrostatics description}
\label{sec:electrostatics}

For large $N$ and for representations $\mathcal{R}$ where the number of copies of each irreducible representation are large, the integral over localization moduli becomes dominated by a saddle-point configuration where the set of moduli $\rho_{si}$ associated with a given representation size $N_s$ are described by a smooth even density $f_s(\rho)$, normalized by $\int_{\mathbb R}\rd\rho\,f_s(\rho)=n_s$. In the continuum approximation, the saddle-point equations for these densities are precisely the equations describing the following 4D axially-symmetric electrostatics problem specified by the matrix model parameters $\{(n_s,N_s)\}$ and $\omega$ \cite{Komatsu:2024bop,Komatsu:2024ydh}.

 Let $z$ be one of the coordinates and let $\rho$ be the radial coordinate in the remaining three directions. We normalize the four-dimensional charge density $\varrho$ by $-\nabla^2V=4\pi^2\varrho$, so a unit point charge has potential $1/r^2$. The configuration is then:
 \begin{itemize}
     \item There is an infinite grounded plane at $z = 0$.
     \item For $s = 1, \dots, q$, there are charged conducting balls (thus at a fixed potential) at
     \begin{equation}
     z_s = \frac{N_s}{2}
    \end{equation}
     with radius $\rho_s$ and having total charge
     \begin{equation}
     Q_s = n_s \; .
    \end{equation}
     \item There is a background potential
     \begin{equation}
     V_{\rm bg}(\rho, z) \equiv - \eta z + \frac{3}{2}\omega^4 \left( z \rho^2 - z^3 \right) \, ,
     \label{Vbg}
     \end{equation}
     where $\eta$ is an arbitrary constant.
     \item The full potential is
     \begin{equation}
     \label{SUGRAsolutions:FullPotential}
     V (\rho, z) = V_{\rm bg}(\rho, z) + \psi (\rho, z) \, ,
     \end{equation}
     where $\psi(\rho, z)$ decays asymptotically, i.e., as $r = \sqrt{\rho^2 + z^2} \to \infty$.
     \item The ball radii $\{\rho_s\}$ are determined jointly in terms of the other parameters by requiring the charge density to vanish at the edge of every ball, $\sigma_s(\rho_s)=0$.
 \end{itemize}
 These constraints fix the radii $\rho_s$, the charge densities $\sigma_s(\rho)$ and the potential $V(\rho,z)$ via Laplace's equation away from the conducting balls,
 \begin{equation}
  V'' + \ddot{V} + \frac{2}{\rho} \dot{V} = 0 \, ,
 \end{equation}
 and
 \begin{equation}
 \sigma_s(\rho) = \frac{1}{4\pi^2}\lim_{\epsilon \to 0^+} \left( -\partial_z V(\rho,z_s+\epsilon) + \partial_z V (\rho,z_s-\epsilon)\right) \, .
 \label{charge_density}
 \end{equation}
 We have defined $V' \equiv \partial_z V$ and $\dot{V} \equiv \partial_{\rho} V$. The density $\sigma_s$ is related to the total charge $Q_s$ on a ball by
 \begin{equation}
 Q_s=4\pi\int_0^{\rho_s}\rd\rho\,\rho^2\sigma_s(\rho).
 \end{equation}
 The signed localization density and the radial charge density are related by
 \begin{equation}
 \sigma_s(\rho)=-\frac{f_s'(\rho)}{2\pi\rho} \, , \qquad 0<\rho<\rho_s \, .
 \end{equation}
We will see below that in the limit where the electrostatics description of the localization integral applies, the protected observables can be read off directly from the asymptotics of the electrostatics potential.
 
\subsection{Dual supergravity solutions}
\label{sec:SUGRAsolutions}

The polarized IKKT matrix model is proposed to be dual to a sector of type IIB supergravity with particular asymptotics associated with the $\alpha' \to 0$ limit of the D-instanton solution \cite{GibbonsGreenPerry1996Instantons,OoguriSkenderis1998DInstantons}. In \cite{Komatsu:2024bop}, the authors showed that solutions of type IIB supergravity with isometries matching the $SO(3) \times SO(7)$ symmetries and supersymmetries of the matrix model can be constructed directly in terms of a single function $V(\rho,z)$ that solves precisely the same electrostatics problem that governs saddles of the localization integral. Thus, via the electrostatics description, we have a one-to-one correspondence between supersymmetric matrix model saddles and supersymmetric smooth Euclidean type IIB
supergravity saddles with the appropriate asymptotics. 

While the original description of the supergravity solutions was presented in Einstein frame, we will give expressions in string frame, since that will make our later discussion about the regime of validity of supergravity cleaner -- see appendix \ref{app:TheoryAndEoM} for our conventions. In terms of the full potential $V(\rho,z)$ corresponding to a set of parameters $\{(n_s,N_s)\}$ and $\omega$, the string frame metric is
\begin{equation}
\label{SUGRAsolutions:Metric}
\frac{\rd s^2}{\alpha'} = A(\rho, z) \left( F(\rho, z) \left( \rd \rho^2 + \rd z^2 \right) + G_6(\rho, z) \, \rd \Omega_6^2 + G_2(\rho,z) \, \rd\Omega_2^2 \right) \, ,
\end{equation}
where $0 \leq \rho,z < \infty$ parametrize the first quadrant over which the $S^6$ and $S^2$ are fibered. The metric functions are given as
\begin{equation}
\label{SUGRAsolutions:MetricFunctions}
\begin{gathered}
A
=
2\pi
\sqrt{\frac{-3\dot V-\rho V''}{\rho(-V'')}} \, ,
\qquad
F=-\frac{V''}{\dot V} \, ,
\qquad
G_6=3\rho \, ,
\\
G_2=-\frac{\rho V''\dot V}{\Delta} \, ,
\qquad
\Delta
=
3\dot V V''
+\rho\left[(\dot V')^2+(V'')^2\right] \, .
\end{gathered}
\end{equation}
Assuming $V$ solves the electrostatics problem described in the previous section, \cite{Komatsu:2024bop} showed that the metric is smooth and positive everywhere.\footnote{In particular, $\dot{V}$, $- V''$, $-3 \dot{V}-\rho V''$ and $\Delta$ are all non-negative throughout the geometry. The dilaton field we write in appendix~\ref{app:ElectrostaticSolutions} is also positive. Note that, for smoothness, the condition to have $\sigma_s(\rho_s) = 0$ is essential in order to avoid diverging electric fields. This is possible because the background potential balances the accumulation of charge that would otherwise happen at the tip of the balls.}  The $S^6$ shrinks smoothly to zero on the axis $\rho=0$, while the $S^2$ shrinks smoothly to zero on every conducting ball and on the grounded plane. Fibering the $S^2$ over an interval between adjacent balls at $z_s$ and $z_{s-1}$ gives a three-cycle that carries NS5 flux, we call this cycle $\Sigma_{3,s}$. Fibering the $S^6$ over a curve that encloses the $s$-th conducting ball and has both endpoints on the axis gives a seven-cycle carrying D1 electric flux, we call this cycle $\Sigma_{7,s}$ \cite{Komatsu:2024bop}. Explicit expressions for these form fields as well as the dilaton as local algebraic functions in $V$ and its derivatives are given
in appendix~\ref{app:ElectrostaticSolutions}. 

In the supergravity picture, changing the parameter $\eta$ from the electrostatics background potential corresponds to a constant axion shift, accompanied by a corresponding shift of $C_2$, leaving the gauge-invariant three-form unchanged; see appendix~\ref{app:ElectrostaticSolutions}.

\section{Supergravity solutions in string frame}
\label{sec:SupergravityStringFrame}

\subsection{Asymptotic behavior}

In this section, we analyze the asymptotic behavior of the supergravity solutions as $r = \sqrt{\rho^2 + z^2} \to \infty$. This was studied in \cite{Komatsu:2024bop}, but here we will emphasize some differences that appear when this region is studied in string frame. In particular, we find that in string frame the asymptotic region of the supergravity solution becomes the neighborhood of a single point, and that the metric is smooth in this neighborhood. 

Since the potential is axisymmetric, it can be written in a multipole expansion far from the sources involving a single series of functions:
\begin{equation}
V(\rho, z) = - \eta z + \frac{3}{2}\omega^4 \left(z \rho^2 - z^3 \right)  + \sum_{l=0}^{\infty} \frac{q_{2l+1}}{(\rho^2 + z^2)^{l+3/2}} U_{2l+1}\left( \frac{z}{\sqrt{\rho^2 + z^2}} \right) \, ,
\label{multipole}
\end{equation}
where $U_{l}(\cos \theta) = \sin((l+1)\theta) / \sin (\theta)$ are the Chebyshev polynomials of the second kind, 
\begin{equation}
q_1 = P = 2 \sum_s z_s Q_s = N
\end{equation}
is the dipole moment of the charge configuration, and higher multipole moments are given by\footnote{The condition at the grounded plate can be enforced by considering image charges that make the configuration antisymmetric under $z \to - z$; the contributions from both the physical charges and their images are taken into account in the multipole moments as presented.}
\begin{equation}
q_{2l+1} = \frac{4 \pi}{l+1} \sum_{s} \int_0^{\rho_s} \rd \rho \, \rho^2 \sigma_s(\rho) (\rho^2 + z_s^2)^{l+1/2} U_{2l + 1}\left( \frac{z_s}{\sqrt{\rho^2 + z_s^2}} \right) \, .
\label{MultipoleDef}
\end{equation}
Here, $\sigma_s(\rho)$ is the charge density in the $s$-th ball, obtained from the potential via (\ref{charge_density}).

This asymptotic expansion can be used to obtain the behavior of the fields at long distances. An interesting feature of the string frame geometry is that the metric can be extended smoothly to the point at infinity, so that we end up with a finite volume region. We can show this from the asymptotic expansion of the potential. Introducing coordinates
\begin{equation}
\tilde{\rho} = \frac{\rho}{r^2} \, , \qquad \tilde{z} = \frac{z}{r^2} \, , \qquad \tilde{r}^2 = \tilde{\rho}^2 + \tilde{z}^2 = \frac{1}{r^2} \, ,
\end{equation}
the asymptotic expansion becomes a series around $\tilde{r} = 0$ of the form
\begin{equation}
V(\tilde{\rho}, \tilde{z}) = - \frac{\eta \tilde{z}}{\tilde{r}^2} + \frac{3\omega^4}{2\tilde{r}^6} \left( \tilde{z}\tilde{\rho}^2 - \tilde{z}^3 \right) + \sum_{l=0}^{\infty} q_{2l+1} \tilde{r}^{2l+3} U_{2l+1} \left( \frac{\tilde{z}}{\tilde{r}} \right) \, .
\end{equation}
Observe now that $\tilde{r}^{2l+1} U_{2l+1}(\tilde{z}/\tilde{r})$ is a homogeneous polynomial of degree $2l+1$ in $(\tilde{\rho}, \tilde{z})$, which is furthermore odd in $\tilde{z}$ and even in $\tilde{\rho}$. We can introduce
\begin{equation}
\mathcal{K}(\tilde{\rho}, \tilde{z}) = \sum_{l=0}^{\infty} q_{2l+1} \frac{\tilde{r}^{2l+1}}{\tilde{z}} U_{2l+1} \left( \frac{\tilde{z}}{\tilde{r}} \right) \, .
\end{equation}
Each term in this series is a polynomial separately even in $\tilde{\rho}$ and $\tilde{z}$. Since the exterior multipole expansion converges for sufficiently large $r$, $\mathcal{K}(\tilde{\rho}, \tilde{z})$ defines an analytic function in a neighborhood of $\tilde{\rho} = \tilde{z} = 0$, separately even in the two coordinates.

Using this, we can go back to the full metric \eqref{SUGRAsolutions:Metric} and evaluate the different components. After some algebra, the result is
\begin{equation}
\label{SUGRAsolutions:MetricAtInfinity}
\begin{aligned}
\frac{\rd s^2}{\alpha'} = & 2\pi \sqrt{\frac{\mathcal{R}}{3\mathcal{X} + \tilde{\rho}^2 \tilde{r}^4 \mathcal{R}}} \left( \frac{3 \mathcal{X} + \tilde{\rho}^2 \tilde{r}^4 \mathcal{R}}{\mathcal{X}} \left( \rd \tilde{\rho}^2 + \rd \tilde{z}^2 \right) + 3 \tilde{\rho}^2 \rd \Omega_6^2 \right. \\ 
& \qquad \qquad \qquad \qquad \qquad \left. + \frac{\mathcal{X} (3 \mathcal{X} + \tilde{\rho}^2 \tilde{r}^4 \mathcal{R})}{(\mathcal{X} + \tilde{z}^2 \tilde{r}^4 \mathcal{Z})^2 + \tilde{z}^2 \tilde{r}^4 \mathcal{R}(3 \mathcal{X} + \tilde{\rho}^2 \tilde{r}^4 \mathcal{R})} \tilde{z}^2 \rd \Omega_2^2 \right) \, ,
\end{aligned}
\end{equation}
where we have introduced the functions
\begin{align}
\mathcal{E} & = \frac{1}{\tilde{\rho}} \partial_{\rho} \mathcal{K} - 4 \mathcal{K} \, , & \mathcal{X} & = 3\omega^4 + \tilde{r}^6 \mathcal{E} \, , & \mathcal{R} & = \frac{1}{\tilde{\rho}} \partial_{\rho} \mathcal{E} - 6 \mathcal{E} \, , & \mathcal{Z} & = \frac{1}{\tilde{z}} \partial_{z} \mathcal{E} - 6 \mathcal{E} \, .
\end{align}
Note that the derivatives of $\mathcal{K}$ and $\mathcal{E}$ are taken with respect to $(z, \rho)$, not the transformed coordinates. The relation is
\begin{equation}
\partial_{\rho} = (\tilde{z}^2 - \tilde{\rho}^2) \partial_{\tilde{\rho}} - 2 \tilde{\rho} \tilde{z} \partial_{\tilde{z}} \, , \qquad \partial_{z} = - 2 \tilde{\rho} \tilde{z} \partial_{\tilde{\rho}} + (\tilde{\rho}^2 - \tilde{z}^2) \partial_{\tilde{z}} \, .
\end{equation}

The main observation is the following. Since $\mathcal{K}$ is analytic and separately even in $\tilde{\rho}$ and $\tilde{z}$, the same is true of $\mathcal{E}$, $\mathcal{X}$, $\mathcal{R}$ and $\mathcal{Z}$ (the factors $1/\tilde{\rho}$ and $1/\tilde{z}$ cancel once the parity of $\mathcal{K}$ is taken into account). These functions take finite, non-zero values at $\tilde{z} = \tilde{\rho} = 0$, and therefore all the metric components are analytic in a neighborhood of that point -- denominators do not vanish. They also inherit the property of being separately even in $\tilde{\rho}$ and $\tilde{z}$. Moreover, \eqref{SUGRAsolutions:MetricAtInfinity} shows that the coefficient of $\rd \tilde{\rho}^2$ differs from that of $\tilde{\rho}^2 \rd \Omega_6^2$ only by terms of $\mathcal{O}(\tilde{\rho}^2)$, and similarly the coefficient of $\rd \tilde{z}^2$ differs from that of $\tilde{z}^2 \rd \Omega_2^2$ only by $\mathcal{O}(\tilde{z}^2)$. Together with analyticity and evenness, this implies that the metric is analytic at the origin. We can actually expand to see the first few orders explicitly
\begin{equation}
\label{SUGRAsolutions:MetricAsymptotic}
\frac{\rd s^2}{\alpha'} = \frac{8\pi\sqrt{3P}}{\omega^2} \left( 1 + \frac{2 q_3}{3 P} (7\tilde{z}^2 - 3 \tilde{\rho}^2) + \dots \right) \left( \rd \tilde{\rho}^2 + \rd \tilde{z}^2 + \tilde{\rho}^2 \rd\Omega_6^2 + \tilde{z}^2 \rd \Omega_2^2 \right) + \dots \, ,
\end{equation}
showing that the geometry is well-behaved at $\tilde{z} = \tilde{\rho} = 0$. Since this point corresponds to the original asymptotic region $r \to \infty$, the metric admits a smooth extension that includes infinity as an ordinary point at finite proper distance. Together with the smooth caps at the charged balls and along the axes, this implies that the completed string-frame geometry is compact and has finite volume. The curvature at the added point is determined by the multipole moments, as we discuss in the next section.

Even if the geometry is smooth at the point at infinity, the full supergravity configuration is still singular. We can show it using the same asymptotic expansion of the potential. The dilaton diverges to $- \infty$ in such a way that the string coupling vanishes asymptotically:
\begin{equation}
\label{SUGRAsolutions:DilatonAsymptotic}
e^{\Phi} = \frac{96\pi P}{\omega^8 (\rho^2 + z^2)^4} \left( 1 + \frac{4 q_3}{3 P} \frac{7z^2 - 3 \rho^2}{(\rho^2 + z^2)^2} + \dots \right) \, .
\end{equation}
Analogous arguments to those used for the metric above show that $e^{\Phi}$ can actually be smoothly continued to the point at infinity. However, the axion diverges there:
\begin{equation}
C_0 = \frac{\omega^8 (\rho^2 + z^2)^4}{96\pi P} \left( 1- \frac{4 q_3}{3 P} \frac{7 z^2 - 3 \rho^2}{(\rho^2 + z^2)^2} + \dots  \right) \, .
\end{equation}
The divergence is such that the D$(-1)$ flux through the asymptotic $S^9$ is $N$, identifying the added point with a source of $N$ D-instantons. One could reasonably ask why this does not produce a large backreaction on the geometry. Even though the field strength from $C_0$ (and from $\Phi$) is becoming large, there is a non-trivial cancellation: the leading $\mathcal{O}(r^2)$ contribution from $\nabla_{\mu} \nabla_{\nu} \Phi$ cancels against the corresponding axion terms multiplied by $e^{2 \Phi}$. This is the asymptotic manifestation of the BPS D-instanton cancellation: even though the full solution is not a pure D-instanton, it approaches the corresponding values for the axion and dilaton near the added point, so that we have
\begin{equation}
\label{SUGRAsolutions:AsymptoticRelationPhiC0}
C_0 = e^{-\Phi} - \frac{\omega^4}{18\pi} \left( \rho^2 + 3 z^2 \right) + \frac{\eta}{54\pi} + \mathcal{O}(r^{-4}) \, .
\end{equation}
Thus, in string frame, the asymptotic region (often associated with the geometry of the dual theory) is a single point, characterized by the divergence of the axion. 

The two-form potentials will not play a very important role, but for the sake of completeness we include their expansions as well:
\begin{align}
B_2 & = - \frac{128\pi \alpha' P}{\omega^4} \frac{z^3}{(\rho^2 + z^2)^4} \left( 1 + \frac{2 q_3}{3P} \frac{13 z^2 - 7 \rho^2}{(\rho^2 + z^2)^2} + \dots \right) \omega_2 \, , \\ 
C_2 & = - \frac{8 \alpha' \omega^4 z^3}{9} \left( 1 - \frac{4 P}{3 \omega^4} \frac{\rho^2 + 61 z^2}{(\rho^2 + z^2)^4} + \dots \right) \omega_2 \, .
\end{align}

\subsection{Regimes of validity}

In order to trust the supergravity description, we must check that both string loop effects and higher-curvature corrections are small. Locally, these conditions are met when
\begin{equation}
g_{\rm loc}(x) = e^{\Phi(x)} \ll 1 \, , \qquad \frac{\alpha'}{\ell_{\rm loc}(x)^2} \ll 1 \, .
\end{equation}
Here, $g_{\rm loc}(x)$ is the local string coupling, and we have defined $\ell_{\rm loc}(x)$ as a length characterizing the smallest scale over which the fields vary. A typical estimate is given by the scalar curvature $\ell_{\rm loc}^{-2} = R$, but this might not be enough if other quantities vary faster -- as an example, in the backgrounds we consider $R \to 0$ asymptotically, but other curvature scales like $\ell_n^{-(2n+4)} = (\nabla^n R_{\mu \nu \rho \sigma})^2$ are finite and nonzero, so that the smallest of them should be taken into account; we discuss this below in more detail. It is perfectly possible that the above conditions are met within a proper subregion of the full geometry: in that case, the supergravity description can only be used there. In some cases, however, the supergravity description might be valid throughout the whole geometry. This is going to depend on the specific parameters $\{(z_s, Q_s)\}$ and $\omega$ -- or $\{(N_s, n_s\}$ and $\omega$ -- characterizing the saddle. Without solving the electrostatic problem, there are few things we can say in general.

One interesting observation we can make is the following. Suppose we have a solution in terms of a potential $V(\rho, z)$, given the parameters $\{(z_s, Q_s)\}$ and $\omega$. We can use the scaling symmetry of the Laplace equation to consider a different potential which also solves Laplace's equation
\begin{equation}
V(\rho, z) \rightarrow c L^3 V \left( \frac{\rho}{L}, \frac{z}{L} \right) \, ,
\end{equation}
where $c$ and $L$ are arbitrary constants. The new potential corresponds to a background with different parameters, namely
\begin{equation}
\label{SUGRASolutions:Scaling}
z_s \rightarrow L z_s \, , \qquad Q_s \rightarrow c L^5 Q_s \, , \qquad \omega^4 \rightarrow c \, \omega^4 \,, \qquad \eta \rightarrow c L^2 \eta \, .
\end{equation}
Scaling all terms in the metric and matter fields, it can be shown that
\begin{align}
\rd s^2 & \rightarrow L \rd s^2 \, , & e^{\Phi} & \rightarrow \frac{1}{c L^2} e^{\Phi} \, , &  C_0 & \rightarrow c L^2 C_0 \, , & B_2 & \rightarrow L B_2 \, , & C_2 & \rightarrow c L^3 C_2 \, .
\end{align}
This implies that the curvature of the new metric is suppressed by powers of $L$ with respect to that of the original one. The same is true for quantities such as $(\partial \Phi)^2$, $H_3^2$, $e^{2 \Phi} (\partial C_0)^2$ or $e^{2 \Phi} \mathcal{F}_3^2$, so sending $L \to \infty$ suppresses higher-derivative corrections. From the form of $e^{\Phi}$, we can also see that sending $c L^2 \to \infty$ suppresses string loop effects. Thus, for any fixed regular electrostatic configuration, taking $L$ and $c L^2$ to be sufficiently large produces a string-frame geometry with parametrically small curvature and string coupling throughout the complete manifold. The same scaling suppresses the matter-field invariants away from the D-instanton point. Note that, in terms of matrix model parameters, this scaling makes both $N_s$ and $n_s$ large.

Another general comment we can make concerns the asymptotic region. The string coupling decays rapidly \eqref{SUGRAsolutions:DilatonAsymptotic}, so loop effects are unimportant. Curvature corrections can be understood from the expansion \eqref{SUGRAsolutions:MetricAsymptotic}. Let us characterize the size of the charge distribution by
\begin{equation}
R_{\rm el} = \max_{s} \sqrt{\rho_s^2 + z_s^2} \, .
\label{Reldef}
\end{equation}
This parameter allows to obtain a bound for the multipole moments: using \eqref{Reldef} and the inequality $|U_{2l+1}(x)|\leq 2(l+1)|x|$ in the definition of the multipole moments, we find\footnote{This inequality assumes that the charge densities are everywhere positive in the charged balls. This is true in all cases explored below; it might be possible to prove it in general using properties of the Laplace equation and the regularity condition at the edge of the balls.}
\begin{equation}
|q_{2l+1}| \leq R_{\rm el}^{2l} P \, ,
\qquad P = 2 \sum_s z_s Q_s \, . 
\end{equation}
Incidentally, this is what guarantees that the multipole expansion \eqref{multipole} converges at sufficiently large radius. The second derivatives of the metric \eqref{SUGRAsolutions:MetricAsymptotic} at the asymptotic point are controlled by $q_3/P$ (once we remove the multiplicative prefactor), so the curvature squared can be estimated to be
\begin{equation}
\frac{P}{\omega^4} \alpha'^2 R_{\mu \nu \rho \sigma} R^{\mu \nu \rho \sigma} \sim  \frac{q_3^2}{P^2} \lesssim R_{\rm el}^4 \, .
\end{equation}
Thus, a sufficient condition for the curvature squared to be small in string units is
\begin{equation}
\label{SUGRAsolutions:ValidityAsymptoticRegion}
\frac{\omega^4  R_{\rm el}^4}{P} \ll 1 \, .
\end{equation}

More generally, we can look at other possible geometric curvature corrections. From the expansion \eqref{SUGRAsolutions:MetricAsymptotic}, it is possible to show that, at infinity:
\begin{equation}
\frac{P^{\frac{n+2}{2}}}{\omega^{2n+4}}\alpha'^{n+2} \nabla_{\alpha_1} \dots \nabla_{\alpha_n} R_{\mu \nu \rho \sigma} \nabla^{\alpha_1} \dots \nabla^{\alpha_n} R^{\mu \nu \rho \sigma} = c_n + \mathcal{O}(r^{-2}) \, ,
\end{equation}
where $c_n$ is defined by the normalization in this equation and is determined from the Taylor series of the asymptotic metric. For even $n$, $c_n$ contains a term quadratic in $q_{n+3}/P$, together with nonlinear combinations of lower multipoles; at sufficiently high orders there are also contributions from the background potential. For odd $n$, $c_n = 0$ by parity. Thus, the full tower of multipole moments is relevant to identify the characteristic curvature length scale asymptotically. Writing $c_{2l}^{\rm mult}$ for the contributions built only from source multipoles, the multipole bound gives
\begin{equation}
\frac{c_{2l}^{\rm mult} \omega^{4l+4}}{ P^{l+1}} \lesssim  \left( \frac{\omega^4 R_{\rm el}^4}{P} \right)^{l+1} \, ,
\end{equation}
where the implicit constants may depend on $l$. Therefore, the $l=0$ condition \eqref{SUGRAsolutions:ValidityAsymptoticRegion} is sufficient to guarantee that these curvature scales are parametrically small at each fixed derivative order.

Similarly, the background potential contributions that appear at $n \geq 4$ contain extra factors of $\omega^4 / P$. Given that $R_{\rm el} \geq z_s \geq 1/2$, we have that \eqref{SUGRAsolutions:ValidityAsymptoticRegion} implies $\omega^4 / P \ll 1$, and thus these terms are also under control. 

In terms of matrix model parameters, the condition \eqref{SUGRAsolutions:ValidityAsymptoticRegion} becomes parametrically
\begin{equation}
{\omega^4 R_{\rm el}^4 \over N} \ll 1 \; .
\label{asympt_mm}
\end{equation}
The parameter $R_{\rm el}$ will itself depend on $\omega$ and $\{(N_s,n_s)\}$ via the solution of the electrostatics problem. We will analyze its behavior in the case of single ball saddles below.

Control over curvature corrections does not by itself imply control of the full $\alpha'$ expansion at the added asymptotic point. The gauge-invariant three-form combinations relevant in string frame are regular: $H_3$ and $e^{\Phi} \mathcal{F}_3$ both vanish at that point. By contrast, as discussed in the previous section, the dilaton gradient and $e^{\Phi} \rd C_0$ diverge separately, but their leading contributions cancel in the two-derivative equations as a consequence of the asymptotic D-instanton relation. It is thus conceivable that this cancellation persists and the condition just stated, \eqref{SUGRAsolutions:ValidityAsymptoticRegion}, is enough to control higher-derivative corrections in the asymptotic region. But this cannot be fully established without understanding the structure of those corrections. 

\section{Matching the supergravity and matrix model on-shell actions}
\label{sec:OnShellMatching}

The correspondence between matrix model saddles and type IIB supergravity backgrounds developed in \cite{Komatsu:2024bop,Komatsu:2024ydh} had a loose end. The on-shell actions computed from supergravity and from the matrix model did not match, even in the regime of validity in which the supergravity description is valid. The result was not conclusive because a proper treatment of the boundary counterterms required to regularize the supergravity action was not developed, but a factor of $2$ mismatch in a term of the form $\sum_s Q_s V_s$ which was not expected to be affected by boundary terms seemed puzzling. In this section, we will carefully reanalyze the action matching, proposing a solution to the factor of $2$ mismatch, together with boundary counterterms that regularize the supergravity action. With this proposal, the on-shell actions perfectly match. Throughout this section we set $\eta=0$.

The result for the matrix model on-shell action obtained in \cite{Komatsu:2024ydh} using the collective field reformulation of the moduli space integrals is, in our conventions,
\begin{equation}
\label{OnShellMatching:MatrixModelAction}
S_{\rm on-shell} = \frac{1}{27} \sum_{s} Q_s V_s - \frac{\omega^4}{36} q_3 = \frac{2}{27} E \, ,
\end{equation}
where $E$ is the electrostatic energy of the configuration. This energy is given by separate contributions from the potential generated by the charges and the background potential, $E = U + U_{\rm bg}$, with
\begin{equation}
U = \frac{1}{2} \sum_s \int_s \rd^3 x \, \sigma_s (x) \psi(x) \, , \qquad U_{\rm bg} = \sum_s\int_s \rd^3 x \, \sigma_s (x) V_{\rm bg}(x) = - \frac{3\omega^4 q_3}{4} \, ,
\end{equation}
where we used that the form of $V_{\rm bg}$ selects the octopole contribution in $U_{\rm bg}$. Furthermore, since $V = \psi + V_{\rm bg} = V_s$ is constant in each conducting ball, $2U + U_{\rm bg} = \sum_s Q_s V_s$. This is enough to prove the identity between on-shell action and electrostatic energy in \eqref{OnShellMatching:MatrixModelAction}.

Let us denote by $\mathcal{I}^{\rm bulk}$ the bulk supergravity action; see appendix \ref{app:TheoryAndEoM} for its form. We show in \eqref{TheoryAndEoM:OnShellAction} that on-shell it is a total derivative
\begin{equation}
\label{OnShellMatching:BulkTotalDerivative}
\mathcal{I}^{\rm bulk}_{\rm on-shell} = \frac{1}{2 \kappa_0^2} \int_M \rd \left[ \frac{9}{2} e^{-2 \Phi} \star \rd \Phi + \frac{1}{4} B_2 \wedge \left( e^{-2 \Phi} \star H_3 + C_0 \star \mathcal{F}_3 \right) - \frac{1}{4} C_2 \wedge \star \mathcal{F}_3 \right] \, .
\end{equation}
We will also add a Gibbons-Hawking-York (GHY) boundary term. Its form in string frame (the counterpart of the usual GHY term in Einstein frame) is
\begin{equation}
\mathcal{I}^{\rm GHY} = - \frac{1}{\kappa_0^2} \int_{\partial M} \sqrt{h} \, e^{-2 \Phi} K \, ,
\end{equation}
with $h$ the induced metric on the boundary and $K$ the trace of the extrinsic curvature computed with respect to the outward-pointing normal.

The geometry and dilaton are regular at the balls $z = z_s$ and at $\rho = 0$, so both the first term in \eqref{OnShellMatching:BulkTotalDerivative} and the GHY term can be ignored there. Said it another way, these are not actual boundaries of the geometry. But at $z = z_s$ the $S^2$ shrinks to zero size, and the two-form potentials are not regular there: they do not integrate to zero. Indeed, we have that
\begin{equation}
B_2 = -2\pi\alpha' z_s \, \omega_2 + \mathcal{O}((z-z_s)^3) \, , \qquad C_2 = \frac{\alpha'}{27} V_s \, \omega_2 + \mathcal{O}((z-z_s)^3) \, ,
\end{equation}
so
\begin{equation}
\label{OnShellMatching:IntegralsOverS2}
\lim_{z \to z_s} \int_{S^2} B_2 = -8\pi^2\alpha' z_s \, , \qquad \lim_{z \to z_s} \int_{S^2} C_2 = \frac{4\pi\alpha'}{27} V_s \, .
\end{equation}
Therefore, we consider a small tube surrounding each ball as part of the boundary of $M$ and evaluate the two last terms in \eqref{OnShellMatching:BulkTotalDerivative}. This tube is topologically $S^2 \times \Sigma_{7,s}$, with $\Sigma_{7,s} \cong S^7$ a $7$-cycle that goes around each ball. The $7$-forms in the last two terms of \eqref{OnShellMatching:BulkTotalDerivative} integrated on $\Sigma_{7,s}$ measure F1 and D1 charges, respectively, so the contribution from each ball is
\begin{equation}
\left. \mathcal{I}_{\rm on-shell} \right|_s = -\pi z_s N_{F1,s} - \frac{1}{54}V_s N_{D1,s} \, .
\end{equation}
One could worry that a similar contribution appears close to the $\rho = 0$ axis, where the $S^6$ collapses. This is not the case because all fields are regular there.

It is useful to rewrite everything in electrostatic variables. For this, we use the results for the charges of the solutions presented in appendix \ref{app:ElectrostaticSolutions}. The D1 charge is identified with the charge of the balls, $N_{D1,s} = Q_s = n_s$. For the F1 charge, we find a non-vanishing charge, contradicting the original analysis of \cite{Komatsu:2024bop} -- this agrees with the more recent analysis of \cite{DHoker:2026eox}. After some manipulations, we show in \eqref{ElectrostaticSolutions:F1Charge} that $N_{F1,s} = F_s/(54\pi)$, with $F_s$ the total vertical force experienced by the $s$-th conducting ball. The contribution to the action from the small tube around that ball is then
\begin{equation}
\left. \mathcal{I}_{\rm on-shell} \right|_s = - \frac{1}{54} \left( z_s F_s + Q_s V_s \right) \, .
\end{equation}

To complete the calculation of the on-shell action, we need also the contribution from the asymptotic region at $r = \sqrt{\rho^2 + z^2} = R \to \infty$. Here the GHY term is important. This calculation was done in \cite{Komatsu:2024bop}, and our result agrees with theirs.\footnote{Up to an overall sign. This sign appears because the authors of \cite{Komatsu:2024bop} work with the Lorentzian action evaluated on the Euclidean solution, but ignore the factor of $-i$ that comes from Wick rotating $t = -i \tau$. Their saddles are thus weighted by $e^{i(-i)S_L} = e^{S_L}$, with $S_L$ the Lorentzian action evaluated on the Euclidean solution, while ours are weighted by $e^{-\mathcal{I}}$.} It is
\begin{equation}
\left. \mathcal{I}_{\rm on-shell} \right|_{\infty} = - \frac{3\omega^8 R^8}{64} + \frac{7\omega^4 P R^2}{45} - \frac{7\omega^4 q_3}{135} \, .
\end{equation}

A final step to compare with \eqref{OnShellMatching:MatrixModelAction} is to eliminate the force term in favor of other quantities of the electrostatic configuration. We can do this via a scaling argument similar to the one in \eqref{SUGRASolutions:Scaling}, but now keeping the charge in each ball fixed. Rescale $(\rho, z) \to (b \rho, b z)$, while simultaneously taking $\omega^4 \to b^{-5} \omega^4$. This gives another configuration with potential $V \to b^{-2} V$ and charge densities $\sigma_s \to b^{-3} \sigma_s$. The energy thus satisfies $E(Q_s, b z_s, b^{-5} \omega^4) = b^{-2} E(Q_s, z_s, \omega^4)$. Differentiating at $b = 1$ and using $\partial_{z_s} E = - F_s$ and $\partial_{\omega^4} E = -3q_3/4$,\footnote{This final derivative requires some care. Given a charge configuration, write the energy functional as $E[\sigma; \omega^4] = (1/2) \sum_{s,s'} \int \sigma_s(x) G(x,y) \sigma_{s'}(y) + \sum_s \int \sigma_s(x) V_{\rm bg}(x)$, with $G(x,y)$ the four-dimensional Green function that includes the image charges to implement the appropriate condition at $z=0$. Note that $\delta E/\delta\sigma_s(x) = V(x) = V_s$, so the total derivative with $\omega^4$ is $\rd E / \rd \omega^4 = \partial_{\omega^4} E + \sum_s V_s \rd Q_s / \rd \omega^4$. Changes of $\rho_s$ do not contribute because $\sigma_s(\rho_s) = 0$. At fixed values of $Q_s$ and $z_s$, the only change with $\omega^4$ is thus through the explicit dependence in $V_{\rm bg}$, and this gives $\left.\partial_{\omega^4} E \right|_{Q_s, z_s} = -3q_3/4$.} we get
\begin{equation}
- \sum_s z_s F_s + \frac{15\omega^4 q_3}{4} = - 2 E \, ,
\end{equation}
or, using the expression for $E$ previously found,
\begin{equation}
\sum_s z_s F_s = \sum_s Q_s V_s + 3\omega^4 q_3 \, .
\end{equation}
Collecting all previous results, we finally write the full supergravity on-shell action as
\begin{equation}
\label{OnShellMatching:OnShellActionRaw}
\mathcal{I}_{\rm on-shell} = - \frac{3\omega^8 R^8}{64} + \frac{7\omega^4 P R^2}{45} -\frac{1}{27} \sum_s Q_s V_s - \frac{29\omega^4 q_3}{270} \, .
\end{equation}

The first two terms diverge as $R \to \infty$ and must be removed by counterterms at the asymptotic region. The two finite terms have the right form, but the coefficients do not match \eqref{OnShellMatching:MatrixModelAction}; in particular, the $\sum_s Q_s V_s$ piece has the wrong sign. So we have turned a factor of $2$ issue into a sign issue. However, there is something we have not taken into account yet. As usual, the action used for the gauge potentials implies we are working in an ensemble where we keep the potentials fixed but the charges can vary -- here, charges mean electric charges associated to the gauge potentials like $N_{D1,s}$ or $N_{F1,s}$, not just $Q_s$. We can explicitly see this as follows. If we vary the Euclidean action, the boundary terms that come with variations of the potentials $C_2$ and $B_2$ are
\begin{equation}
\delta \mathcal{I}^{\rm bulk} \supset \frac{1}{2 \kappa_0^2} \int_M \rd \left[  - \delta C_2 \wedge \star \mathcal{F}_3 + \delta B_2 \wedge \left( e^{-2 \Phi} \star H_3 + C_0 \star \mathcal{F}_3 \right) \right] \, .
\end{equation}
The full variation is presented in appendix \ref{app:TheoryAndEoM}. With no additional boundary terms, having a stationary action requires to fix the boundary values of $C_2$ and $B_2$.

We will argue that this is not the appropriate ensemble for the compact RR modes describing a fixed matrix-model saddle. These modes are associated with nontrivial cycles in the interior of the geometry, rather than with the asymptotic boundary. In particular, the $S^2$ shrinks smoothly on the conducting balls, but the two-form potentials need not admit a globally regular gauge choice. Just like before, we deal with this by introducing artificial cuts separating different gauge patches -- these cuts are not physical boundaries, and the final result can equivalently be expressed in terms of periods over compact cycles, as we will later show. To see which boundary conditions are naturally imposed, evaluate the variation above on our family of solutions; using \eqref{OnShellMatching:IntegralsOverS2}, we obtain
\begin{equation}
\label{OnShellMatching:CompactVariation}
\delta \mathcal{I}^{\rm bulk} \supset -\sum_s \left(\frac{2}{27}N_{D1,s} \, \delta V_s + 4\pi N_{F1,s} \, \delta z_s\right) \, .
\end{equation}
Although this has been evaluated on the space of solutions, its interpretation as a canonical variation follows from the boundary terms in the variation of the original action, before imposing the equations of motion. In particular, we are not inferring the ensemble by differentiating the on-shell action in \eqref{OnShellMatching:OnShellActionRaw}.

Equation \eqref{OnShellMatching:CompactVariation} shows that the original action naturally fixes the potentials $V_s$, which are conjugate to the D1 charges $N_{D1,s} = n_s$. However, when computing the contribution of a given matrix-model saddle, the multiplicities $n_s$ are prescribed, while the potentials $V_s$ are determined dynamically by the electrostatic problem. We must therefore change the ensemble of these compact RR modes, from fixed potential to fixed D1 electric flux. This is implemented by adding the Legendre-transform term
\begin{equation}
\label{OnShellMatching:LegendreTerm}
\mathcal I_{\rm Leg} = \frac{2}{27}\sum_s N_{D1,s} V_s = \frac{2}{27} \sum_s Q_s V_s \, .
\end{equation}
After adding this term, the resulting variational principle is compatible with fixing the D1 charges, as required. Notice that we have not changed the ensemble for $B_2$. Its conjugate electric charges $N_{F1,s}$ are not independent labels of the matrix-model saddle, whereas the quantities $z_s=N_s/2$ are prescribed by the representation. Thus the natural choice is to keep the corresponding NS potential fixed, rather than the F1 charges. 

It is worth emphasizing that \eqref{OnShellMatching:LegendreTerm} should not be understood as an integral over new physical boundaries at the conducting balls. More invariantly, we can define the compact RR periods and their conjugate electric fluxes by
\begin{equation}
\mathfrak{a}_s = \frac{1}{2\pi\alpha'} \int_{\hat{\Sigma}_{3,s}} \rd C_2 = \frac{2}{27} V_s \, , \qquad N_{D1,s} = \frac{1}{(2\pi)^6 \alpha'^3} \int_{\Sigma_{7,s}} \star \mathcal{F}_3 = Q_s \, ,
\end{equation}
where $\hat{\Sigma}_{3,s}$ is the $3$-cycle between the grounded plane and the $s$-th charged ball, while $\Sigma_{7,s}$ is the seven cycle enclosing the $s$-th ball. Then the Legendre transform takes the form
\begin{equation}
\mathcal I_{\rm Leg} = \sum_s N_{D1,s} \mathfrak{a}_s \, .
\end{equation}
This is the pairing between the compact RR three-form periods and their dual electric fluxes. It makes the ensemble change independent of the particular choice of gauge patches or artificial cuts. 

Adding \eqref{OnShellMatching:LegendreTerm} to the on-shell action, we get
\begin{equation}
\mathcal{I}_{\rm on-shell} + \mathcal{I}_{\rm Leg} = - \frac{3\omega^8 R^8}{64} + \frac{7\omega^4 P R^2}{45} + \frac{1}{27} \sum_s Q_s V_s - \frac{29\omega^4 q_3}{270} \, .
\end{equation}
The coefficient of $\sum_sQ_sV_s$ now agrees with the matrix-model result. The remaining discrepancy, involving $q_3$, must be resolved by examining the counterterms at the asymptotic boundary.

We will introduce local counterterms evaluated on the nine-dimensional surface $\Sigma_R=\{r=R\}$ surrounding the asymptotic point. It will be simplest to express the results in terms of the Einstein frame boundary metric and a combination of the dilaton and axion:
\begin{equation}
  h^{(E)}_{ab}=e^{-\Phi/2}h^{(s)}_{ab} \, ,
  \qquad
  W=e^{-\Phi}-C_0 \, .
  \label{eq:local-ct-definitions}
\end{equation}
The curvature scalar of the Einstein-frame boundary metric will appear repeatedly below; it is related to that of the string-frame metric by
\begin{equation}
\mathcal{R}[h^{(E)}] =  e^{\Phi/2}
  \left[
    \mathcal R[h^{(s)}]
    +4D_s^2\Phi
    -\frac{7}{2}(D_s\Phi)^2
  \right] \, ,
\end{equation}
where in these expressions we view $\Phi$ and $C_0$ as scalars restricted to $\Sigma_R$, $D_a$ denotes the covariant derivative built from $h_{ab}$, and an $s$ subscript or superscript denotes the corresponding string-frame quantities.

The asymptotic induced metric associated with any electrostatics solution takes the
form
\begin{equation}
  h^{(E)}_{ab}
  =\alpha'\sqrt{2\pi}\omega^2 R^2
    \bigl[\gamma_{ab}+O(R^{-6})\bigr],
  \label{eq:local-ct-metric}
\end{equation}
with $\gamma_{ab}$ the metric on a unit $S^9$. We note that the
$O(R^{-2})$ correction in the string-frame metric cancels against the dilaton Weyl factor. Since $\mathcal R[\gamma]=72$, it will be convenient work with a rescaled curvature
\begin{equation}
\chi = {\mathcal{R}[h^{(E)}] \over 72}  \; .
\end{equation}
We will see that we can  properly regulate the divergences and match the finite matrix model result using the following local counterterm action:\footnote{For the last term in brackets, we could alternatively use 
\begin{equation}
\frac{11}{280\sqrt{\chi}}
\left(
D_a\Phi D^a\Phi
-e^{2\Phi}D_aC_0 D^aC_0
\right) \; .
\end{equation} as explained in Appendix \ref{app:counterterms}}
\begin{equation}
\mathcal I_{\mathrm{ct}}^{(9)}
  =
  \frac{1}{\kappa_0^2}
  \int_{\Sigma_R} \rd^9x\,\sqrt{h^{(E)}}\,
  \left[
    9\sqrt{\chi}
    -2e^\Phi\sqrt{\chi}\,W
    -\frac{11}{140}\frac{e^\Phi}{\sqrt{\chi}}\,
      D_aW D^a\Phi
  \right] ,
  \label{eq:local-ct-action}
\end{equation}
with covariant derivatives built and contracted using the Einstein frame boundary metric. The non-polynomial dependence on curvature is a well known feature of gravitational counterterm actions. In particular, the first term here including the coefficient is the standard counterterm that regularizes the Gibbons--Hawking--York boundary term for a spherical boundary in asymptotically flat spacetime \cite{Lau:1999dp,Mann:1999pc,Kraus:1999di,Mann:2005yr}.

To evaluate \eqref{eq:local-ct-action}, let
$\mathcal I_{\mathrm{ct},i}$ denote the contribution of its $i$th term.
Using $2\kappa_0^2=(2\pi)^7\alpha'^4$ and our earlier expressions for the asymptotic fields gives
\begin{align}
  \mathcal I_{\mathrm{ct},1}
  &=
  \frac{3\omega^8R^8}{64}
  -\frac{\omega^4PR^2}{15}
  +\frac{31\omega^4q_3}{180}
  +o(1) \, ,
  \label{eq:local-ct-expansion-1}
  \\
  \mathcal I_{\mathrm{ct},2}
  &=
  -\frac{4\omega^4PR^2}{45}
  -\frac{7\omega^4q_3}{135}
  +o(1) \, ,
  \label{eq:local-ct-expansion-2}
  \\
  \mathcal I_{\mathrm{ct},3}
  &=
  -\frac{11\omega^4q_3}{270}
  +o(1) \, .
  \label{eq:local-ct-expansion-3}
\end{align}
Summing the three contributions,
we obtain
\begin{equation}
  \mathcal I_{\mathrm{ct}}^{(9)}
  =
  \frac{3\omega^8R^8}{64}
  -\frac{7\omega^4PR^2}{45}
  +\frac{43\omega^4q_3}{540}
  +o(1).
  \label{eq:local-ct-total}
\end{equation}

Combining all the contributions to the gravitational action, we now find
\begin{equation}
  \begin{split}
    \mathcal I_{\mathrm{ren}}
    &=
    \lim_{R\to\infty}
    \left[
      \mathcal I_{\mathrm{on-shell}}(R)
      +\mathcal I_{\mathrm{Leg}}
      +\mathcal I_{\mathrm{ct}}^{(9)}(R)
    \right]
    \\
    &=
    \frac{1}{27}\sum_s Q_sV_s
    -\frac{\omega^4q_3}{36},
  \end{split}
  \label{eq:local-ct-renormalized-action}
\end{equation}
which agrees with the matrix-model result \eqref{OnShellMatching:MatrixModelAction}.

The coefficients for $\mathcal{I}_{\mathrm{ct},1}$ and $\mathcal{I}_{\mathrm{ct},2}$ are fixed by demanding that the divergences cancel. While $\mathcal{I}_{\mathrm{ct},3}$ only gives a finite contribution, its coefficient is  
fixed by a variational principle, requiring the boundary variation to vanish for arbitrary
$\delta q_3$ at fixed $\omega$, $P$, and $\eta$ (corresponding to the parameters of the matrix model), as we show in Appendix~\ref{app:counterterms}. Thus, the agreement for the $q_3$ term between the matrix model and supergravity is non-trivial.

To close, we note that we have only checked that the counterterm prescription works for configurations of the type described by the electrostatic problem. It is unclear whether it will work for other configurations satisfying the same asymptotic behavior, for example perturbations of the fields that take us out of the family of electrostatic solutions. Further, as usual there is some degree of arbitrariness in choosing finite counterterms: for example, we are always free to add a function of fixed data, say $\omega^4$ or $P=N$.  It would be useful to perform a more systematic holographic renormalization analysis, along the lines of those developed for non-conformal D$p$-brane backgrounds in \cite{Kanitscheider:2008kd,Wiseman:2008qa}. 

\section{An exact Coulomb field reformulation of the localization integral}
\label{sec:scalar-localization}

In this section, we point out an exact reformulation of the localization integral as the partition function of a statistical gas of rigid, dipole-like charge distributions interacting in two dimensions via a Coulomb field. There is one such object for each irreducible block of
$\mathcal R$: its dimension fixes the height of the object, while its
localization modulus gives its horizontal position. Related observations in the context of simpler matrix models have appeared in \cite{Kazakov:1998ji}. The electrostatics description that appears in the supergravity picture appears in a long-wavelength limit where we introduce a collective distribution of the dipoles.

\subsection{The 1D localization gas}
Consider a representation
$\mathcal R$ with $n_s$ copies of the $N_s$-dimensional irreducible
representation, with $N=\sum_s n_s N_s$. Let
$a=(s,i)$ label the individual copies, let $\rho_a=\rho_{si}$ be the modulus corresponding to representation $a$ and $b_a=N_s$. In this case, the localization integral (\eqref{MatrixSetup:FixedSectorObservable} with $f=1$) takes the form of a partition function for a 1D gas of particles, 
\begin{eqnarray}
 Z_{\mathcal R} 
 &=&C_{\mathcal R}\int\prod_a\rd \rho_a \exp^{-\beta E(\{\rho_a\})} \, , \cr
E(\{\rho_a\}) &=& \sum_a U_{b_a}(\rho_a)
             + \sum_{a<a'}g_{b_ab_{a'}}(\rho_a-\rho_{a'}) \, , \qquad \qquad \beta = {2 \over 27} \; .
 \label{eq:sl-gas}
\end{eqnarray}
We have different particle types associated with the different representation sizes $b_a$. We have potential energy
\begin{equation}
 U_b(\rho)=\frac{3\omega^4}{16}(12b\rho^2-b^3+b) \, ,
 \label{eq:sl-one-body}
\end{equation}
for particles of type $b$, 
and a pairwise interaction energy
\begin{equation}
 g_{bb'}(\rho)=-{27 \over 2} \sum_{J=|b-b'|/2}^{(b+b'-2)/2}
 \log\frac{\big[(J+2/3)^2+\rho^2\big]^3
                 \big[J^2+\rho^2\big]}
                {\big[(J+1/3)^2+\rho^2\big]^3
                 \big[(J+1)^2+\rho^2\big]} \; ,
 \label{eq:sl-pair}
\end{equation}
between particles of type $b$ and $b'$. We have made the choice $\beta = 2/27$ since we will later make contact with the 4D electrostatics picture and we saw in the previous section that the action and 4D electrostatics energy are related by $S = (2/27)E$.

The normalization constant is \cite{Komatsu:2024ydh}
\begin{equation}
 \begin{gathered}
 C_{\mathcal R}=A_N\prod_{s=1}^{q}\frac{w_{N_s}^{n_s}}{n_s!},
 \qquad
 A_N=\frac{(2\pi)^{5N^2+N/2}}{\prod_{k=1}^{N-1}k!},
 \\
 w_b=\frac{4}{\pi b}
 \prod_{J=1}^{b-1}\left(\frac{3J+2}{3J+1}\right)^3.
 \end{gathered}
 \label{eq:sl-weights}
\end{equation}
The full partition function is $Z=\sum_{\mathcal R}Z_{\mathcal R}$ as
before. 

\subsection{The 2D dipole gas}

A key observation is that the Fourier transform of the pair interaction potential
has a very simple and suggestive form. Using the Fourier transform convention
$\widehat f(p)=\int_{\mathbb R}\rd \rho\,e^{-ip\rho}f(\rho)$, and defining
$z_b=b/2$, we can Fourier transform \eqref{eq:sl-pair} and sum the resulting
finite geometric series to give
\begin{equation}
 \widehat g_{bb'}(p)
 =d(p)^2\widehat G_D(p;z_b,z_{b'}) \, ,
 \label{eq:sl-fourier}
\end{equation}
where 
\begin{equation}
 \widehat G_D(p;z,z')=\frac{\pi}{|p|}
 \left(e^{-|p||z-z'|}-e^{-|p|(z+z')}\right) \, , \qquad d(p)=\frac{6\sinh(p/6)}
 {\sqrt{1+\frac43\sinh^2(p/6)}} \, .
 \label{eq:sl-green}
\end{equation}
The latter expression is exactly the Green function for a 2D Coulomb field in the presence of a grounded line at $z=0$, after Fourier transforming the horizontal coordinate. In position space, we have:
\begin{equation}
 G_D(x,z;x',z')=\frac12
 \log\frac{(x-x')^2+(z+z')^2}{(x-x')^2+(z-z')^2},
 \qquad -\Delta_XG_D(X,Y)=2\pi\delta^{(2)}(X-Y).
 \label{eq:sl-green-position}
\end{equation}
We can also interpret $G_D(X,Y)$ as the Coulomb interaction energy for a pair of point charges at $X$ and $Y$. The numerator here can be interpreted as arising from interaction between a charge and the image charge for the other charge.

\begin{figure}[tb]
 \centering
 \includegraphics[width=0.6\linewidth]{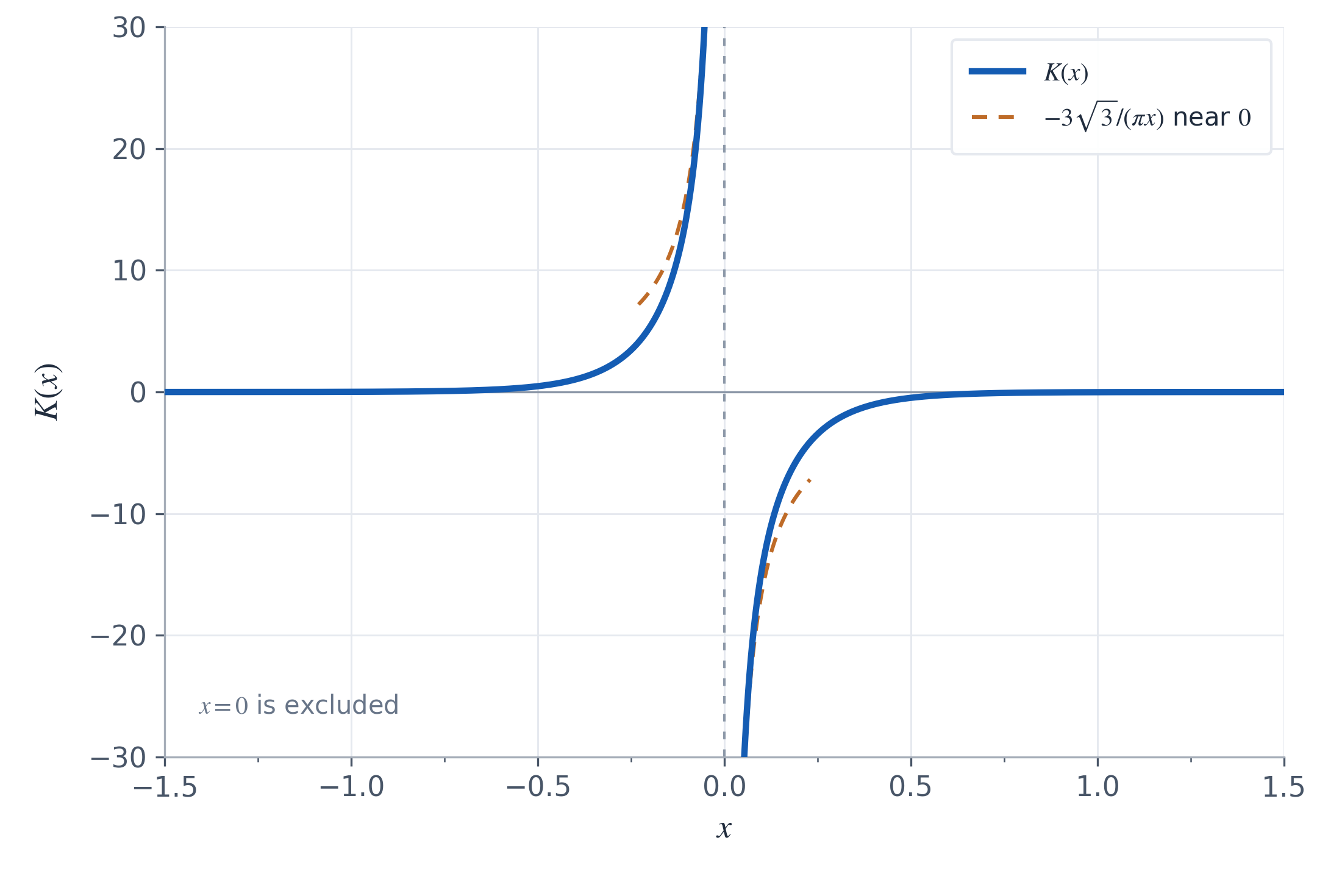}
 \caption{Left: Dipole charge distribution $K(x)$ (solid) and its asymptote near $x=0$.}
 \label{fig:sl-kernel}
\end{figure}

The interaction potential $g_{bb'}(x,x')$ can then be interpreted as the interaction potential between a pair of identical charge distributions at heights $z_b$ and $z_{b'}$ with horizontal positions $x$ and $x'$,
\begin{equation}
g_{b b'}(x_b-x_b') = \int_{\mathbb H}\rd^2X\int_{\mathbb H}\rd^2Y\,
 j_a(X)G_D(X,Y)j_{a'}(Y) \, ,
 \label{eq:sl-cross-energy}
\end{equation}
where $\mathbb H$ denotes the upper half-plane and the charge distribution is 
\begin{equation}
 j_a(\rho,z)=K(\rho_a-\rho)\delta(z-z_{b_a}) \, , \qquad  K(x)=i\int_{\mathbb R}\frac{\rd p}{2\pi}\,e^{ip x} d(p) \, .
 \label{eq:sl-charge-profile}
\end{equation}
The charge is confined to height $z_b$ and spread out horizontally in a distribution that we can write explicitly as
\begin{equation}
 K(x)=-9\sqrt3\,
 \frac{\operatorname{Re}{}_2F_1(-3ix,1+3ix;1;1/4)}
 {\sinh(3\pi x)}.
 \label{eq:sl-kernel-closed}
\end{equation}
The function is plotted in Figure \ref{fig:sl-kernel}.
Its limiting behavior is\footnote{When integrating $K$ against another function, we use the principle value prescription to deal with the singularity at $x=0$.}
\begin{equation}
 K(x)=-\frac{3\sqrt3}{\pi x}+O(x)\quad(x\to0),
 \qquad
K(x)=O\!\left(|x|^{-1/2}e^{-2\pi|x|}\right)
 \quad(|x|\to\infty).
 \label{eq:sl-kernel-asymptotics}
\end{equation}
This distribution is plotted in Figure \ref{fig:sl-kernel}. The small $p$ behavior of $d(p)$,
\begin{equation}
d(p) = p - {p^3 \over 72} + {\cal O}(p^5)
\label{dexp}
\end{equation}
and specifically $d(0)=0$ and $d'(0)=1$ imply that
\begin{equation}
 \int_{\mathbb R} \rd x\,K(x)=0,
 \qquad
 \int_{\mathbb R} \rd x\,x K(x)=-1.
 \label{eq:sl-dipole-moment}
\end{equation}
so each distribution has zero total charge and unit dipole moment.

The one-body potential $U(\rho)$ in \eqref{eq:sl-gas} also has an exact electrostatics interpretation. Defining the background potential
\begin{equation}
 W_\omega(\rho,z)
 =\frac32\omega^4(z\rho^3-z^3\rho)+\frac{\omega^4}{4}z\rho \; ,
 \label{eq:sl-background}
\end{equation}
we can check that this 
vanishes at the grounded boundary and obeys
\begin{equation}
\int_{\mathbb H}\rd^2X\,j_a(X)W_\omega(X)
 =U_{b_a}(\rho_a)
 \label{eq:sl-background-identity}
\end{equation}
where again we can use the small $p$ behavior of $d$ to integrate $K$ against the polynomial $W_{\omega}$.

All together, we have that the energy appearing in \ref{eq:sl-gas} is exactly the electrostatic energy of a gas of identical dipoles with horizontal position $\rho_a$ and height $z_{b_a}$ sitting above a grounded plane at $z=0$ with a background potential $W_{\omega}$. This excludes any self-interaction energy for the dipoles. 

For a given configuration of dipoles, we can write the full Coulomb potential as
\begin{equation}
 W(X)=W_\omega(X)+\int_{\mathbb H}\rd^2Y\,G_D(X,Y)J_{\mathcal R}(Y),
 \qquad J_\mathcal{R} = \sum_a j_a
 \label{eq:sl-coulomb-potential}
\end{equation}
where the first term is the background potential and the second term (that we will call $W_{\text{src}}$) is the part sourced by the dipoles. The function $W(X)$ satisfies Laplace's equation everywhere away from the horizontal lines $z = z_{b_a}$.

\subsection{Emergence of the conducting ball description}
\label{subsec:sl-electrostatic-limit}

We can understand the electrostatics picture of
section~\ref{sec:electrostatics} as a limit where we consider scales much larger than the characteristic size of a single dipole object and where the number of dipoles associated with each representation size is large so that we can approximate their distribution by a continuous density. In the appropriate limit that we will explain, the entropy of configurations in our statistical model is suppressed relative to the energy and the integral is dominated by a saddle-point that is simply the energy minimizing configuration. 

We consider a limit where the number of copies of each representation is large, and 
replace the configuration of dipoles at height $z_s=N_s/2$
by a smooth density $f_s(\rho)$, normalized by
$\int_{\mathbb R}\rd\rho\,f_s(\rho)=n_s$. 

If we consider a limit
\begin{equation}
 n_s=L\bar n_s \qquad L \to \infty
 \label{eq:sl-saddle-scaling}
\end{equation}
with $\bar{n}_s$ and $N_s$ fixed, the scaling
\begin{equation}
\omega^4 = L \bar{\omega}^4
\end{equation}
with $\bar\omega$ fixed, results in both terms in the energy scaling as $L^2$ for a given distribution while the entropy associated with a distribution (based on the volume of microscopic configuration space giving a particular distribution) scales as $L\log L$. Thus, the entropy term is suppressed in this limit and the integral will be dominated by the energy-minimizing saddle which has a distribution for which $f_s/L$ has a limit with finite width and some fixed shape. 

In terms of the continuous distribution, we can write the source as
\begin{equation}
 J(\rho,z)=\sum_s\int_{\mathbb R}\rd\hat\rho\,
 f_s(\hat\rho)K(\hat\rho-\rho)\delta(z-z_s),
 \label{eq:sl-collective-source}
\end{equation}
and the potential $W$ is given by~\eqref{eq:sl-coulomb-potential}
with $J_{\mathcal R}$ replaced by $J$. The energy of a
dipole of type $s$ at horizontal position $x$ in this field is
\begin{equation}
 \varepsilon_s(x)=\int_{\mathbb R}\rd\rho\,
 K(x-\rho)W(\rho,z_s).
 \label{eq:sl-insertion-energy}
\end{equation}
Requiring that such a dipole is in equilibrium, i.e. that the horizontal force on the dipole vanishes, gives
$\partial_x\varepsilon_s=0$ in the support of $f_s$. Thus
\begin{equation}
 \varepsilon_s(\rho)=\lambda_s
 \qquad\hbox{on the support of }f_s.
 \label{eq:sl-field-boundary}
\end{equation}
For a minimizing configuration, moving a dipole outside
the support cannot lower the energy, so we must have
$\varepsilon_s(\rho)\geq\lambda_s$ there. 

We can now consider a further limit where the fields and densities vary a negligible amount on the scale of a dipole (whose size is order 1). This will be true if the widths of the various distributions are much larger than 1.

Using the small-$p$ expansion of $d(p)$ in \eqref{dexp}, we find that 
\begin{equation}
 \int_{\mathbb R}\rd\rho\,K(x-\rho)F(\rho)
 =F'(x)+\frac1{72}F'''(x)+\cdots
 \label{eq:sl-profile-expansion}
\end{equation}
for a slowly varying function $F$. Applying this to $\varepsilon_s$, we have
\begin{equation}
\varepsilon_s(\rho)\simeq\partial_\rho W(\rho,z_s)\, ,
 \label{eq:sourceapprox}
\end{equation}
so our equilibrium condition becomes
\begin{equation}
 \partial_\rho W(\rho,z_s)=\lambda_s
 \qquad\hbox{on the support of }f_s.
 \label{eq:sl-constant-field}
\end{equation}
The horizontal electric field is constant along each occupied
interval. This has a direct physical interpretation: a dipole only feels
a force when the electric field varies along its direction.

For the saddle configuration, the density $f_s$ is even and $W$ is odd in $\rho$ (since we are considering a distribution of dipoles). Thus, we can define
\begin{equation}
 W(\rho,z)=\rho V(\rho,z)\, ,\qquad \rho\geq0 \, ,
 \label{eq:sl-electrostatic-dictionary}
\end{equation}
such that $V$ is non-singular at $\rho = 0$. Away from the sources, the two-dimensional Laplace equation becomes
\begin{equation}
 0=(\partial_\rho^2+\partial_z^2)(\rho V)
 =\rho\left(\partial_\rho^2+\frac2\rho\partial_\rho
 +\partial_z^2\right)V.
 \label{eq:sl-laplace}
\end{equation}
which is exactly the axially-symmetric 4D Laplace equation that we considered earlier. On the support $[-\rho_s, \rho_s]$ of the dipole distribution at height $z_s$, we have $W(\rho,z_s)=\lambda_s\rho$, so $V$ is constant on $z = z_s$, $\rho < \rho_s$; this can be interpreted as a 4D conducting ball. The grounded line at $z=0$ becomes the
grounded plane of the four-dimensional problem.

Using $-\Delta W=2\pi J$, the jump in the normal derivative determines
the charge density on the ball:
\begin{equation}
 \sigma_s(\rho)
 =-\frac{[\partial_zV]^+_-}{4\pi^2}
 =-\frac{[\partial_zW]^+_-}{4\pi^2\rho}
 =-\frac{f_s'(\rho)}{2\pi\rho}.
 \label{eq:sl-density-dictionary}
\end{equation}
Since $f_s$ vanishes at its endpoints, the total charge is
\begin{equation}
Q_s=4\pi\int_0^{\rho_s}\rd\rho\,\rho^2\sigma_s(\rho)
 =-\int_{-\rho_s}^{\rho_s}\rd\rho\,\rho f_s'(\rho)
 =n_s.
 \label{eq:sl-ball-charge}
\end{equation}
The vanishing of $\sigma_s$ at the ball's edge follows naturally from the dipole picture, where the edge of the distribution is allowed to vary dynamically. A non-vanishing charge density at the edge of the conducting ball generically has
an inverse-square-root singularity in the charge density, which corresponds to 
$f_s(\rho)\sim A_s\sqrt{\rho_s-\rho}$. We can check that in this case, the condition that $\partial_\rho W\geq\lambda_s$ past the edge of the distribution is violated and the 2D electrostatic energy can be lowered by moving a dipole infinitesimally from inside the distribution to outside of it.

Finally, the background potential~\eqref{eq:sl-background}
gives
\begin{equation}
 V_{\rm bg}(\rho,z)
 =\frac32\omega^4(z\rho^2-z^3)+\frac{\omega^4}{4}z.
 \label{eq:sl-matched-background}
\end{equation}
The linear term corresponds to $\eta=-\omega^4/4$ in
\eqref{Vbg}.\footnote{In section \ref{sec:OnShellMatching}, we established the action matching with $\eta=0$, following the leading collective-field calculation of \cite{Komatsu:2024ydh}. The exact dipole-gas description instead selects $\eta_{\star} = -\omega^4/4$; this difference, together with the finite-dipole smearing correction, accounts for a contribution $\omega^4N/72$ to the one-body action. This term is common to all saddles at fixed $N$ and is subleading when the representation dimensions are large, so it does not affect the leading action matching. More generally, shifting $\eta_\star$ to an arbitrary $\eta$ amounts to adding $-(\eta-\eta_\star)\Tr(\mathbf{1})/27$ to the matrix-model action. This dependence can be reproduced on the supergravity side by replacing $W=e^{-\Phi}-C_0$ with $W+\eta/(54\pi)$ in the second counterterm of \eqref{eq:local-ct-action}.}

We emphasize that there are three separate assumptions required for the dipole gas model to be well approximated by the 4D electrostatics picture \cite{Komatsu:2024ydh}. For each representation type, the width of the distribution of dipoles should be much larger than 1, the density of dipoles in this distribution should be large, and the entropy should be suppressed relative to energy. For the case where we have $n_\star$ copies of the $N_\star$-dimensional irreducible representation, we need, in addition to the entropy condition above, that $1 \ll \rs \ll n_\star$, where $\rs$ is the ball radius. We can use the results of Section \ref{sec:OneBallQuantitative} below to express these directly in terms of matrix model parameters.

\subsection{Integrating in the Coulomb field}
\label{subsec:sl-vertex-operators}

We can give an alternative representation of the localization integral by explicitly integrating in
the Coulomb field $W$ produced by the dipoles. We define a path integral 
\begin{equation}
 \begin{gathered}
 \langle\mathcal F\rangle_0
 =\frac{1}{\mathcal Z_0}
 \int_{\substack{W\in i\mathbb R\\W|_{z=0}=0}}[\rd W]\,
 e^{-S_0[W]}\mathcal F[W],
 \\
 S_0[W]=-\frac{\beta}{4\pi}
 \int_{\mathbb H}\rd^2X\,(\nabla W)^2 \; ,
 \qquad \beta=\frac{2}{27}.
 \end{gathered}
 \label{eq:sl-scalar-measure}
\end{equation}
where the integration contour runs over imaginary values of $W$ (as we would expect for a Coulomb field in Euclidean signature). With these definitions, we have
\begin{equation}
 \langle W(X)W(Y)\rangle_0
 =-\frac{1}{\beta}G_D(X,Y) \; .
 \label{eq:sl-field-covariance}
\end{equation}
Then, using the dipole charge distribution $K$ to define a smearing operator 
\begin{equation}
  \mathcal D F(x,z)
 \equiv\int_{\mathbb R}\rd\rho\,K(x-\rho)F(\rho,z),
\end{equation}
and a normal-ordered vertex\footnote{Here, normal ordering means that we don't allow self-contractions of the fields appearing in $\mathcal{V}$. Explicitly, if $A$ is linear in $W$, we can define $:\!e^A\!:_0=e^{-\langle A^2\rangle_0/2}e^A$ after introducing a UV regulator.}
\begin{equation}
 \mathcal V(x,z; W)
 \equiv:\!\exp[-\beta\mathcal D W(x,z)]\!:_0 \; .
 \label{eq:sl-vertex}
\end{equation}
Then using Wick's theorem $\langle e^{\sum_aA_a}\rangle_0
=e^{\frac12\sum_{a,a'}\langle A_aA_{a'}\rangle_0}$ for $A$ linear in $W$, we find
\begin{equation}
 \left\langle\prod_a\mathcal V(x_a,z_{b_a} ; W)\right\rangle_0
 =\exp\!\left[-\beta \sum_{a < a'}\int_{\mathbb H}\rd^2X\int_{\mathbb H}\rd^2Y\,
 j_a(X)G_D(X,Y)j_{a'}(Y)\right].
 \label{eq:sl-wick}
\end{equation}
Introducing the background field into the vertex operator exactly reproduces the background field term in the localization integral: using 
\begin{equation}
 e^{-\beta\int_{\mathbb H}\rd^2X\,j_a(X)W_\omega(X)} \mathcal V(x,z_b ; W)
 =:\!\exp[-\beta\mathcal D(W+W_\omega)(x,z_b)]\!:_0 = V(x,z_b ; W + W_\omega) ,
 \label{eq:sl-background-dressing}
\end{equation}
we have
\begin{equation}
 Z_{\mathcal R}
 =C_{\mathcal R} \left\langle \prod_a \int dx_a \mathcal V(x_a,z_{b_a} ; W +  W_\omega)\right\rangle_0 \, ,
 \label{eq:sl-field-partition}
\end{equation}
recovering the localization integral exactly. 

\subsection{The rank generating function and the axion source}
\label{subsec:rank-fugacity-axion}

In this section, we return to the exact finite $N$ description and note that the generating function for partition
functions at different ranks takes a particularly simple form.

In defining this generating function, it will be important to be careful about the overall normalization of the partition function. For this, we treat $U(N)$ as a gauge symmetry and thus divide by the volume of the gauge group. For both this gauge group volume and the bosonic integral, we will make a specific natural choice for the measure. In a Hermitian
basis with $\Tr(T^AT^B)=\delta^{AB}$, define the normalized partition function as
\begin{equation}
 \widehat Z_N\equiv
 \frac{1}{\operatorname{Vol}_0 U(N)}
 \int\!\prod_{I=1}^{10}\prod_{A=1}^{N^2}
       \frac{\dd X_I^A}{\sqrt{2\pi}}
       \prod_{\alpha=1}^{16}\prod_{A=1}^{N^2}\dd\psi_\alpha^A\,
       e^{-S_\omega} \, ,
 \label{eq:rank-normalized-integral}
\end{equation}
where the $1/\sqrt{2 \pi}$ is chosen so that $e^{-(X_I^A)^2/2}$ integrates to one. For the gauge group volume integral, we normalize so that if $U=e^{i\theta^A T^A}$ parametrizes $U(N)$ near the identity, the integration measure locally takes the form $\dd\theta^A/\sqrt{2\pi}$. This leads to 
\begin{equation}
 \operatorname{Vol}_0 U(N)
 =\frac{(2\pi)^{N/2}}{\prod_{k=1}^{N-1}k!} \, .
 \label{eq:rank-group-measure}
\end{equation}
With these choices, the bosonic measure factors and gauge group volume exactly cancel the quantity $A_N$ appearing in the normalization $\mathcal{C}_\mathcal{R}$ in \ref{eq:sl-gas}.

Now, we define the simple generating function
\begin{equation}
\mathcal Z(\mu) = \sum_{N=0}^\infty e^{- \mu N} \hat{Z}_N \; ,
\end{equation}
where $\mu$ is a chemical potential for gauge group rank (or D-instanton number) and we set $\hat{Z}_0 = 1$. This should be understood as a formal power series, as it will generally have zero radius of convergence.

Using our results above, we have
\begin{eqnarray*}
\mathcal{Z}(\mu) &=& \sum_{N=0}^\infty e^{- \mu N}  \sum_{\{n_b\}|\sum_b b n_b = N}  \mathcal{C}_{\{n_b\}}  \hat{Z}_{\{n_b\}} \cr 
&=& \sum_{\{n_b\}} e^{- \mu \sum_b b n_b}   \left(\prod_{b\geq1}\frac{w_b^{n_b}}{n_b!}\right)    \left\langle
 \prod_{b\geq1}\prod_{i=1}^{n_b}
 \int_{\mathbb R}\rd x_{bi}\,
 \mathcal V(x_{bi},z_b; W + W_\omega)
 \right\rangle_0 \cr 
 &=& \left\langle \exp \left(\sum_{b \ge 1} e^{-\mu b} w_b \int \rd x \mathcal{V}(x,z_b ; W + W_\omega) \right) \right\rangle_0 \, .
\end{eqnarray*}

The exponential involving the chemical potential can be absorbed into a modification of the background potential:
\begin{equation}
 W_{\mu}(\rho,z)=27\mu\,z\rho,
 \qquad
 \beta\mathcal D W_{\mu}(x,z_b)
 =-b\log\mathfrak q=\mu b.
 \label{eq:sl-q-background}
\end{equation}
Defining 
\[
W_{\omega,\eta}(\rho,z) = {3 \over 2} \omega^4(z \rho^3 - z^3 \rho) - \eta z \rho \, ,
\]
we have finally that
\begin{equation}
\boxed{
\mathcal{Z}(\mu) = \left\langle \exp \left(\int \rd^2X \mathcal{J}(z) \mathcal{V}(x,z ; W + W_{\omega, -\omega^4/4 - 27 \mu}) \right) \right\rangle_0} \; ,
\end{equation}
where we have defined
\[
\mathcal{J}(z) = \sum_{b \ge 1} w_b \delta \left(z - {b \over 2} \right)
\]
This corresponds to a formal deformation of the action for the Coulomb field path integral by the vertex operator with a $z$-dependent coupling. 

In the 4D electrostatics language, adding the chemical potential gives rise to a shift of $\eta$
\begin{equation}
 \delta\eta=-27\mu \, .
 \label{eq:rank-axion-shift}
\end{equation}
In supergravity language, this corresponds to a shift 
\begin{equation}
\delta C_0 = -{\mu \over 2 \pi} \, , \qquad \qquad \delta C_2=\delta C_0 B_2 \, .
\end{equation}
This leaves the local gauge-invariant field strengths, metric, and dilaton unchanged.

\section{The matrix resolvent and holomorphic electrostatics}
\label{sec:ResolventHolomorphic}

In this section, we show that the averaged trace of the resolvent of the
protected matrix $\phi=X_3-iX_{10}$ determines a  holomorphic function
whose imaginary part gives the two-dimensional Coulomb potential from the previous section. This
provides a direct map from  the matrix model to the 2D and 
4D electrostatics systems and thus to the on-shell supergravity fields. The expectation value of the protected observable
$\Tr(\phi^{2k})$ can be read off from the asymptotic expansion of
the resolvent function and gives the multipole moment $q_{2k+1}$ of the charge distribution in the 4D electrostatics picture. Adding sources for these protected operators corresponds to
a modification of the background potential in the electrostatics problem.

The Coulomb field reformulation of the localization integral in the
preceding section was exact at finite $N$. Here we will work in the
continuum, long-wavelength regime in which the dipole distributions
give the conducting balls. 

\subsection{The holomorphic potential and the matrix resolvent}
\label{ProtectedResolvent:Subsection}

A basic observation that we saw already in the previous section is that the axially-symmetric 4D Laplace equation
\begin{equation}
\partial_z^2 V+\partial_\rho^2 V+
\frac{2}{\rho}\partial_\rho V=0
\end{equation}
becomes, after writing $W(\rho,z)=\rho V(\rho,z)$, the
2D Laplace equation
\begin{equation}
\partial_\rho^2 W+\partial_z^2 W=0.
\end{equation}
Here, $W$ is the 2D Coulomb potential from the previous section:
\begin{equation}
W(\rho,z)=\rho V(\rho,z) \; .
\label{wW}
\end{equation}

The real and imaginary parts of any analytic function automatically
satisfy the Laplace equation, thus motivating a reformulation of the
electrostatics problem in terms of a complex function
$\mathcal W(z+i\rho)$. We will choose to encode $W=\rho V$ in the
imaginary part of this function:
\begin{equation}
W(\rho,z)=\operatorname{Im}\mathcal W(z+i\rho),
\qquad \zeta=z+i\rho.
\label{HolomorphicElectrostatics:Reconstruction}
\end{equation}
Here, $\mathcal{W}$ is unique up to an additive real constant.

The background potential $V_{\rm bg}=-\eta z+\frac32\omega^4(z\rho^2-z^3)$ arises from the polynomial background\footnote{For the background potential arising from the exact reformulation in section \ref{sec:scalar-localization}, we have $\eta = -\omega^4/4$.}
\begin{equation}
\mathcal W_{\rm bg}(\zeta)
=-\frac{\eta}{2}\zeta^2-\frac38\omega^4\zeta^4.
\label{HolomorphicElectrostatics:Background}
\end{equation}
We fix the additive real constant in $\mathcal W$ by requiring
$\mathcal W-\mathcal W_{\rm bg}=O(\zeta^{-2})$ at infinity. We will now show that the holomorphic function $\mathcal{W}$ is related directly to the matrix model resolvent, an operator-valued function 
\begin{equation}
(\zeta I-\widehat\phi)^{-1}
\end{equation}
of the complex variable $\zeta$, where $\widehat\phi$ is a rescaled version of the protected operator, 
\begin{equation}
\widehat\phi=\frac{\phi}{\omega} = \frac{1}{\omega} \left(X_3 - i X_{10}\right),
\end{equation}
and $I$ is the identity matrix.

Specifically, we will show that, with appropriate analytic continuations
\begin{equation}
\boxed{\quad
\mathcal W(\zeta)-\mathcal W_{\rm bg}(\zeta)
=\partial_\zeta G_{\mathcal R,\mathrm{cont}}(\zeta),
\qquad
G_{\mathcal R}(\zeta)
\equiv \left\langle\operatorname{Tr}
 (\zeta I-\widehat\phi)^{-1}\right\rangle_{\mathcal R} \; .}
\label{ResolventHolomorphic:MainIdentity}
\end{equation}
Here the subscript $\mathrm{cont}$ denotes the continuum saddle
approximation in the localization integral described in detail in the previous section. 

To show this directly, recall that the protected matrix has eigenvalues
on the localization locus
\begin{equation}
\phi^{\rm eig}_{s,i,\ell}
=\omega(\ell-i\rho_{si}),
\qquad \ell=-j_s,\ldots,j_s,
\qquad j_s=\frac{N_s-1}{2}.
\label{ProtectedMultipoles:Eigenvalues}
\end{equation}
Here $i=1,\ldots,n_s$ indicates the copy number of the representation
$N_s$ and $\ell$ labels the $L_3$ eigenvalue within that copy.
When the moduli in the $s$-th block form a dense saddle, the density
\begin{equation}
f_s(\rho)=\sum_{i=1}^{n_s}\delta(\rho-\rho_{si})
\end{equation}
can be well approximated by a smooth function normalized by
$\int\rd \rho\,f_s(\rho)=n_s$. For the symmetric saddles considered
here, $f_s(\rho)$ is even. For large $N_s$, the normalized
$L_3$ eigenvalues fill the interval $[-z_s,z_s]$, where the
electrostatic dictionary is
\begin{equation}
z_s=\frac{N_s}{2},
\qquad Q_s=n_s,
\qquad \widehat\phi=\frac{\phi}{\omega}.
\label{ResolventHolomorphic:Dictionary}
\end{equation}
Thus the resolvent evaluated on the saddle $\mathcal{R}$ in the continuum approximation is\footnote{This expression is holomorphic outside the union of rectangles $[-z_s,z_s] \times [-\rho_s,\rho_s]$ but can be analytically continued to the entire region outside the support of the dipole distributions. The formula below for the derivative is already holomorphic on this full region.} 
\begin{equation}
G_{\mathcal R,\mathrm{cont}}(\zeta)
=\sum_s\int\rd\rho\,f_s(\rho)
\int_{-z_s}^{z_s}\frac{\rd y}{\zeta-y+i\rho}.
\label{ResolventHolomorphic:ContinuumResolvent}
\end{equation}
Differentiating with respect to $\zeta$ (which is equivalent to a $-y$ derivative inside the integral) reduces the integral over $y$
to its endpoints:
\begin{equation}
\partial_\zeta G_{\mathcal R,\mathrm{cont}}
=\sum_s\int\rd\rho\,f_s(\rho)
\left[\frac{1}{\zeta+z_s+i\rho}
-\frac{1}{\zeta-z_s+i\rho}\right].
\label{ResolventHolomorphic:EndpointKernels}
\end{equation}
We can identify this expression directly with the 2D Coulomb potential
in the continuum dipole picture. Recall that
\begin{equation}
G_D(\rho,z;\hat\rho,z_s)
=\frac{1}{2}
\log\frac{(\rho-\hat\rho)^2+(z+z_s)^2}
         {(\rho-\hat\rho)^2+(z-z_s)^2},
\end{equation}
and that the continuum charge density is
$J(\rho,z)=-\sum_s f_s'(\rho)\delta(z-z_s)$. The charge-generated part of the
two-dimensional potential is therefore
\begin{equation}
W_{\rm src}(\rho,z)
=-\sum_s\int\rd\hat\rho\,
G_D(\rho,z;\hat\rho,z_s)f_s'(\hat\rho).
\end{equation}

Since $f_s$ is even, we can replace $\hat\rho$ by $-\hat\rho$
in \eqref{ResolventHolomorphic:EndpointKernels}. Setting
$\zeta=z+i\rho$, we then observe that
\begin{equation}
\operatorname{Im}\left[
\frac{1}{\zeta+z_s-i\hat\rho}
-\frac{1}{\zeta-z_s-i\hat\rho}
\right]
=\partial_{\hat\rho}G_D(\rho,z;\hat\rho,z_s).
\end{equation}
Integrating by parts, with no boundary term because the densities
vanish at the endpoints of their supports, gives
\begin{align*}
\operatorname{Im}
\partial_\zeta G_{\mathcal R,\mathrm{cont}}(\zeta)
&=\sum_s\int\rd\hat\rho\,
f_s(\hat\rho)\partial_{\hat\rho}G_D(\rho,z;\hat\rho,z_s)\\
&=-\sum_s\int\rd\hat\rho\,
f_s'(\hat\rho)G_D(\rho,z;\hat\rho,z_s)\\
&=W_{\rm src}(\rho,z).
\end{align*}
Thus, we conclude that
\begin{equation}
\quad
W_{\rm src}(\rho,z) = \rho\,\psi(\rho,z)
=\operatorname{Im}
\partial_\zeta G_{\mathcal R,\mathrm{cont}}(z+i\rho).
\quad
\label{ProtectedResolvent:Reconstruction}
\end{equation}
Holomorphicity then gives
\eqref{ResolventHolomorphic:MainIdentity}, up to a real constant
fixed to zero by the asymptotic normalization.

Equivalently, in the original matrix normalization,
\begin{equation}
\boxed{
\rho\,\psi(\rho,z)
=-\operatorname{Im}
\left\langle\operatorname{Tr}
\left((z+i\rho)I-\phi/ \omega\right)^{-2}
\right\rangle_{\mathcal R,\mathrm{cont}}.}
\label{ProtectedResolvent:OriginalNormalization}
\end{equation}

\subsection{Protected observables from Laurent coefficients or multipole moments}

The asymptotic expansion of (\ref{ResolventHolomorphic:MainIdentity}) directly relates the expectation values of the protected observables $\langle \Tr(\phi^{2k})\rangle$ to the Laurent coefficients of $\mathcal{W}$ and then to the multipoles $q_{2k+1}$ defined earlier.

Starting from (\ref{ResolventHolomorphic:MainIdentity}), we find
\begin{equation}
\mathcal W-\mathcal W_{\rm bg}
=-\sum_{k=0}^\infty\frac{2k+1}{\omega^{2k}}
\frac{\langle\Tr\phi^{2k}\rangle_{\mathcal R,\mathrm{cont}}}
     {\zeta^{2k+2}} \, ,
\end{equation}
where we used the vanishing of the expectation values of odd powers
of $\phi$. So the Laurent coefficients of $\mathcal{W}$ give the protected observables directly. 

Taking the imaginary part of this expression and dividing by $\rho$ gives
\begin{equation}
V-V_{\rm bg}
=\sum_{k=0}^\infty\frac{2k+1}{\omega^{2k}}
\frac{U_{2k+1}\!\left(z/\sqrt{\rho^2+z^2}\right)}
     {(z^2+\rho^2)^{k+3/2}}
\langle\Tr\phi^{2k}\rangle_{\mathcal R,\mathrm{cont}} \, ,
\end{equation}
where we have used the result
\begin{equation}
\operatorname{Im}\!\left[(z+i\rho)^{-(2k+2)}\right] = -{\rho \over  (z^2 + \rho^2)^{k + 3/2}} U_{2k+1}\left(z \over \sqrt{\rho^2 + z^2} \right) \; ,
\end{equation}
that follows from the definition of the Chebyshev polynomials.

Comparing to the expression (\ref{multipole}) for the multipole expansion of the potential $V$, we 
see that protected observable expectation values are directly proportional to the multipoles defined in (\ref{MultipoleDef}),
\begin{equation}
\boxed{
\left\langle\Tr\phi^{2k}\right\rangle_{\mathcal R,\mathrm{cont}}
=\frac{\omega^{2k}}{2k+1}q_{2k+1}
}\, .
\label{ProtectedMultipoles:MainResult}
\end{equation}
As a check, we note that for $k=0$, the moment $q_1$ is just the
dipole moment in the $z$ direction including image charges. The
right hand side here is then 
\begin{equation}
q_1=2\sum_s z_sQ_s=\sum_s N_sn_s=N,
\end{equation}
in agreement with the left side. 

The direct relation between protected observables and electrostatic multipoles also leads to a direct relation to the asymptotic metric in the corresponding supergravity solutions, for example, writing $r^2=\rho^2+z^2$, the asymptotic metric can be expressed as
\begin{equation}
\label{SUGRAsolutions:MetricAsymptoticOriginal}
\begin{aligned}
\frac{\rd s^2}{\alpha'}
={}& \frac{8\pi\sqrt{3P}}{\omega^2 r^4}
\bigg[
1+\frac{2q_3}{3P}\frac{7z^2-3\rho^2}{r^4}
\\
&\qquad
+\frac{q_5}{P}\frac{21z^4-30\rho^2z^2+5\rho^4}{r^8}
-\frac{2q_3^2}{9P^2}\frac{(7z^2-3\rho^2)^2}{r^8}
\bigg]
\\
&\qquad\times
\left(
\rd\rho^2+\rd z^2
+\rho^2\rd\Omega_6^2+z^2\rd\Omega_2^2
\right)
+\dots \, .
\end{aligned}
\end{equation}
so the multipoles and thus the protected observable expectation values can be read off the metric asymptotics, as usual in AdS/CFT. 

\subsection{Sources as deformations of the holomorphic background}
\label{SourceDeformedElectrostatics:Subsection}

In this subsection, we show that formally deforming the matrix model
by sources for the protected operators $\Tr(\phi^{2k})$ deforms the
electrostatics problem by changing the background potential while
leaving the conductor positions and charges fixed.  

Deforming the matrix model action by sources as
\begin{equation}
S_{\rm eff}\mapsto S_{\rm eff}
+\sum_{k\geq1}\tau_k\Tr\phi^{2k}
\end{equation}
gives, using the relation (\ref{ProtectedMultipoles:MainResult}) and the definition (\ref{MultipoleDef}) of the multipoles,
\begin{equation}
\begin{aligned}
\delta S_{\rm eff}
=\sum_{k\geq1}
\frac{\tau_k\omega^{2k}}{(k+1)(2k+1)}
\sum_s\int_0^{\rho_s}\rd\rho\,4\pi\rho^2
\sigma_s(\rho)\mathcal H_k(\rho,z_s) \, ,
\end{aligned}
\label{ResolventHolomorphic:SourceFunctional}
\end{equation}
where $\mathcal H_k(\rho,z)$ is a polynomial in $\rho$ and $z$
with integer coefficients defined by
\begin{equation}
\mathcal H_k(\rho,z)
=\frac{1}{\rho}\operatorname{Im}(z+i\rho)^{2k+2}.
\label{defH}
\end{equation}
We can check that
\begin{equation}
\left(\partial_\rho^2+\frac{2}{\rho}\partial_\rho+
\partial_z^2\right)\mathcal H_k=0,
\qquad \mathcal H_k(\rho,0)=0,
\label{SourceDeformedElectrostatics:HarmonicProperties}
\end{equation}
so $\mathcal H_k$ is a source-free axially symmetric solution of the
4D Laplace equation vanishing at $z=0$.

With the present normalization of the relation in \cite{Komatsu:2024ydh}, the localization action and the
corresponding electrostatics energy functional are related in the
continuum, $\eta=0$ convention by
\begin{equation}
S_{\rm Elec}=E=\frac{27}{2}S_{\rm eff}.
\end{equation}
The deformation by sources therefore corresponds to a modification
of the electrostatics energy functional by
\begin{equation}
\delta S_{\rm Elec}
=\sum_s\int_0^{\rho_s}\rd\rho\,4\pi\rho^2
\sigma_s(\rho)\delta V(\rho,z_s),
\end{equation}
where
\begin{equation}
\delta V(\rho,z)
=\frac{27}{2}\sum_{k\geq1}
\frac{\tau_k\omega^{2k}}{(k+1)(2k+1)}\mathcal H_k(\rho,z).
\label{ResolventHolomorphic:SourcePotential}
\end{equation}
The effect on the saddle point equations determining the charge
distribution is thus to modify the background potential by
$\delta V$. In terms of the associated holomorphic function, this is simply
\begin{equation}
\boxed{
\delta\mathcal W_{\rm bg}(\zeta)
=\frac{27}{2}\sum_{k\geq1}
\frac{\tau_k\omega^{2k}}{(k+1)(2k+1)}\zeta^{2k+2}.
}
\label{ResolventHolomorphic:SourcePolynomial}
\end{equation}
The positions $z_s$ and total charges $Q_s$ remain fixed by the
representation $\mathcal R$, while the charge distributions, radii,
potentials, and multipole moments respond to the sources.

As examples, the first two cases give
\begin{equation}
\mathcal H_1(\rho,z)=4(z^3-z\rho^2),
\qquad
\mathcal H_2(\rho,z)=6z\rho^4-20z^3\rho^2+6z^5.
\end{equation}
Consequently, a quadratic source shifts the coefficient of the
original cubic background,
\begin{equation}
\frac32\omega^4
\longmapsto\frac32\omega^4-9\tau_1\omega^2.
\label{SourceDeformedElectrostatics:QuadraticSource}
\end{equation}
whereas a quartic source adds the independent quintic harmonic function
\begin{equation}
\delta_{\tau_2}V_{\rm bg}
=\frac{27}{5}\tau_2\omega^4
\left(z\rho^4-\frac{10}{3}z^3\rho^2+z^5\right).
\label{SourceDeformedElectrostatics:QuarticSource}
\end{equation}
For $k>1$, the source therefore changes qualitatively the asymptotic
boundary behavior of the potential rather than only changing the
coefficient already present in the undeformed electrostatics problem.
Sources and expectation values are thus encoded in the growing and
decaying coefficients in the asymptotic expansion of $\mathcal W$.

We can also understand the effect of sources in the exact dipole gas reformulation of the localization integral discussed in section \ref{sec:scalar-localization}. Since the source terms are linear in the dipole charge distributions, the modification can again be captured by a shift in the background potential. For a dipole associated with a block of dimension
$b$, localization replaces the protected matrix by $\phi=\omega(L_3-i\rho I_b)$, so the source term corresponds to a shift
\begin{equation}
\delta U_b(\rho)
=\frac{27}{2}\sum_{k\geq1}\tau_k\omega^{2k}
\sum_{\ell=-(b-1)/2}^{(b-1)/2}(\ell-i\rho)^{2k}.
\label{SourceDeformedElectrostatics:ExactOneBody}
\end{equation}
in the 1D potential. This corresponds to a shift in the background potential $W_{bg}$ determined by
\begin{equation}
\int_{\mathbb R}\rd\hat\rho\,
K(\rho-\hat\rho)\,
\delta W_{\rm bg}(\hat\rho,b/2)
=\delta U_b(\rho).
\label{SourceDeformedElectrostatics:ExactBackgroundCondition}
\end{equation}
As above, we can take $W_{\rm bg} = \text{Im}\mathcal{W}_{\rm bg}$; for the $\Tr \phi^{2k}$ deformation, $\delta \mathcal{W}_{\rm bg}$ is an even $(k+2)$th order polynomial whose coefficients are determined by \eqref{SourceDeformedElectrostatics:ExactBackgroundCondition}. As examples, for the quadratic and quartic sources, respectively, the required polynomials are
\begin{equation}
\begin{aligned}
\delta_{\tau_1}\mathcal W_{\rm bg}(\zeta)
&=\tau_1\omega^2
\left(\frac94\zeta^4-\frac34\zeta^2\right),\\
\delta_{\tau_2}\mathcal W_{\rm bg}(\zeta)
&=\tau_2\omega^4
\left(\frac9{10}\zeta^6-\frac34\zeta^4
+\frac{11}{60}\zeta^2\right).
\end{aligned}
\label{SourceDeformedElectrostatics:ExactExamples}
\end{equation}
Their highest-degree terms reproduce
\eqref{ResolventHolomorphic:SourcePolynomial} in the 4D electrostatics approximation.

The deformation of the original matrix integral by quartic or higher order sources results in a divergent integral since $\text{Re}\,\Tr(X_3 - i X_{10})^{2k}$ is unbounded below. However, for appropriate sign choices of the sources, the corresponding electrostatic problem and the exact 2D dipole model remain well-behaved. For example, with the quartic source, the corresponding potential deformation at fixed $z$ grows as $\rho^4$ when the source has a positive sign. The interpretation is that even though the matrix model integral itself does not converge (with the original integration contour), the integral about each saddle of the localization integral converges to a well-defined result.\footnote{An interesting point is that there are cases with a positive quartic source and a large negative quadratic source where the saddle-point distribution will have densities whose support is a pair of disjoint intervals.} The divergent parts of the full integral correspond to parts of the integration domain that decouple from these finite parts. Thus, we expect that there is a well-defined dipole partition function over a particular source domain where the potential at each $z$ is bounded below. Derivatives of this expression about the origin (taken as limits from within the domain where the sources make sense) should then determine the various correlators of observables in the matrix model. 

\section{Solving the electrostatics problem via conformal mapping}
\label{sec:ConformalMappintElectrostaticsProblem}

In order to calculate the protected observables directly or find the supergravity solutions explicitly, it is necessary to solve the 4D electrostatics problem. We study the solution in detail below for the case where $\mathcal{R}$ corresponds to $n_{\star}$ copies of the $N_{\star}$-dimensional irreducible representation. In this section, we note that the electrostatics problem has a nice reformulation in terms of the holomorphic function $\mathcal{W}$ whose imaginary part is $W=\rho V$, and that this reformulation can be solved using standard conformal mapping techniques (though numerics are still required).

The following constraints on $\mathcal{W}$ will be shown to give a $V$ satisfying the electrostatics problem. Let $\zeta=z+i\rho$ and consider a region of the complex $\zeta$ plane consisting of the first quadrant with the vertical slits
$\Gamma_s=\{z_s+i\rho:0<\rho\leq\rho_s\}$ removed.
The slit lengths $\rho_s$ are initially unknown. We seek a holomorphic
function $\mathcal W(\zeta)$ on this domain, together with real constants
$V_s$, obeying
\begin{equation}
\begin{aligned}
\operatorname{Im}\mathcal W&=0
&&\text{on both coordinate axes},\\
\operatorname{Im}(\mathcal W_\pm - V_s \zeta) &= 0
&&\text{on both faces of }\Gamma_s,\\
\mathcal W(\zeta)&=\mathcal W_{\rm bg}(\zeta)
+\mathcal O(\zeta^{-2})
&&\text{as }|\zeta|\to\infty,
\end{aligned}
\label{HolomorphicElectrostatics:BoundaryData}
\end{equation}
where
\begin{equation}
\mathcal W_{\rm bg}(\zeta)
=-\frac{\eta}{2}\zeta^2-\frac{3}{8}\omega^4\zeta^4.
\label{HolomorphicElectrostatics:MappingBackground}
\end{equation}
Here $+$ and $-$ mean approach from $z>z_s$ and $z<z_s$, respectively.
We require continuous boundary limits along the coordinate axes and separately on each slit face, and boundedness near all finite boundary points, including the origin, slit attachments, and tips. These conditions exclude additional boundary singularities.

From the middle equation in (\ref{HolomorphicElectrostatics:BoundaryData}), we see that $\operatorname{Re}\mathcal W'_\pm=V_s$, so the function $j_s$ defined by
\begin{equation}
j_s(\rho)=i\bigl[
\mathcal W'_+(z_s+i\rho)-\mathcal W'_-(z_s+i\rho)\bigr].
\label{HolomorphicElectrostatics:Jump}
\end{equation}
is real. We impose 
\begin{equation}
\frac{1}{\pi}\int_0^{\rho_s}d\rho\,\rho\,j_s(\rho)=Q_s,
\qquad
\lim_{\rho\to\rho_s^-}j_s(\rho)=0,
\label{HolomorphicElectrostatics:Charges}
\end{equation}
with $j_s(\rho)/\rho$ smooth and even at the origin. These constraints determine $V_s$ and $\rho_s$ as part of the solution.

To recover the electrostatic description, we have already seen that $V = W / \rho$ will satisfy the 4D Laplace equation. The first condition in (\ref{HolomorphicElectrostatics:BoundaryData}) makes $V$ vanish at $z=0$. Away from the conductors, the Schwarz reflection formula $\mathcal W(\bar\zeta)=\overline{\mathcal W(\zeta)}$ implies that $W$ is odd in $\rho$, and locally
\begin{equation}
W(\rho,z)=\rho\,\mathcal W'(z)-\frac{\rho^3}{6}\mathcal W'''(z)+\mathcal O(\rho^5),
\end{equation}
so $V=W/\rho$ is regular and even at the radial axis away from the conductors.
The polynomial $\mathcal W_{\rm bg}$ gives precisely
$V_{\rm bg}=-\eta z+\frac32\omega^4(z\rho^2-z^3)$, while the remainder
gives a decaying potential.  Each slit becomes a ball at the constant
potential $V_s$.

Finally, letting $r$ be the real part of $\mathcal{W}$, the fact that $\mathcal{W}$ is a function of the combination $z + i \rho$ gives the Cauchy-Riemann equations 
\begin{equation}
\partial_{\bar\zeta} \mathcal{W} = \frac12(\partial_z + i \partial_\rho) (r + iW) = 0  \qquad \implies \qquad \partial_z r = \partial_\rho W  \qquad \partial_\rho r = - \partial_z W\; 
\end{equation}
which allow us to express the derivatives of the real part of $\mathcal{W}$ in terms of the derivatives of $W$. This allows us to write
\begin{equation}
\mathcal W'=\partial_\rho(\rho V)+i\rho\,\partial_zV,
\qquad \frac{j_s(\rho)}{\rho} = 
-\bigl[\partial_zV\bigr]^+_- = 4\pi^2\sigma_s(\rho) \, ,
\label{HolomorphicElectrostatics:Density}
\end{equation}
where the second formula follows from the definition of $j$ in terms of $\mathcal{W}'$.
Consequently, \eqref{HolomorphicElectrostatics:Charges} prescribes the
total charges and makes the density vanish at each ball's edge.  These
are exactly the four-dimensional electrostatic boundary conditions.

Conversely, for any solution $V$ of the electrostatics problem, the harmonic function $\rho V$ has a harmonic conjugate on the simply connected slit quadrant. This defines $\mathcal W$ up to an additive real constant, which is chosen to satisfy the asymptotic normalization in \eqref{HolomorphicElectrostatics:BoundaryData}. The resulting function satisfies all the conditions on $\mathcal W$.

\subsection{Solution via conformal mapping}
\label{subsec:HalfPlaneElectrostatics}

Using a conformal mapping, we can reformulate the problem of finding $\mathcal{W}$ on the first quadrant with slits removed to one of finding a holomorphic function on the upper half-plane with boundary conditions on the real axis.

\begin{figure}[t]
\centering
\setlength{\unitlength}{1pt}
\begin{picture}(410,108)
\small
\put(18,20){\vector(1,0){135}}
\put(18,20){\vector(0,1){70}}
\put(158,17){$z$}
\put(12,95){$\rho$}
\put(12,7){$0$}
\put(86,20){\line(0,1){56}}
\put(80,35){\vector(0,1){25}}
\put(92,62){\vector(0,-1){25}}
\put(82,7){$z_s$}
\put(91,78){$z_s+i\rho_s$}
\put(39,97){$\zeta$-plane}
\put(239,20){\vector(1,0){163}}
\put(380,6){$\operatorname{Re}u$}
\put(258,17){\line(0,1){6}}
\put(255,6){$0$}
\put(296,17){\line(0,1){6}}
\put(325,17){\line(0,1){6}}
\put(363,17){\line(0,1){6}}
\put(290,6){$a_s$}
\put(319,6){$b_s$}
\put(357,6){$c_s$}
\put(292,68){$\operatorname{Im}u>0$}
\put(296,97){$u$-plane}
\put(235,61){\vector(-1,0){61}}
\put(199,68){$Z$}
\end{picture}
\caption{Boundary correspondence for one slit.  The interval
$(a_s,b_s)$ maps upward along the face with $z<z_s$, and $(b_s,c_s)$
maps downward along the face with $z>z_s$.  The tip has preimage $b_s$;
the attachment has distinct preimages $a_s$ and $c_s$.}
\label{fig:SlitUniformization}
\end{figure}
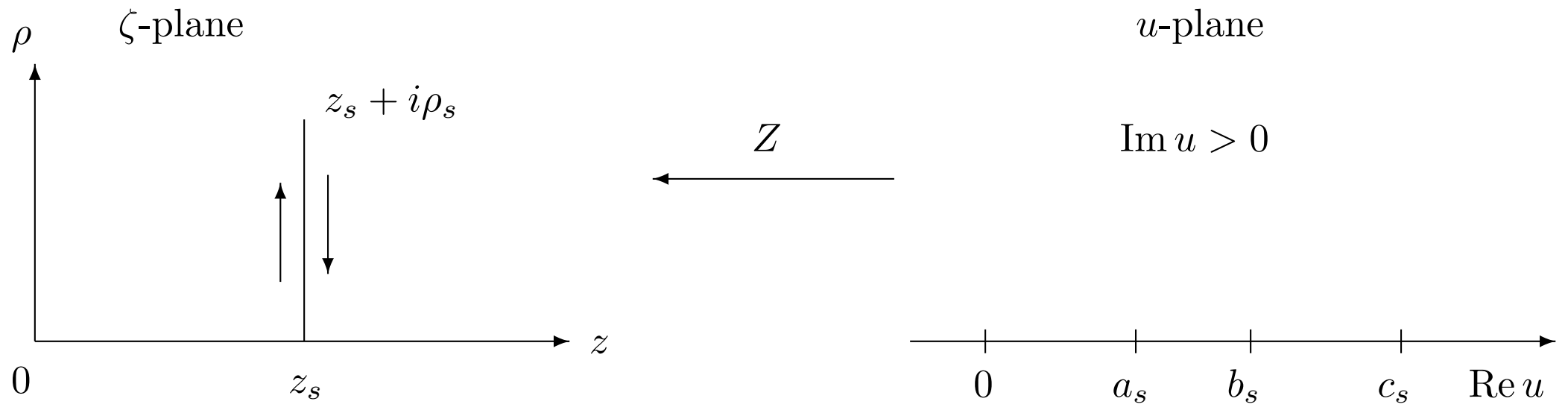

The Riemann mapping theorem guarantees that there is a biholomorphic map $Z$ from the upper half-plane to the first quadrant with slits,
\begin{equation}
\zeta=Z(u),\qquad F(u)=\mathcal W(Z(u)),\qquad \operatorname{Im}u>0.
\label{HalfPlaneElectrostatics:Map}
\end{equation}
We can fix the upper-half-plane parametrization by taking $Z(0)=0$, $Z(\infty)=\infty$ and $a_1=1$. The first two conditions leave only the rescaling $u\mapsto ku$, $k>0$, and the third fixes this freedom. The two faces of slit $s$
are the images of adjacent real intervals $(a_s,b_s)$ and $(b_s,c_s)$,
with
\begin{equation}
Z(a_s)=Z(c_s)=z_s,\qquad Z(b_s)=z_s+i\rho_s.
\label{HalfPlaneElectrostatics:Prevertices}
\end{equation}
The example with a single slit is illustrated in  Figure~\ref{fig:SlitUniformization}. Since the slit domain has a polygonal boundary, the explicit form of $Z$ can be given via a Schwarz--Christoffel transformation, defined by
\begin{equation}
Z'(u)=\frac{C}{\sqrt u}
\prod_{s=1}^q\frac{u-b_s}{\sqrt{u-a_s}\sqrt{u-c_s}},
\qquad C>0,\qquad Z(0)=0.
\label{ConformalAppendix:SC}
\end{equation}
Here the conductors are ordered by increasing $z_s$, so that
\begin{equation}
0<a_1<b_1<c_1<a_2<b_2<c_2<\cdots<a_q<b_q<c_q,
\end{equation}
and all branch cuts are taken to the left. Boundary values are taken from above. 

Let $I_s=(a_s,c_s)$ and define $\rho(t)=\operatorname{Im}Z(t)$ on these
intervals. According to (\ref{HolomorphicElectrostatics:BoundaryData}), the boundary values of $F$ are
\begin{equation}
\operatorname{Im}F(t)=
\begin{cases}
V_s\rho(t),&t\in I_s,\\
0,&t\notin\bigcup_s I_s.
\end{cases}
\label{HalfPlaneElectrostatics:BoundaryValues}
\end{equation}
We can write an explicit solution for $F(u)$ by using the Schwarz integral formula that reconstructs an analytic function in the upper half-plane from its imaginary boundary values. Specifically, the function $f(u)\equiv F(u)-\mathcal W_{\rm bg}(Z(u))$ has continuous boundary values, is bounded near finite boundary points, and decays as $u\to\infty$. The Schwarz formula therefore gives
\begin{equation}
f(u) = {1 \over \pi} \int_{\mathbb{R}} {\text{Im} f(t) \over t - u} \rd t \, .
\end{equation}
Noting that $\operatorname{Im}\mathcal W_{\rm bg}(\zeta)$ vanishes on the real and imaginary $\zeta$ axes, and using \eqref{HalfPlaneElectrostatics:BoundaryValues}, the Schwarz formula gives
\begin{equation}
\displaystyle
F(u)=\mathcal W_{\rm bg}(Z(u))
+\frac1\pi\sum_s\int_{a_s}^{c_s}\frac{g_s(t)}{t-u}\,\rd t \, .
\label{HalfPlaneElectrostatics:SchwarzIntegral}
\end{equation}
where
\begin{equation}
 g_s(t)=V_s\rho(t)
-\operatorname{Im}\mathcal W_{\rm bg}(Z(t)),\qquad t\in I_s.
\end{equation}
On the real axis, $F(t)$ and $F'(t)$ denote limits from the upper half-plane. For $x\in I_s$, away from the attachment prevertices, differentiation and integration by parts give the boundary formula
\begin{equation}
\begin{aligned}
F'(x+i0)={}&\mathcal W'_{\rm bg}(Z(x))Z'(x)\\
&+\frac1\pi\sum_{r=1}^q\operatorname{PV}\int_{a_r}^{c_r}
\frac{g_r'(t)}{t-x}\,dt+i\,g_s'(x),
\end{aligned}
\label{HalfPlaneElectrostatics:BoundaryDerivative}
\end{equation}
where only the integral with $r=s$ requires a Cauchy principal value. The endpoint terms vanish because $g_r(a_r)=g_r(c_r)=0$.

So far, the construction has unknown parameters $(a_s,b_s,c_s, \rho_s, V_s)$ and $C$. These are fixed by the geometric conditions
\eqref{HalfPlaneElectrostatics:Prevertices}, the required charges, and regularity at the endpoints.  Near a tip,
$Z(u)-Z(b_s)=A_s(u-b_s)^2+\cdots$, with $A_s\ne0$.
Thus $\mathcal W'(Z(u))=F'(u)/Z'(u)$ would have a singularity at $u=b_s$, leading via (\ref{HolomorphicElectrostatics:Density}) to a diverging charge density at the corresponding conductor edge, unless
\begin{equation}
F'(b_s)=0.
\label{HalfPlaneElectrostatics:Tip}
\end{equation}
The boundary values
already give $\operatorname{Im}F'(b_s)=V_s\operatorname{Im}Z'(b_s)=0$ so only the real part is an independent condition. To see that this condition makes the density vanish, note that $F-V_s Z$ has real boundary values near $b_s$ and hence extends analytically by Schwarz reflection. Setting $x=u-b_s$, we then have
\begin{equation}
Z'(u)=2A_sx+\mathcal O(x^2),\qquad
F'(u)=F''(b_s)x+\mathcal O(x^2),
\end{equation}
and therefore
\begin{equation}
\mathcal W'(Z(u))=\frac{F''(b_s)}{2A_s}+\mathcal O(x).
\end{equation}
The two faces approach the same limit, with $|x|=\mathcal O(\sqrt{\rho_s-\rho})$, so the jump in $\partial_zV$ vanishes and
$\sigma_s(\rho)=\mathcal O(\sqrt{\rho_s-\rho})$.
For a fixed map, the explicit equations above for $F$ make the tip conditions linear equations for the potential parameters $V_s$.

The charges take the form
\begin{equation}
\displaystyle
Q_s=-\frac1\pi\int_{a_s}^{c_s}\rho(t)\operatorname{Re}F'(t)\,\rd t \, .
\label{HalfPlaneElectrostatics:Charge}
\end{equation}
To see this, note that moving from $a_s$ to $c_s$ along the real axis, we have $Z(t)=z_s+i\rho(t)$ and
\begin{equation}
F'(t)=i\rho'(t)\mathcal W'(Z(t)),\qquad
\operatorname{Re}F'(t)=-\rho(t)\rho'(t)\partial_zV.
\end{equation}
The interval $(a_s,b_s)$ traverses the $z<z_s$ face from the attachment
to the tip, whereas $(b_s,c_s)$ traverses the $z>z_s$ face in reverse.
Thus,
\begin{equation}
\begin{aligned}
\int_{a_s}^{c_s}\rho(t)\operatorname{Re}F'(t)\,dt
&=\int_0^{\rho_s}d\rho\,\rho^2
\bigl(\partial_zV|_+-\partial_zV|_-\bigr)\\
&=-4\pi^2\int_0^{\rho_s}d\rho\,\rho^2\sigma_s(\rho)
=-\pi Q_s.
\end{aligned}
\end{equation}
as we claimed.

These results reduce the solution of the electrostatics problem to a finite-dimensional parameter problem involving functions defined by various integrals. After choosing trial ball radii, the geometric conditions determine the Schwarz--Christoffel parameters. For a fixed map, the tip conditions are linear equations for the potentials $V_s$. In the final step, we vary the radii to solve $Q_s^{\rm calc}(\vec\rho)=Q_s$, where $\vec\rho=(\rho_1,\ldots,\rho_q)$, by root finding or minimization of the residual $\sum_s[Q_s^{\rm calc}(\vec\rho)-Q_s]^2$. In Appendix~\ref{sec:OneDiskConformal}, we employ this method to provide an explicit solution in the case of a single ball. We have checked that the results match with those of the integral equation method employed in the next section. 

To close this section, we note that parameterizing the solutions to the electrostatics problem via the pair of functions $Z$ and $\mathcal{W}$ lines up with the recent supergravity analysis in \cite{DHoker:2026eox}, where the solutions where similarly parameterized in terms of a pair of holomorphic functions rather than a solution to the Laplace equation.

\section{Quantitative analysis of the one-ball family}
\label{sec:OneBallQuantitative}

In this section, we perform a detailed analysis for the class of saddles associated with representations $\mathcal{R}_{\star}$ having $n_{\star}$ copies of the $N_{\star}$-dimensional irreducible representation.

For these, the electrostatics configuration contains a single conducting ball above the grounded plane (or equivalently, a 4D parallel plate capacitor with ball-shaped plates). Using numerics, we solve the electrostatics problem explicitly, and then use the relation between protected observable $\Tr(\phi^{2k})$ expectation values and electrostatic multipoles to compute the $\mathcal{R}_{\star}$-sector contribution to these observables. While these expectation values depend on three parameters $(\omega,n_\star,N_\star)$, their values in the electrostatics approximation depends non-trivially on only a single parameter, which can be taken to be the ratio $\hzs$ of the ball height above the conducting plane to the ball radius, or alternatively the combination 
\begin{equation}
\xi \equiv \frac{n_{\star}}{\omega^4 N_{\star}^5} \; .
\end{equation}
We analyze the solution to the electrostatics problem and the corresponding results for $\Tr(\phi^{2k})$ expectation values in this sector over the full range of $\xi$. We first obtain analytic results at the two
ends of the family, then describe the numerical results for intermediate values of $\xi$.

\subsection{Solutions for products of a single irreducible representation}
\label{subsec:OneBallsolutions}

In this section, we specialize to the family of saddles that correspond to $SU(2)$ representations with some number $n_{\star}$ of copies of a single irreducible representation of dimension $N_{\star}$. In the electrostatics picture, these correspond to the case where we have a single conducting ball with charge $\Qs = n_{\star}$ at a vertical position $\zs = N_{\star} / 2$ above the infinite conducting plane.

We describe the general method to solve the electrostatics problem in appendix \ref{app:ElectrostaticProblem}; here we collect the main results and present the general logic.
The first step is to use the scale invariance and linearity of the Laplace equation to set up a normalized electrostatic problem which is more convenient for the analysis. Before imposing that the charge density vanishes at $\rho = \rs$, the problem is characterized by the position $\zs$ and size $\rs$ of the ball, as well as the total charge $\Qs$ and the background potential parameter $\omega$. Introduce a normalized potential
\begin{equation}
\hV(\hr, \hz) \equiv \frac{1}{V_0} V(\rs \hr, \rs \hz) \equiv \hat{V}_{\rm bg}(\hr, \hz) + \hpsi(\hr, \hz) \, ,
\end{equation}
where the constant $V_0$ will be fixed momentarily, and the splitting of the potential into background and ball-generated terms follows \eqref{SUGRAsolutions:FullPotential}. This normalized potential also solves the Laplace equation, but now with a charged ball of radial size $1$ at a vertical position $\hzs = \zs / \rs$ (there is also the infinite grounded plane at $\hz = 0$). If the value of the original potential at the charged ball is $V_{\star}$, we choose $V_0$ so that the normalized potential generated by the ball satisfies
\begin{equation}
\hpsi(\hr, \hzs) = 1 - \nu \hr^2 \qquad {\rm if}\; 0 < \hr < 1 \, ,
\end{equation}
which means we take
\begin{equation}
\label{SUGRAsolutions:V0FromNu}
\nu = \frac{3\omega^4 \zs \rs^2}{2 V_0} = \frac{3\omega^4 \zs^3}{2 \hzs^2 V_0} \, , \qquad V_0 = V_{\star} + \eta \zs + \frac{3}{2}\omega^4 \zs^3 \, .
\end{equation}

The normalized problem before imposing the charge density to vanish at the edge of the ball is thus characterized by the two parameters $\hzs$ and $\nu$. As discussed in detail in appendix \ref{app:ElectrostaticProblem}, the extra regularity condition fixes uniquely $\nu$ in terms of $\hzs$, giving a function $\nu (\hzs)$. This function cannot be obtained analytically for all $\hzs$, but two limits can be deduced
\begin{equation}
\lim_{\hzs \to \infty} \nu(\hzs) = \frac{2}{3} \, , \qquad \lim_{\hzs \to 0} \nu(\hzs) = 1 \, .
\end{equation}

Note however that we have not specified yet how $\hzs$ is fixed from the original problem. This is done by computing the total charge induced in the ball by the solution to the normalized problem, and demanding it to be $\Qs = n_{\star}$. From the normalized potential, we can get a normalized charge density and total charge
\begin{equation}
\hat{\sigma} (\hr) = - \frac{1}{4\pi^2}\left( \left. \partial_{\hz} \right|_{\hzs^+} \hpsi - \left. \partial_{\hz} \right|_{\hzs^-} \hpsi \right) = \frac{\zs}{\hzs V_0} \sigma(\rs \hr) \, , \qquad \hat{Q}_{\star} = 4 \pi \int_0^1 \rd \hr \, \hr^2 \hat{\sigma}(\hr) \, .
\end{equation}
The original total charge is then
\begin{equation}
\Qs = 4 \pi \int_0^{\rs} \rd \rho\, \rho^2 \sigma(\rho) = \frac{3\omega^4 \zs^5}{2} \frac{\hat{Q}_{\star}}{\hzs^4 \, \nu(\hzs)}  \, ,
\end{equation}
where we have used the value of $V_0$ in terms of $\nu$ given in \eqref{SUGRAsolutions:V0FromNu}. Note that $\hat{Q}_{\star}$ is also a function of $\hzs$ alone, so this equation gives a direct relation between the parameters of the original problem and $\hzs$:
\begin{equation}
\label{SUGRAsolutions:XiZStarRelation}
\xi \equiv \frac{n_{\star}}{\omega^4 N_{\star}^5} = \frac{\Qs}{32\omega^4 \zs^5} = \frac{3}{64} \frac{\hat{Q}_{\star}(\hzs)}{\hzs^4 \, \nu(\hzs)}  \, ,
\end{equation}
where we have emphasized the sole dependence on $\hzs$ in the right-hand side.

\begin{figure}[t!]
    \centering
    \includegraphics[width = 1.0 \linewidth]{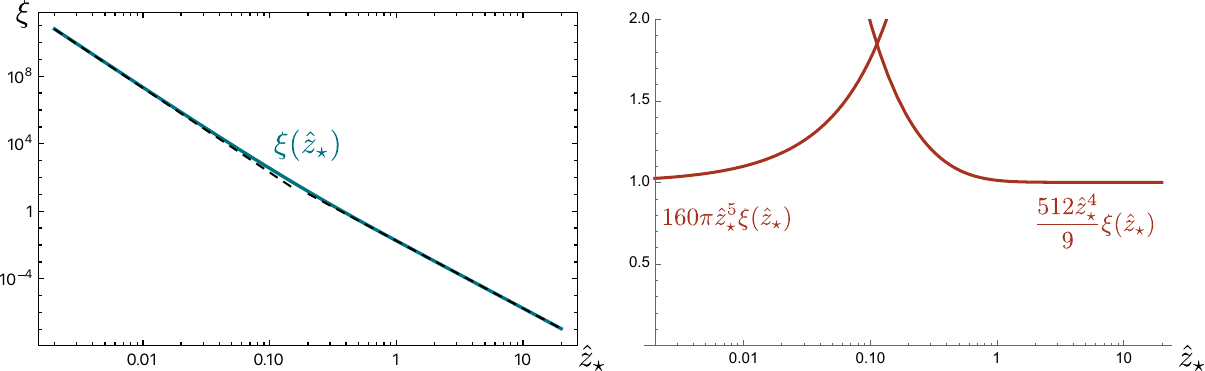}
    \caption{Left: Relation between $\xi$ and $\hzs$ as given by equation \eqref{SUGRAsolutions:XiZStarRelation}. The black dashed lines are the analytic approximations in \eqref{SUGRAsolutions:XiApproximations}. Right: Quotient between the true value $\xi(\hzs)$ and the analytic approximations.}
    \label{fig:PlotXiZStar}
\end{figure}

The logic is the following: from the parameters $(n_{\star}, N_{\star}, \omega)$ (or, equivalently, $(\Qs,\zs, \omega)$) the value of $\xi$ is fixed, and inverting the above relation we get the relevant value of $\hzs$ (this immediately gives $\rs = \zs / \hzs$). The normalized problem is then solved, obtaining $\hpsi(\hr, \hz)$, and from it we get the potential
\begin{equation}
\label{SUGRAsolutions:RelationNormalizedPotential}
V(\rho, z) = \frac{3\omega^4 \zs^3}{2 \hzs^2 \nu(\hzs)} \hat{V} \left( \frac{\hzs \rho}{\zs}, \frac{\hzs z}{\zs} \right) \, ,
\end{equation}
where $\hV$ is given by
\begin{equation}
\label{SUGRAsolutions:NormalizedPotential}
\hV(\hr, \hz) = - \frac{2 \eta \hzs \nu(\hzs)}{3\omega^4 \zs^2} \hz + \frac{\nu(\hzs)}{\hzs} \left( \hz \hr^2 - \hz^3 \right)  + \hpsi(\hr, \hz) \, .
\end{equation}
In figure \ref{fig:PlotXiZStar}, we show the relation between $\xi$ and $\hzs$: in general we need the numerical solution to get the $\hzs$ associated to a given $\xi$, but in the two limits $\xi \ll 1$ and $\xi \gg 1$ the analytic results obtained in appendix \ref{app:ElectrostaticProblem} work well,
\begin{align}
\label{SUGRAsolutions:XiApproximations}
& \xi \approx \frac{9}{512} \frac{1}{\hzs^4} \quad {\rm if}\; \hzs \gg 1 \, , & \xi \approx & \frac{1}{160\pi} \frac{1}{\hzs^5} \quad {\rm if}\; \hzs \ll 1 \, .
\end{align}

Working with the normalized potential is a convenient trick to solve the electrostatics problem, but in the context of saddles having a single irreducible representation (possibly repeated $n_{\star}$ times) it is also useful to rewrite the supergravity solution in terms of $\hV(\hr, \hz)$. We use \eqref{SUGRAsolutions:RelationNormalizedPotential} to rewrite the string frame metric as 
\begin{equation}
\label{SUGRAsolutions:MetricNormalized}
\frac{\rd s^2}{\alpha'} = \frac{\zs}{\hzs} \hat{A}(\hr, \hz) \left( \hat{F}(\hr, \hz) \left( \rd \hr^2 + \rd \hz^2 \right) + \hat{G}_6(\hr, \hz) \, \rd \Omega_6^2 + \hat{G}_2(\hr, \hz) \, \rd\Omega_2^2 \right) \, ,
\end{equation}
where now
\begin{equation}
\begin{gathered}
\hat{A}(\hr, \hz) = 2\pi \sqrt{\frac{( - 3 \partial_{\hr} \hV - \hr \partial_{\hz}^2 \hV )}{\hr (- \partial_{\hz}^2 \hV)}} \, , \qquad \hat{F}(\hr, \hz) = - \frac{\partial_{\hz}^2 \hV}{\partial_{\hr} \hV} \, , \qquad  \hat{G}_6(\hr, \hz)  = 3 \hr \, , \\
\hat{G}_2(\hr, \hz) = - \frac{\hr \partial_{\hz}^2 \hV \partial_{\hr} \hV}{\hat{\Delta}} \, , \qquad \hat{\Delta} = 3 \partial_{\hr} \hV \partial_{\hz}^2 \hV + \hr ( (\partial_{\hz} \partial_{\hr} \hV)^2 + (\partial_{\hz}^2 \hV)^2 ) \, .
\end{gathered}
\end{equation}
Notice that none of these functions depends on the first term proportional to $\eta$ of the normalized potential \eqref{SUGRAsolutions:NormalizedPotential}. Thus, up to the overall scale set by $\zs = N_{\star} / 2$, the metric is entirely characterized by $\hzs$ or, equivalently, $\xi$. Thinking in terms of matrix model parameters, if we change $N_{\star}$ and adjust $n_{\star}$ or $\omega$ to leave $\xi$ invariant, the metric changes by an overall constant factor only; while if we scale $n_{\star}$ and $\omega$ but keeping $\xi$ and $N_{\star}$ fixed, the metric does not change at all.

We can also rewrite the remaining fields in terms of the normalized potential. We collect them here for completeness
\begin{subequations}
\label{SUGRASolutions:FieldsNormalized}
\begin{align}
e^{\Phi} & = - \left( \frac{36\pi}{\omega^4 \zs^2} \right) \hzs \nu(\hzs) \frac{3 \partial_{\hr} \hV + \hr \partial_{\hz}^2 \hV}{\hr \sqrt{3 \hat{\Delta} \partial_{\hr} \hV (- \partial_{\hz}^2 \hV)}} \, , \\
C_0 & = - \left( \frac{\omega^4 \zs^2}{36\pi} \right) \frac{1}{\hzs \nu(\hzs)} \frac{3 \partial_{\hr} \hV (\partial_{\hz} \hV + \hr \partial_{\hz} \partial_{\hr} \hV) + \hr \partial_{\hz} \hV \partial_{\hz}^2 \hV}{3 \partial_{\hr} \hV + \hr \partial_{\hz}^2 \hV} \, , \\
B_2 & = - 2\pi\alpha'\zs \frac{1}{\hzs} \left( \hz - \frac{\hr \partial_{\hr} \hV \partial_{\hz} \partial_{\hr} \hV}{\hat{\Delta}} \right) \omega_2 \, , \\
C_2 & = \frac{\alpha' \omega^4 \zs^3}{18} \frac{1}{\hzs^2 \nu(\hzs)} \left( \hV - \hr \frac{\partial_{\hr} \hV}{\hat{\Delta}} \left( \partial_{\hz} \hV \partial_{\hz} \partial_{\hr} \hV + 3 \partial_{\hr} \hV (- \partial_{\hz}^2 \hV) \right) \right) \omega_2 \, .
\end{align}
\end{subequations}
In each case, we have isolated the prefactor that depends on specific data of the matrix saddle beyond $\xi$. Note also that $C_0$ and $C_2$ depend on the term proportional to $\eta$ of the normalized potential \eqref{SUGRAsolutions:NormalizedPotential}, and thus it is not true that, other than their prefactor, they are entirely fixed by $\hzs$ of $\xi$. But the dependence is simply the one given by the axion shift symmetry discussed in appendix \ref{app:ElectrostaticSolutions}.

This rewriting in which the fields depend only on $\xi$ up to prefactors is useful to discuss when the supergravity approximation is valid for the single ball solutions. At $\xi \sim \mathcal{O}(1)$ (equivalently, $\hzs \sim \mathcal{O}(1)$), the curvature scale in string units is set by $\alpha' R \sim z_{\star}^{-1}$. Thus, the requirement to have small curvatures is equivalent to $z_{\star} \sim N_{\star} \gg 1$. Similarly, the local string coupling is controlled by $e^{\Phi} \sim \omega^{-4} z_{\star}^{-2} \sim Q_{\star}^{-1} z_{\star}^{3}$, where the last step uses that we are at fixed, finite $\xi$. Requiring this to be small gives $\Qs \gg \zs^3$. Both conditions can be summarized as
\begin{equation}
n_{\star} \gg N_{\star}^3 \gg 1 \, .
\label{oneballvalidity}
\end{equation}
This guarantees the single-ball supergravity solution is valid everywhere whenever $\xi \sim \mathcal{O}(1)$.

We can also check when the sufficient condition (\ref{asympt_mm}) for the smallness of curvature invariants in the asymptotic region holds. We find up to order one numerical factors
\begin{equation}
{\omega^4 R_{el}^4 \over N} \sim {(1 + \hat{z}_{\star}^{-2})^2 \over \xi N_{\star}^2 }
\label{asymp_zs}
\end{equation}
so for $\xi \sim {\cal O}(1)$, the condition $N_{\star} \gg 1$ is sufficient. 

There may be additional parameter regimes where some S-dual description is weakly curved and weakly coupled at least locally, but here we would expect the string coupling to become large asymptotically. 

\subsection{Near and far ball limits}
\label{subsec:OneBallEndpoints}

In this section, we describe analytic results in the limits $\hzs\gg1$ (small-$\xi$) and 
$\hzs\ll1$ (large-$\xi$).

The two limits have simple physical interpretations.  When $\hzs\gg1$, the
image conductor is far from the ball on the scale of its radius, and the
leading solution is that of an isolated conducting ball in a quadratic
external field.  When $\hzs\ll1$, the ball and its image approach a
parallel-plate configuration over most of their extent. In appendix \ref{app:ElectrostaticProblem}, we find the following leading behaviors in the two limits:
\begin{equation}
\begin{array}{c|c|c|c|c}
\text{regime}
&\nu
&\widehat\sigma(\hr)
&\widehat Q_{\star}
&\xi
\\[2pt]\hline
\hzs\gg1
&\dfrac{2}{3}
&\dfrac{\sqrt{1-\hr^2}}{\pi^2}
&\dfrac14
&\dfrac{9}{512}\hzs^{-4}
\\[9pt]
\hzs\ll1
&1
&\dfrac{1-\hr^2}{4\pi^2\hzs}
&\dfrac{2}{15\pi\hzs}
&\dfrac{1}{160\pi}\hzs^{-5}
\end{array}
\label{OneBallQuantitative:EndpointTable}
\end{equation}
The
small-$\hzs$ approximation is valid outside an edge layer of width
$1-\hr=\cO(\hzs)$ where it crosses over to the square-root behavior required for a
smooth ten-dimensional solution. 

For $\xi \ll 1$, the smallness of (\ref{asymp_zs}) that guarantees the smallness of asymptotic curvatures holds in the window
\begin{equation}
{1 \over N_{\star}^2} \ll \xi \ll 1 \; ,
\end{equation}
while for large $\xi$, the smallness of (\ref{asymp_zs}) is always satisfied. 

Demanding that the asymptotic curvature condition is satisfied, that the prefactor $z_{\star}$ in front of the metric (\ref{SUGRAsolutions:MetricNormalized}) is large and the prefactor $1/(z_{\star}^2 \omega^4)$ in front of the dilaton is small give the necessary conditions
\begin{equation}
n_{\star} \gg N_{\star} \gg 1 \qquad \qquad {1 \over N_{\star}^{1 \over 2}} \ll \omega \ll {n_{\star}^{1 \over 4} \over N_{\star}^{3 \over 4}} \; .
\label{genconstraints}
\end{equation}
which are somewhat weaker than the earlier constraints (\ref{oneballvalidity}) since we were previously assuming $\xi = {\cal O}(1)$.

\subsection{Numerical interpolation}
\label{subsec:OneBallNumerics}

For intermediate values of $\hzs$, the electrostatics solution is obtained numerically. In 
Appendix~\ref{app:ElectrostaticProblem} we introduce a transformation
\begin{equation}
h(t) = 4\pi \int_t^1 \rd x {x \hat{\sigma}(x) \over \sqrt{x^2 - t^2}}
\end{equation}
and show that the electrostatics problem reduces to a Fredholm equation of the second kind,
\begin{equation}
h(1) = 0 \, , \qquad h(x)-\int_0^1\rd t\,t\,
\mathcal K_{\hzs}(x,t)h(t)
=1-\frac{3\nu}{2}x^2,
\qquad 0\leq x\leq1,
\label{OneBallQuantitative:FredholmEquation}
\end{equation}
where the kernel is given in (\ref{K2}). This simultaneously determines the parameter $\nu(\hzs)$ and the function $h(t)$. The analytic expressions
\begin{equation}
h(t)\longrightarrow1-t^2
\quad(\hzs\to\infty) \, ,
\qquad
h(t)\sim\frac{2}{3\pi\hzs}(1-t^2)^{3/2}
\quad(\hzs\to0) \, ,
\label{OneBallQuantitative:HEndpoints}
\end{equation}
provide checks of the numerical results.

Once $h$ is known, the remaining electrostatics quantities can be reconstructed, for example,
\begin{equation}
\hat{\sigma}(x) = -{1 \over 2\pi^2 x} {\rd \over \rd x} \int_x^1 \rd t {t h(t) \over \sqrt{t^2 - x^2}} \, , \qquad \qquad 
\widehat Q_{\star}
=\int_0^1\rd t\,t h(t) \; .
\label{OneBallQuantitative:QFromH}
\end{equation}
From $\hat{\sigma}$, we can directly calculate the multipole moments $q_{2k+1}$ according to 
\begin{equation}
q_{2k+1} = {4 \pi V_0 \rho_{\star}^2 \over k+1} \int_0^1 \rd x \, x^2 \hat{\sigma}(x) \mathcal{H}_k(\rho_{\star} x, z_{\star})
\end{equation}
where $\mathcal{H}_k$ is the polynomial defined in (\ref{defH}). We can then finally use (\ref{ProtectedMultipoles:MainResult}) to calculate the expectation value of $\langle \Tr \phi^{2k} \rangle$.

To present the results, it will be useful to define normalized quantities
\begin{equation}
{\cal T}_k = {(2 k + 1) \over n_{\star} N_{\star}} \left({2 \over \omega N_{\star}}\right)^{2k} \langle \Tr \phi^{2k} \rangle = {q_{2k+1} \over 2 Q_{\star} z_{\star}^{2k+1}} \; .
\end{equation}
The normalization is chosen so that in the electrostatics approximation, the left side is a function only of $\xi$ and furthermore approaches 1 for $\xi \to 0$. In the electrostatics language, we are normalizing the multipoles by their value for a point charge $Q_{\star}$ at the height $z_{\star}$ together with its image charge.

\begin{figure}[h!]
  \centering
\includegraphics[width= \textwidth]{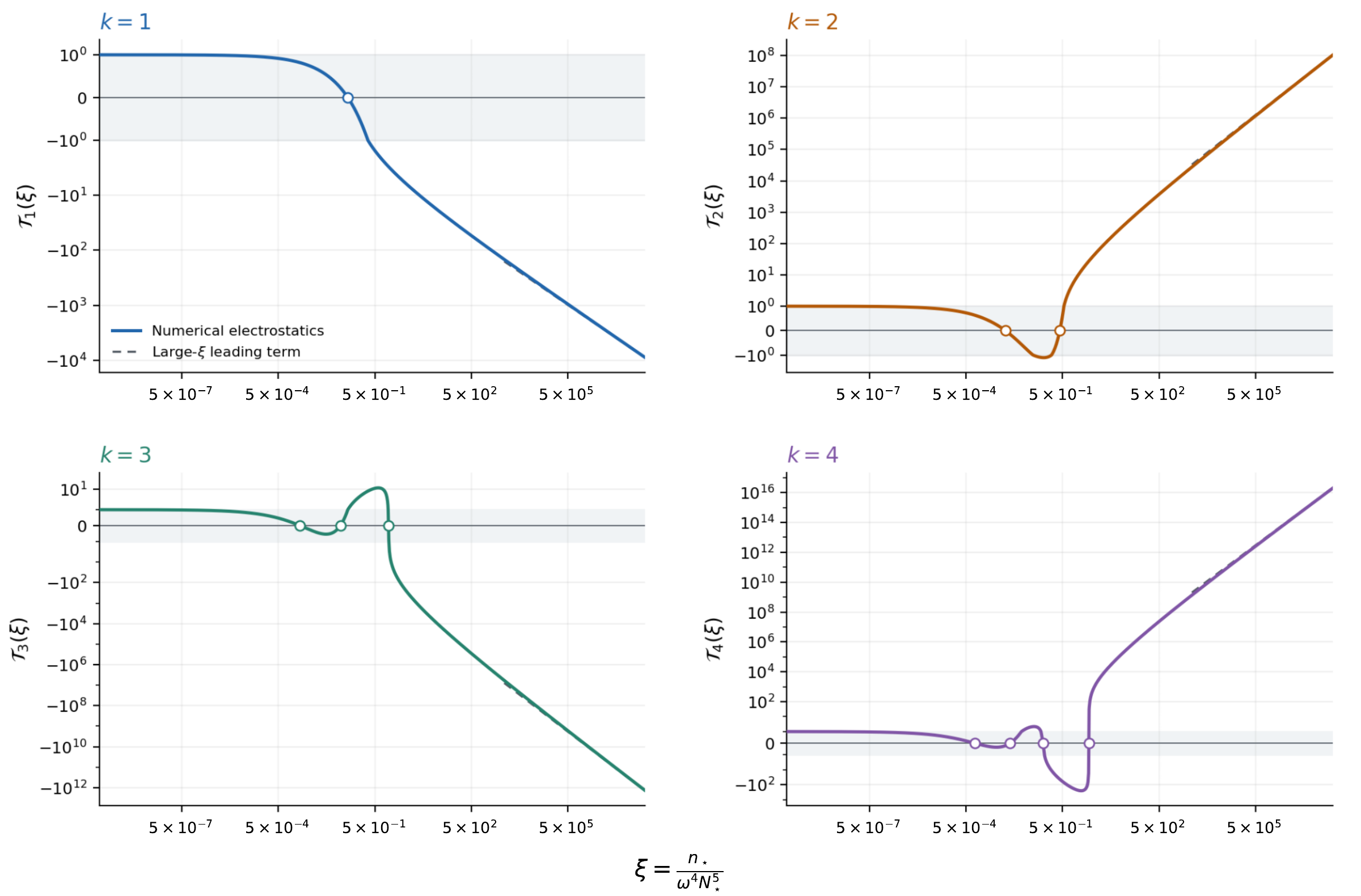}
  \caption{Normalized traces ${\cal T}_k = {\cal N}_k \langle \Tr(\phi^{2k}) \rangle$ versus $\xi$ for single-ball sectors in the electrostatics approximation. The vertical axis is linear in the shaded region and logarithmic outside.}
  \label{fig:traces}
\end{figure}

The numerical results for ${\cal T}_k(\xi)$ for $k=1,2,3,4$ are shown in Figure \ref{fig:traces}. For these examples, ${\cal T}_k$ has $k$ zeros whose corresponding $\xi$ values span several orders of magnitude for $k=3,4$.

To exhibit the additional dependence of $\mathcal{T}_k$ on $N_{\star}$ and $n_{\star}$ away from the limits where electrostatics is valid, we also evaluated the localization integral for $\mathcal{T}_1$ directly in sectors $(n_{\star},N_{\star})$ at fixed total matrix size $N=n_{\star}N_{\star}$ equal to $N=32$ and $N=128$. We retained the exact finite-$N_{\star}$ one-loop determinant and the discrete $L_3$ spectrum, and integrated the Gaussian $U(1)$ center-of-mass modulus analytically. The $n_{\star}=1$ sector is then completely analytic, while the $n_{\star}=2$ sector reduces to a one-dimensional integral evaluated numerically. Results for the remaining sectors were obtained using a Monte Carlo algorithm designed by ChatGPT6.\footnote{The maximum estimated relative standard error in the positive density moment is $\pm 0.84 \%$.} Each sector is plotted against its own $\xi=n_{\star}/(\omega^4N_{\star}^5)$, so equal horizontal coordinates correspond to different values of $\omega$ in different sectors.

\begin{figure}[h!]
  \centering
\includegraphics[width= \textwidth]{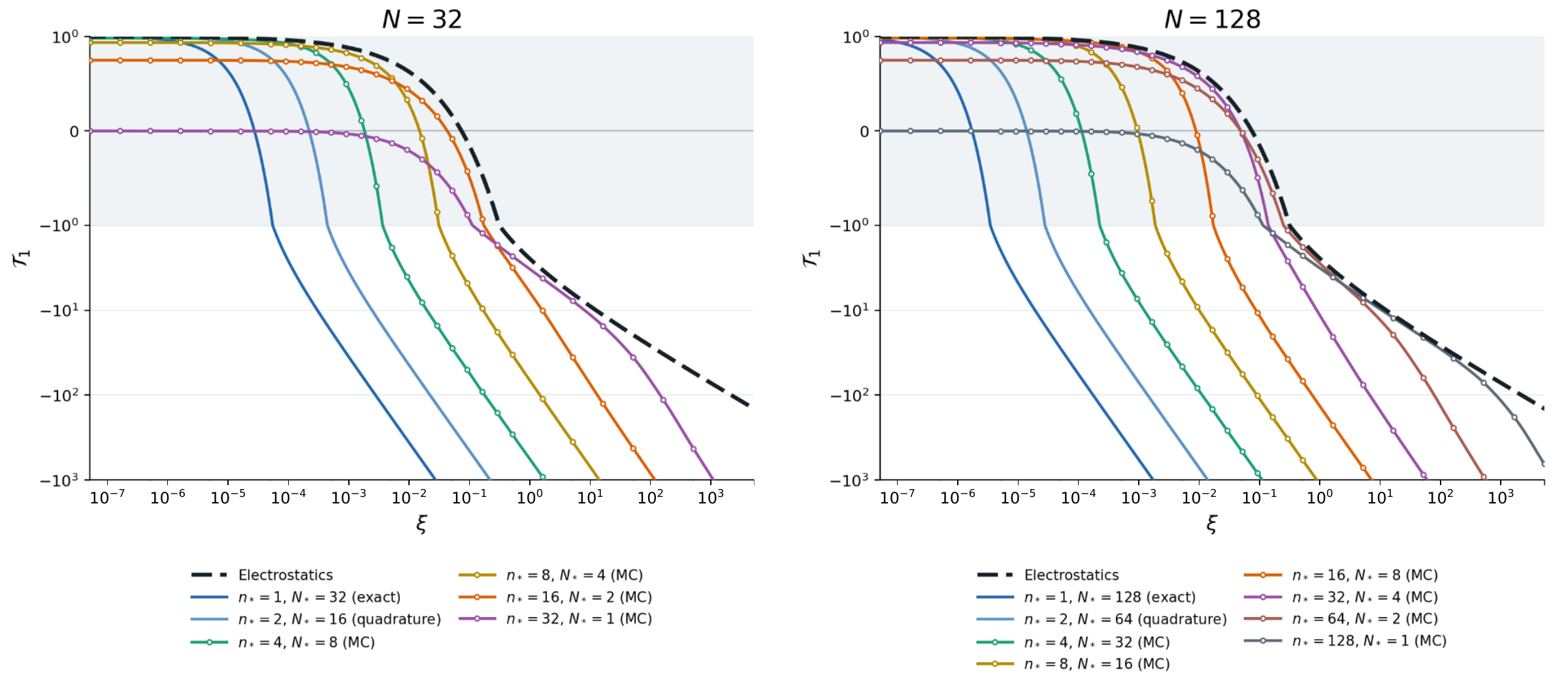}
  \caption{Full localization results for normalized traces ${\cal T}_k = {\cal N}_k \langle \Tr(\phi^{2k}) \rangle$ versus $\xi$ for  $(n_{\star},N_{\star})$ sectors with fixed $N= n_{\star} N_{\star}$. Each sector is plotted against its own $\xi=n_{\star}/(\omega^4N_{\star}^5)$, so equal horizontal coordinates correspond to different values of $\omega$ in different sectors.}
  \label{fig:traces1}
\end{figure}

\section{Saddle-decoupling limits}
\label{sec:Limits}

We will now study the saddles containing several copies of a single irreducible representation in different limits in which some region of the geometry decouples from the rest. 

\subsection{Large $\zs$: the D1-brane limit}
\label{subsec:D1}

In the electrostatics problem, if we send $\zs \to \infty$ while keeping the radial size of the charged ball fixed we land in a situation in which there is a single charged ball in a background potential, but no trace of the infinite grounded plane. We will now study the effects of this limit in the supergravity background. In terms of the normalized parameter $\hzs$, we are taking $\hzs \to \infty$, so according to the results of Section \ref{subsec:OneBallsolutions}
\begin{equation}
\nu(\hzs) \longrightarrow \frac{2}{3} \, , \qquad \xi(\hzs) \sim \frac{9}{512} \frac{1}{\hzs^4} \, , \qquad \hat{Q}_{\star} \longrightarrow \frac14 \, .
\end{equation}
We look at the region with coordinate $z = \zs + x$ for $x$ fixed as we make $\zs$ large (so $x/\zs \to 0$). Since we want to keep $\rs$ fixed, the total charge of the ball scales as
\begin{equation}
\Qs = \frac{9}{16} \omega^4 \zs \rs^4  \, ,
\end{equation}
so in this limit we are keeping $\Qs / (\omega^4 \zs)$ fixed.

The background potential in terms of the coordinate $x$ becomes
\begin{equation}
V_{\rm bg}(\rho, x) = - \eta \zs - \frac32\omega^4 \zs^3 - \left( \eta + \frac92\omega^4 \zs^2 \right) x + \frac32\omega^4 \zs \left( \rho^2 - 3 x^2\right) + \frac32\omega^4 \left( x\rho^2 - x^3 \right) \, .
\end{equation}
Notice that the metric depends only on $V''$, $\dot{V}$, and $\dot{V}'$. This means that the constant piece and the $\cO(x)$ terms of the background potential do not affect the geometry. In order for the other pieces not to blow up, we take $\omega^4 \zs$ fixed as $\zs \to \infty$ (which implies $\Qs$ is also fixed). The last term in $V_{\rm bg}$ can then be ignored, but note that the constant and $\cO(x)$ terms diverge. Even though this does not affect the metric (or the NSNS-fields $\Phi$ and $B_2$), the RR-fields $C_0$ and $C_2$ are affected by these diverging pieces. A constant axion shift accompanied by the corresponding shift of $C_2$ leaves the local gauge-invariant field strengths unchanged. If we redefine
\begin{equation}
C_0' = C_0 + \frac{\zeta}{54\pi} \, , \qquad C_2' = C_2 + \frac{\zeta}{54\pi} B_2 \, ,
\end{equation}
it is simple to check that the primed fields are equivalent to those obtained from a shifted background potential
\begin{equation}
V_{\rm bg}'(\rho, x) = V_{\rm bg}(\rho, x) - \zeta \zs - \zeta x \, .
\end{equation}
We can choose $\zeta = - \eta - \frac92\omega^4 \zs^2$ to cancel the term linear in $x$. This still leaves a diverging constant piece, but we can clearly add a constant to the potential without changing the electrostatic problem at all. At the level of the background, this just adds a term proportional to $\omega_2$ to $C_2$, which leaves the local field strengths invariant.  Note that $B_2$ also has a constant divergence once we write $z = \zs + x$; this can be removed in an identical way.

In summary, the supergravity background can be obtained in this limit from a potential $V(\rho, x) = V_{\rm bg}(\rho, x) + \psi(\rho, x)$, where
\begin{equation}
V_{\rm bg}(\rho, x) = W_0 (\rho^2 - 3 x^2) \, ,
\end{equation}
and $\psi(\rho, x)$ is the potential generated by a charged ball of total charge $\Qs$ at $x = 0$, with radial size $\rs$ adjusted to have vanishing charge density at the edge. We emphasize that $W_0 = \frac32\omega^4 \zs$ is a finite parameter in this limit. In terms of matrix model parameters, we are looking at a saddle with $n_{\star}$ copies of the $N_{\star}$-dimensional irreducible representation, and we are taking
\begin{equation}
N_{\star} \to \infty \, , \qquad n_{\star} = Q_{\star} \;\; {\rm fixed} \, , \qquad \omega^4 N_{\star} = \frac43 W_0 \;\; {\rm fixed} \, .
\end{equation}
From the matrix model perspective, this limit turns the fuzzy sphere associated with the large-$N_{\star}$ irreducible representation into a commutative $S^2$, with an effective $U(n_\star)$ gauge theory describing the $n_\star$ coincident copies. This suggests an interpretation in terms of Euclidean D1-branes wrapping $S^2$.

In order to test this and understand better the supergravity background, we can look at its asymptotic region $r=\sqrt{x^2 + \rho^2} \to \infty$. The potential $\psi(\rho, x)$ can be expanded in multipoles much like before, but now the charge distribution is invariant under $x \to -x$, so we only get even multipoles
\begin{equation}
V(\rho, x) = W_0 (\rho^2 - 3 x^2) + \frac{\Qs}{\rho^2 + x^2} + \sum_{l=1}^{\infty} \frac{q_{2l}}{(\rho^2 + x^2)^{l+1}} U_{2l}\left( \frac{x}{\sqrt{\rho^2 + x^2}} \right) \, ,
\end{equation}
where
\begin{equation}
q_{2l} = (-1)^l \frac{4 \pi}{2 l + 1} \int_0^{\rs} \rd \rho \, \rho^{2l+2} \sigma(\rho) \, .
\end{equation}
In appendix \ref{app:ElectrostaticProblem}, we obtain explicitly the charge density of the ball in the limit we are working, $\hzs \gg 1$. From those results, it is immediate to obtain all the multipole moments:
\begin{equation}
q_{2l} = (-1)^l \frac{3}{2^{2l+2}} \frac{(2l)!}{l!(l+2)!} W_0 \rs^{2l+4} \, ,
\end{equation}
where $\rs$ is fixed in terms of the charge (or number of copies of the representation) by
\begin{equation}
\Qs = \frac{9}{16} \omega^4 \zs \rs^4 = \frac38 W_0 \rs^4 \, .
\end{equation}

Using the previous expansion, we can read the asymptotic form of the metric
\begin{equation}
\label{Limits:D1Asymptotics}
\frac{\rd s^2}{\alpha'} = \frac{6\pi \rs^2}{\sqrt{2} r^3} \left( \rd r^2 + r^2 \rd \Omega_7^2 \right) + \frac{2\pi \sqrt{2}}{3 \rs^2}  r^3 \rd \Omega_2^2 + \dots \, ,
\end{equation}
where $r = \sqrt{\rho^2 + x^2}$, and the subleading terms break the $SO(8) \times SO(3)$ symmetry into $SO(7) \times SO(3)$ because they depend separately on $x$ and $\rho$. This is the asymptotic geometry expected for a stack of $n_\star$ spherical D1-branes wrapping $S^2$, as studied in \cite{BobevEtAl2018SphericalBranes}. Indeed, defining $g_s=3/(2W_0)$ and $H(r)=18\pi^2\rho_\star^4/r^6=32\pi^2g_sn_\star/r^6$, the metric takes the standard near-horizon D1-brane form,
\begin{equation}
\frac{\rd s^2}{\alpha'} = H^{1/2} (\rd r^2 + r^2 \rd\Omega_7^2) + (2\pi)^2 H^{-1/2} \rd \Omega_2^2 + \dots \, ,
\end{equation}
with the expected dilaton profile $e^\Phi=g_sH^{1/2} + \dots$. Furthermore, the D1-charge can be read from the leading term of $C_2$ ($C_0 H_3$ gives a subleading contribution to $\cF_3$)
\begin{equation}
C_2 = - \frac{4 \alpha' W_0}{27 \rs^4} r^6 \omega_2 + \dots \; \Rightarrow \; N_{D1} = \frac{1}{(2 \pi)^6 \alpha'^3} \int_{S^7_{\infty}} \star\, \cF_3 =  \Qs = n_{\star} \, .
\end{equation}

Note that, after taking the limit, the original asymptotic D($-1$) region has disappeared: it is at $x , \rho \sim \mathcal{O}(z_{\star})$, which is a scale we are not accessing in this limit. We are looking at a region of the full geometry of typical size $\rho_{\star} \ll z_{\star}$. In fact, the condition \eqref{SUGRAsolutions:ValidityAsymptoticRegion} for the validity of the original D($-1$) asymptotic region is badly violated
\begin{equation}
\frac{\omega^4R_{\rm el}^4}{P}
=\frac{\omega^4(\zs^2+\rs^2)^2}{2\zs\Qs}
\sim\frac{W_0\zs^2}{3\Qs}
=\frac89\frac{\zs^2}{\rs^4}
\longrightarrow\infty.
\end{equation}
Despite this, the D1 region we just discussed can be a valid supergravity background. The scaling properties of the potential imply that, in this region, the curvature scale and string coupling behave as
\begin{equation}
\alpha' R \sim \frac{1}{\rs} \, , \qquad e^{\Phi} \sim \frac{1}{W_0 \rs} \, ,
\end{equation}
where we have used that $W_0 \to c L W_0$ under the scaling. Requiring $\rs \gg 1$ and $W_0 \rs \gg 1$, we then conclude that the D1 background is valid in the regime
\begin{equation}
n_{\star} \gg \omega^4 N_{\star} \gg n_{\star}^{-1/3} \, .
\end{equation}
This makes the background valid in any fixed, finite region, but it cannot avoid potential asymptotic issues. In fact, as expected for a D1 background, the curvature of the $S^7$ is blowing up asymptotically in \eqref{Limits:D1Asymptotics}. This produces a large curvature as $r \to \infty$:
\begin{equation}
\alpha' R \sim \frac{r}{\rs^2} \, .
\end{equation}

From the matrix model point of view, this limit corresponds to one where the non-commutative field theory defining the physics of the $n_{\star}$ coincident fuzzy spheres becomes a commutative $U(n_{\star})$ Euclidean supersymmetric field theory on $S^2$ associated with spherical Euclidean D1-branes. The analogous limit for the BMN model, giving a 2+1 dimensional theory on $S^2 \times \mathbb{R}$ was described in \cite{Maldacena:2002rb,Ling:2006up}. 

\subsection{Small $\zs$: the NS5-brane limit}
\label{subsec:NS5}

We now consider the opposite limit in which $\zs$ becomes small compared to $\rs$, so that $\hzs \to 0$. It is perhaps more natural to think of this regime as sending $\rs \to \infty$ at fixed $\zs$. We will keep $\rho$ fixed and $0 < z < \zs$, so that we look at the region between the infinite grounded plate at $z = 0$ and a charged ball at fixed potential $V_{\star}$ at $z = \zs$. This ball is becoming effectively another infinite plate as $\rs \to \infty$, but we need to understand how this regime is approached from the finite solution in order to keep a well-behaved supergravity background. Note that, for $0 < \rho < \rs$ and $0 < z < \zs$, any solution to the Laplace equation satisfying the above boundary conditions together with regularity at $\rho = 0$ is of the form
\begin{equation}
\label{Limits:PotentialSineExpansion}
V(\rho, z) = \frac{V_{\star}}{\zs} z + \sum_{n=1}^{\infty} c_n \frac{\zs}{n \pi \rho} \sinh\left( \frac{n \pi \rho}{\zs} \right) \sin \left( \frac{n \pi z}{\zs} \right) \, ,
\end{equation}
for some constants $c_n$. If we know the full solution, these constants can be obtained from an inverse sine transform at any fixed $\rho \in [0, \rs]$ as
\begin{equation}
c_n = \frac{2}{\zs^2} \frac{n \pi \rho}{\sinh(n \pi \rho / \zs)} \int_0^{\zs} \rd z\, \left( V(\rho,z) - \frac{V_{\star}}{\zs} z  \right) \sin \left( \frac{n \pi z}{\zs} \right) \, .
\end{equation}

A general solution at finite $\rs$ will have all the $c_n$ non-zero. But, if $\omega$ and $\zs$ are fixed, the conformal mapping analysis in appendix \ref{app:SmallZsLimit} shows that $c_1 \to 0$ as $\rs \to \infty$. The bound below then shows that all the $c_n$ vanish and we are left with just the standard constant-field parallel plate solution, $V(\rho, z) = V_{\star} z / \zs$. This does not give a valid supergravity background, so we must take a different scaling limit in which some other parameter scales with $\rs$.

Note however the following: at finite $\rs$, we know that $-  V'' > 0$ \cite{Komatsu:2024ydh} (this is required to have a real, Euclidean supergravity solution). Using the sine expansion \eqref{Limits:PotentialSineExpansion} for $V''$ and inverting to obtain the coefficients $c_n$ from $V''$, it is immediate to show that $c_1 \geq 0$. Furthermore, using $|\sin(nx)| \leq n \sin(x)$ for $0 \leq x \leq \pi$ and $n \in \mathbb{N}$, we get
\begin{equation}
|c_n| \leq c_1 \frac{\sinh(\pi \rho/ \zs)}{\sinh(n \pi \rho/\zs)} \, , \qquad 0 \leq \rho \leq \rs \, .
\end{equation}
Evaluating this bound at $\rho=\rs/2$ and sending $\rs \to \infty$ for $n \geq 2$, if $c_1$ stays finite the right-hand side is exponentially suppressed, so $c_n \to 0$ for all $n \geq 2$. On any fixed compact region of the gap, the radial factors grow at most exponentially in $n$ with a fixed exponent, so the same bound makes the sum of the higher modes vanish. Thus, the only finite limiting solution we can obtain from our original finite-$\rs$ potential is of the form
\begin{equation}
\label{Limits:PotentialLimitC1}
V(\rho, z) = \frac{V_{\star}}{\zs} z + c_1 \frac{\zs}{\pi \rho} \sinh\left( \frac{\pi \rho}{\zs} \right) \sin \left( \frac{\pi z}{\zs} \right) \, , \qquad 0 < \rho < \infty \, ,
\end{equation}
with $c_1 \geq 0$.

The question is then how to scale the parameters in order to keep $c_1 > 0$ finite as $\rs \to \infty$, so that in the limit we get a valid supergravity background. We give a detailed analysis of how to do this in appendix \ref{app:SmallZsLimit}; the end result is that to keep $c_1$ fixed as $\rs \to \infty$, we must take
\begin{equation}
\label{Limits:LambdaScalingSmallZstar}
\omega^4 = \frac{\pi e c_1}{6\zs \rs^2} e^{\pi \rs / \zs} [1+o(1)] \to \infty \, .
\end{equation}
We also obtained that in this regime, keeping $\zs$ fixed, we must have
\begin{equation}
\xi = \frac{Q_{\star}}{32\omega^4 \zs^5} \approx \frac{1}{160\pi} \frac{\rs^5}{\zs^5} \, .
\end{equation}
The leading charge--radius relation is not sufficiently accurate to substitute directly into the exponential in \eqref{Limits:LambdaScalingSmallZstar}. Defining
\begin{equation}
\label{Limits:ChargeRadiusScale}
R = \left(\frac{5\pi \Qs}{\omega^4}\right)^{1/5} \, ,
\end{equation}
the exact conformal solution gives
\begin{equation}
\label{Limits:RadiusCorrectionSmallZstar}
\rs = R - \frac{\zs}{\pi}\left(\log\frac{2\pi R}{\zs}-\frac12\right)
+\cO\left(\frac{\zs^2}{R}\right) \, .
\end{equation}
The logarithmic correction changes the power multiplying the exponential. As derived in appendix \ref{app:SmallZsLimit}, the first coefficient is equivalently
\begin{equation}
\label{Limits:C1ChargeScaling}
c_1 = \frac{12\omega^4 R^3}{e^{3/2}}e^{-\pi R/\zs}
\left[1+\cO\left(\frac{\zs}{R}\right)\right] \, .
\end{equation}
With these results we can read the limit that produces the solution \eqref{Limits:PotentialLimitC1} for the potential. It is 
\begin{equation}
\begin{gathered}
\omega \to \infty \, , \quad \Qs \to \infty \, , \quad \zs \;\;{\rm fixed} \, , \\
Q_{\star}^{3/5} \omega^{8/5} \exp \left( - A \frac{\Qs^{1/5}}{\omega^{4/5}}\right)
= \frac{\pi^{3} e^{3/2} c_1}{12A^3 \zs^3} \; \;{\rm fixed}\,,
\end{gathered}
\end{equation}
with $A^5 = 5\pi^6 / \zs^5$. In terms of matrix model parameters, this is
\begin{equation}
\begin{gathered}
\omega \to \infty \, , \quad n_{\star} \to \infty \, , \quad N_{\star} \;\; {\rm fixed} \, , \\
n_{\star}^{3/5} \omega^{8/5} \exp \left( - A \frac{n_{\star}^{1/5}}{\omega^{4/5}} \right)
= \frac{\pi^3 e^{3/2} c_1}{12 A^3 \zs^3} \; \; {\rm fixed} \, ,
\end{gathered}
\end{equation}
with $A^5 = 5\pi^6 / \zs^5 = 160\pi^6/N_{\star}^5$.

What is the geometry and matter configuration that results from the limiting potential \eqref{Limits:PotentialLimitC1}? It is convenient to write it using modified spherical Bessel functions of the first kind, defined recursively as
\begin{equation}
i_0(x) = \frac{\sinh(x)}{x} \, , \qquad i_{n+1}(x) = x^n \frac{\rd }{\rd x} \left( x^{-n} i_n(x) \right) \;\;{\rm for}\; n \in \mathbb{N} \, .
\end{equation}
These capture the $\rho$ dependence of the potential, and produce the string frame geometry
\begin{equation}
\frac{\rd s^2}{\alpha'} = 2\zs \sqrt{\frac{i_2}{i_0}} \left( \frac{i_0}{i_1} \left( \rd y^2 + \rd \theta^2 \right) + 3 y \, \rd \Omega_6^2 + \frac{i_0 i_1 \sin^2(\theta)}{i_0 i_2 \sin^2(\theta) + i_1^2 \cos^2(\theta)} \rd \Omega_2^2 \right) \, ,
\end{equation}
where we have introduced rescaled coordinates
\begin{equation}
y = \frac{\pi}{\zs} \rho \in (0, \infty) \, , \qquad \theta = \frac{\pi}{\zs} z \in (0, \pi) \, ,
\end{equation}
and all the $i_n$ are evaluated at $y$, $i_n \equiv i_n(y)$. The remaining fields are
\begin{subequations}
\begin{align}
e^{\Phi} & = \frac{54\zs}{c_1} \frac{i_2}{\sqrt{3 y \, i_0 i_1 \left( i_0 i_2 \sin^2(\theta) + i_1^2 \cos^2(\theta) \right) }} \, , \\
C_0 & = - \frac{V_{\star}}{54\pi\zs} - \frac{c_1}{54\zs} \left( i_0 - 3 \frac{i_1^2}{i_2} \right) \cos(\theta) \, , \\
B_2 & = - 2\alpha'\zs \left( \theta - \frac{i_1^2 \sin(\theta) \cos(\theta)}{i_0 i_2 \sin^2(\theta) + i_1^2 \cos^2(\theta)} \right) \omega_2 \, , \\
\nonumber C_2 & = \frac{\alpha'}{27} \left( \frac{V_{\star}}{\pi} \theta + c_1 i_0 \sin(\theta) \right. \\
& \qquad \qquad \quad \left. - c_1 \frac{i_1^2 \sin(\theta) \left( \frac{V_{\star}}{c_1 \pi} \cos(\theta) + i_0 (\cos^2 (\theta) + 3 \sin^2(\theta)) \right)}{i_0 i_2 \sin^2(\theta) + i_1^2 \cos^2(\theta)} \right) \omega_2 \, .
\end{align}
\end{subequations}

In order to show that this solution corresponds to a NS5 background, we can study it asymptotically. The geometry becomes
\begin{equation}
\frac{\rd s^2}{\alpha'} = 2\zs \left( \rd y^2 + \rd \Omega_3^2 + 3 y \, \rd \Omega_6^2 \right) + \cO(y^{-1}) \, , \qquad \rd \Omega_3^2 \equiv \rd \theta^2 + \sin^2(\theta) \rd \Omega_2^2 \, .
\end{equation}
This suggests we have Euclidean NS5-branes wrapping the $S^6$ -- note the cylindrical transverse space $\rd y^2 + \rd \Omega_3^2$. Classical supergravity backgrounds describing such spherical NS5-branes were constructed in \cite{BobevEtAl2018SphericalBranes}. The interpretation is further supported by the asymptotically linear dilaton,
\begin{equation}
e^{\Phi} = \frac{54\zs}{c_1} \sqrt{\frac{4 y}{3}} e^{-y} \left(1 + \cO(y^{-1} ) \right) \, ,
\end{equation}
together with the appropriate $H_3$ flux through the $\mathbb{S}^3$:
\begin{equation}H_3 = - 4\alpha'\zs \sin^2(\theta) \rd \theta \wedge \omega_2 \, , \qquad N_{NS5} = \frac{1}{4 \pi^2 \alpha'} \int_{\mathbb{S}^3_{\rm NS5}} H_3 = 2\zs = N_{\star} \, .
\end{equation}
With the same orientation, we can see that the form of the  solution presented contains also D5 charge, $N_{D5} = - V_{\star} / (27 \pi)$. However, this can be removed by using the axion shift symmetry:
\begin{equation}
C_0 \rightarrow C_0 + \frac{V_{\star}}{54 \pi \zs} \, , \qquad C_2 \rightarrow C_2 + \frac{V_{\star}}{54 \pi \zs} B_2 \, ,
\end{equation}
which removes all dependence on $V_{\star}$ from the RR potentials. Note that this corresponds to shifting the potential by $V \rightarrow V - V_{\star} z / \zs$.

Much like in the D1 case, here we are looking at a region of size $\zs \ll \rs$, so we have lost the original asymptotic D$(-1)$ part. In this case, the decoupling asymptotic region becomes weakly curved
\begin{equation}
 \frac{\omega^4R_{\rm el}^4}{P}
 \sim\frac{1}{\zs\rs}\longrightarrow0,
 \qquad P=2\zs\Qs,\quad R_{\rm el}\sim\rs .
\end{equation}
The potential scaling now determines
\begin{equation}
\alpha' R \sim \frac{1}{\zs} \, , \qquad e^{\Phi} \sim \frac{\zs}{c_1} \, ,
\end{equation}
so the condition to have a valid NS5 solution is
\begin{equation}
c_1 \gg \zs \gg 1 \, .
\end{equation}
The geometry is also regular at $y=0$: the $S^6$ shrinks smoothly, the transverse three-dimensional geometry remains finite, and the dilaton approaches a finite value. Thus the conditions above control the supergravity approximation throughout the limiting geometry.

\subsection{Large $\Qs$: the 't Hooft limit}
\label{subsec:tHooft}

We can study one more limit in the electrostatic problem, in which we take
\begin{equation}
\zs \; \; {\rm fixed} \, , \qquad \rs \;\;{\rm fixed} \, , \qquad \Qs \to \infty \, .
\end{equation}
This means the normalized parameter $\hzs = \zs / \rs$ is fixed and finite, so that the solution can be found numerically as described in appendix \ref{app:ElectrostaticProblem}. Since we are keeping $\xi$ finite, this also means that $\Qs/\omega^4$ is fixed, so $\omega \to \infty$ appropriately. In terms of matrix model parameters, 
\begin{equation}
N_{\star} \; \;{\rm fixed} \, , \qquad n_{\star} \to \infty \, , \qquad \omega^4 \to \infty \, , \qquad \frac{n_{\star}}{\omega^4} \;\;{\rm fixed} \, .
\end{equation}

Regarding the supergravity background, to analyze this limit it is convenient to use the normalized form in \eqref{SUGRAsolutions:MetricNormalized} and \eqref{SUGRASolutions:FieldsNormalized}. The metric and dilaton are
\begin{equation}
\frac{\rd s^2}{\alpha'} = \zs \rd \hat{s}^2 \, , \qquad e^{\Phi} = - \frac{36\pi}{\omega^4 \zs^2} \hat{\mathcal{G}}(\hzs) \, ,
\end{equation}
where both $\rd \hat{s}^2$ and $\hat{\mathcal{G}}$ depend only on $\hzs$, and thus are finite. Since we are keeping $\zs$ fixed, the geometry is fixed, but the string coupling $e^{\Phi}$ goes to zero as $\omega^{-4} \to 0$. This is thus a limit in which we have weakly-coupled string theory on the previous background.

\section{Supergravity saddles from matrix model basins?}
\label{sec:Basins}

The polarized IKKT model provides a particularly fascinating example of a holographic duality since the gravity theory is a full-fledged gravitational model whose low-energy description includes (Euclidean) Einstein gravity in higher dimensions while the matrix model itself is defined as a simple matrix integral without time or spatial coordinates. 

We have seen that protected observables in the matrix model, computed by localization, split exactly into a sum of contributions associated with the various supersymmetric saddles. The contribution associated with a representation $\mathcal{R}$ is naturally associated with an observable in the gravitational saddle labelled by the same representation. For example, we have seen that matrix model free energy $\log Z_\mathcal{R}$ matches the supergravity action evaluated on the solution $M_\mathcal{R}$ for the corresponding saddle, and the matrix model expectation value $\langle \Tr(\phi^{2n}) \rangle_\mathcal{R}$ can be read off from the asymptotic supergravity fields in the solution $M_\mathcal{R}$.

It would be ideal to be able to establish evidence that the duality holds for observables that cannot be computed by supersymmetric localization, for example $\Tr(X_3^2)$ or $\Tr(X_{10}^2)$. In principle, we could compute these directly via a Monte Carlo evaluation of the original matrix integral. The $SU(2)$ results for such a calculation were already reported in \cite{Chou:2025rwy}. Using Claude, we have performed  calculations for the $SU(3)$ and $SU(4)$ models. In each case, we can compare with the large $\omega$ and small $\omega$ perturbative results, and for the localization calculations of $\langle \Tr(\phi^2) \rangle = \langle\Tr(X_3^2) - \Tr(X_{10}^2) \rangle$. After integrating out the fermions, the resulting Pfaffian is complex for $N>2$, but the variation in the phase $e^{i \theta}$ from 1 is small for these small values of $N$ so there is no significant sign problem.

\begin{figure}
    \centering
\includegraphics[width=0.7\linewidth]{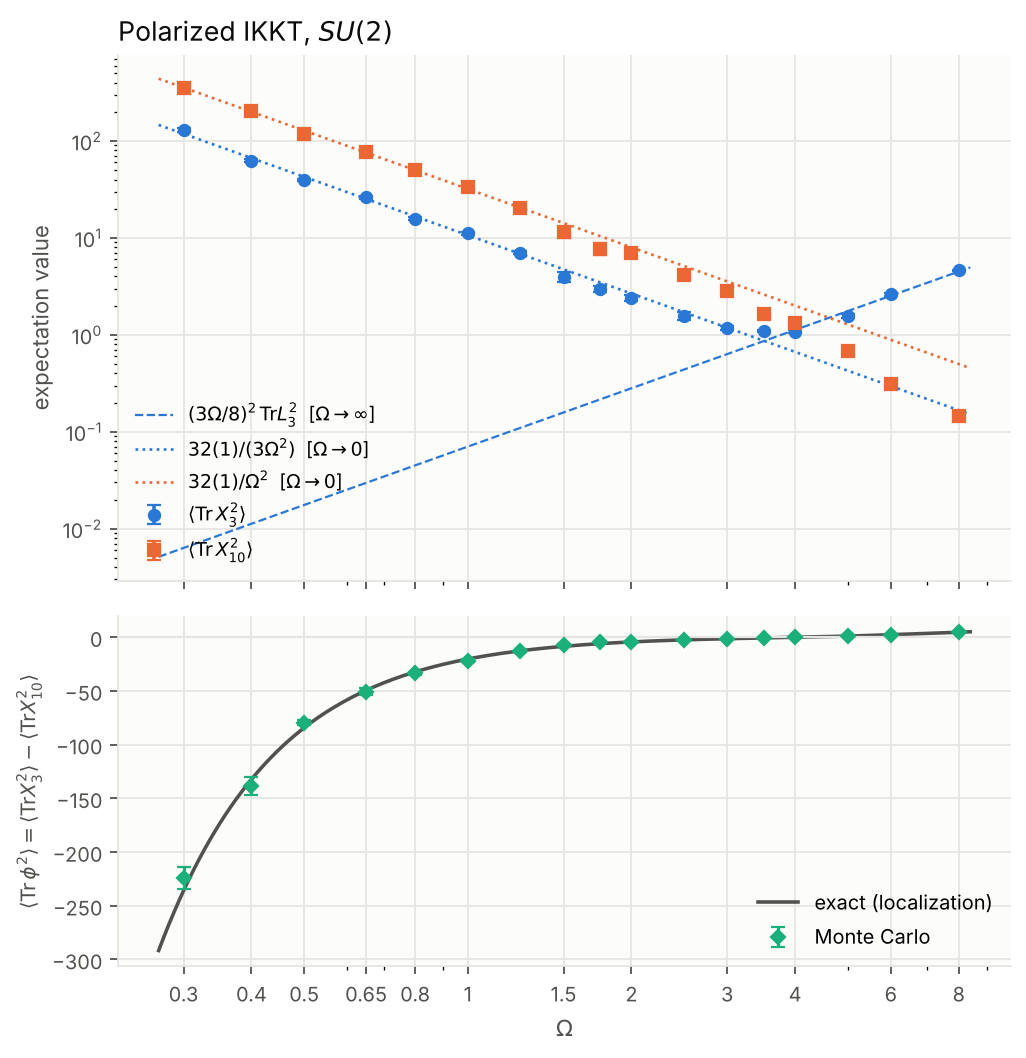}
    \caption{Monte Carlo results for the evaluation of $\langle \Tr(X_3^2)\rangle$ and $\langle \Tr(X_{10}^2)\rangle$ in the $SU(2)$ polarized IKKT matrix model, as a function of $\Omega = 8/3 \omega$. These are compared with perturbative results at small and large $\omega$ (see section \ref{sec:Background}) and with the localization results for $\langle \Tr(X_3^2 -X_{10}^2)\rangle$.}
    \label{fig:su2_traces}
\end{figure}

\begin{figure}
    \centering
\includegraphics[width=0.7\linewidth]{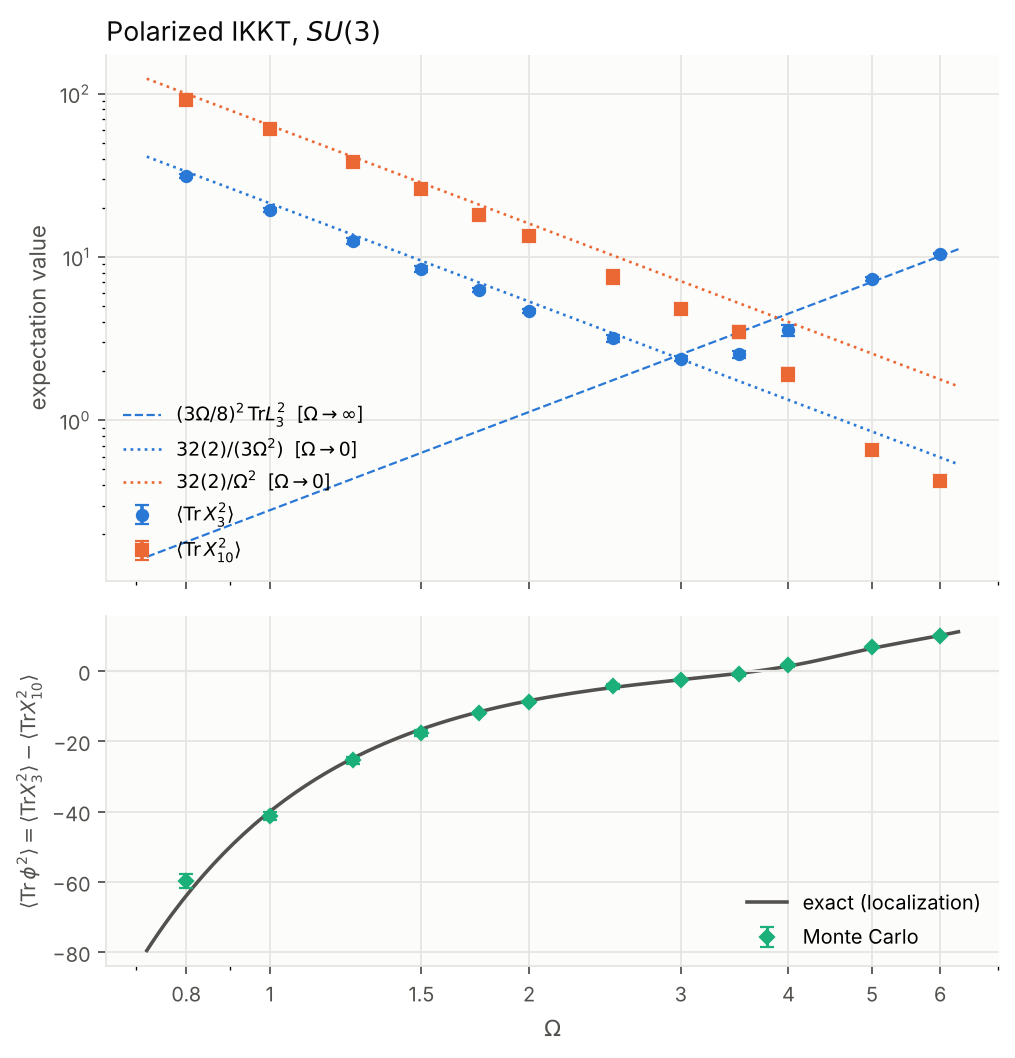}
    \caption{Monte Carlo results for the evaluation of $\langle \Tr(X_3^2)\rangle$ and $\langle \Tr(X_{10}^2)\rangle$ in the $SU(3)$ polarized IKKT matrix model, as a function of $\Omega = 8/3 \omega$. These are compared with perturbative results at small and large $\omega$ (see section \ref{sec:Background}) and with the localization results for $\langle \Tr(X_3^2-X_{10}^2)\rangle$.}
    \label{fig:su3_traces}
\end{figure}

\begin{figure}
    \centering
\includegraphics[width=0.7\linewidth]{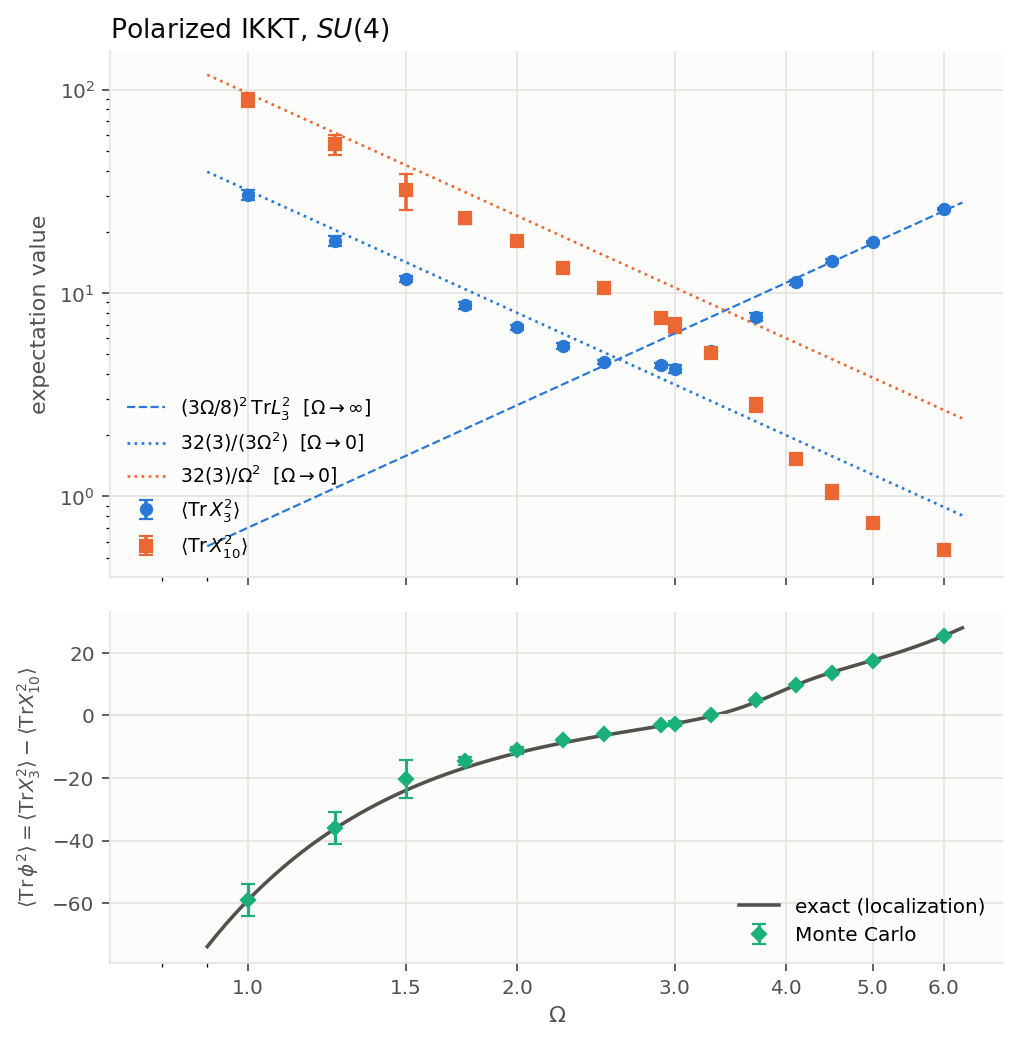}
    \caption{Monte Carlo results for the evaluation of $\langle \Tr(X_3^2)\rangle$ and $\langle \Tr(X_{10}^2)\rangle$ in the $SU(4)$ polarized IKKT matrix model, as a function of $\Omega = 8/3 \omega$. These are compared with perturbative results at small and large $\omega$ (see section \ref{sec:Background}) and with the localization results for $\langle \Tr(X_3^2-X_{10}^2)\rangle$.}
    \label{fig:su4_traces}
\end{figure}

Our results for the Monte Carlo evaluations of $\Tr(X_3^2)$ or $\Tr(X_{10}^2)$ are shown in Figures \ref{fig:su2_traces}, \ref{fig:su3_traces}, and \ref{fig:su4_traces}. We see that the direct calculations produce excellent agreement both with the perturbative results and with the localization calculation for $\Tr(\phi^2)$. A Claude analysis indicates that using reasonable compute power, a similar analysis could be extended to $N \sim 12–16$. Specifically, the low $N$ analysis found that the phase average $\langle e^{i \theta} \rangle$ fell from 1 (exactly) at $N=2$ to 0.95 at $N = 3$ and 0.86 at $N = 4$ (worst at $\omega \approx 1.125–1.3125$). The estimate of maximum $N$ for Monte Carlo simulations assumes $\log \langle e^{i \theta} \rangle$ continues to change with the number of degrees of freedom at the same rate and uses that the statistics needed to overcome phase fluctuations tend to grow as the inverse square of the $e^{i \theta}$ average \cite{Troyer_2005}. 

Being able to obtain results for arbitrary observables at $N \sim 12-16$ might already be enough for making an interesting comparison with supergravity, especially for quantities whose finite $N$ corrections scale as $1/N^2$. However, there is a significant conceptual issue to overcome. The Monte Carlo results we have described correspond to the full polarized IKKT path integral, which gets contributions from all saddles. However, there is no regime of parameters where supergravity is valid for every gravitational saddle. For example, the result (\ref{genconstraints}) from our analysis of the $(n_{\star},N_{\star})$ saddles indicates that for the supergravity solution corresponding to this $(n_{\star},N_{\star})$ saddle to be reliable everywhere, we need
\begin{equation}
n_{\star} \gg N_{\star} \gg 1 \, , \qquad n_{\star} \ll \omega^2 N \ll n_{\star} \left({n_{\star} \over N_{\star}} \right)^{1 \over 2} \; .
\end{equation}
However, in section \ref{sec:Background}, we have seen that for $\omega^2 N \gg 1$,  the irreducible saddle with $n_{\star} = 1$ should completely dominate the integral, and the backreacted supergravity description of \cite{Komatsu:2024bop} is not valid for this saddle.

Thus, observables related to the $(n_{\star},N_{\star})$ supergravity solution in its regime of validity will at best correspond to some highly exponentially suppressed contribution to a matrix model observable.

\paragraph{Basin-restricted observables}

Despite the apparent difficulties we have just described, there may be a natural way to compare supergravity observables to the matrix model even when these supergravity observables represent highly suppressed contributions. 

The basic idea is to divide the integration domain in the path integral into regions associated with the various classical saddles so that each observable can be written as a sum of contributions labeled by $SU(2)$ representations $\mathcal{R}$
as for the localization results. We would then conjecture that the partial contribution labeled by representation $\mathcal{R}$ corresponds to an observable in the gravitational saddle labelled by the same representation. 

If the path integral were purely bosonic, the obvious way to split the contributions into sectors would be to assign each point in the integration domain to a basin of the potential $V_{bos}(X)$. For a configuration $\{X^i\}$, we could consider the gradient flow from $\{X^i\}$ based on the potential $V_{bos}(X)$; for all points except a measure zero set, this will take us to one of the  minima $X_\mathcal{R}$ of $V$ labeled by a representation $\mathcal{R}$, and we would define the basin associated with $\mathcal{R}$ to be all points whose flows lead to $X_\mathcal{R}$.

In our case with fermions, we could similarly assign basins to each bosonic configuration using the bosonic potential $V_{bos}(X)$, but in this case, it seems more natural to use the effective potential obtained after integrating out the fermions. Since this effective potential is generally complex, we take the real part that determines the magnitude of $e^{-S}$. In terms of the  Pfaffian obtained via integrating out the fermions this is
\begin{equation}
V_{eff}(X) = S_{bos}(X) - \text{Re} \log |\text{Pf} \mathcal{M}(X)| \; .
\end{equation}
Making use of this effective potential leads to an additional complication: the saddles of the effective potential are not the same as the saddles of the bosonic potential. It is still true that $V_{eff}$ is bounded below\footnote{The magnitude of the logarithm of the Pfaffian only grows logarithmically with $X$ for large $X$ while the bosonic action grows quadratically.} and that each bosonic configuration outside a measure zero set flows to a minimum of $V_{eff}$, so observables can be precisely written as a sum of terms associated with the various minima. But we need to understand how to associate minima of the effective potential to minima of the bosonic potential (and thus to representations $\mathcal{R}$).

The simplest prescription is the following: for any minimum of the effective potential, associate this minimum to representation $\mathcal{R}$ if it goes over smoothly to the minimum $X_\mathcal{R}$ of the bosonic potential for $\omega \to \infty$ (where the effects of the Pfaffian become negligible). The basin associated with $\mathcal{R}$ is then the set of basins for $V_{eff}$ associated with minima that approach $X_\mathcal{R}$ for large $\omega$.\footnote{A complicating feature is that certain saddles may exist only for some bounded range of $\omega$, so one would need a separate prescription for these.}

In practice, we can usually assign a representation to a minimum of the effective potential in a simpler way. In appendix \ref{sec:EffectiveActionSaddles}, we argue that as $\omega$ decreases from large values, the minima of the effective potential associated with a representation $R$ are deformed to an extremal manifold $M_R$ with configurations of the form\footnote{We can prove that there are extrema of this form and that they are minima in the sector of configurations that do not break any additional symmetry. We have not shown that they are local minima without the symmetry restriction.}
\begin{equation}
X^i = \oplus_a \alpha_a L_{(N_a)}^i \, , \qquad X^{p} = \oplus_a m^p_a I_{N_a} \; .
\label{mForm}
\end{equation}
The parameters $\alpha_a$ determine the fuzzy sphere radii, while the $m^p_a$ are coordinates of the fuzzy-sphere centers in the transverse seven-dimensional space. For a given $\mathcal{R}$, these are nonzero below some critical value $\omega_\mathcal{R}$ due to a repulsive pairwise potential coming from the Pfaffian term. The nonzero $\vec{m}_a$ break the $SO(7)$ symmetry so instead of single extremum, we have $SO(7)$ orbits of extrema. 

Because of the form (\ref{mForm}), we can easily read off the $\{N_a\}$ by noting that 
\begin{equation}
X^i X^i = \oplus_a \alpha_a^2 {N_a^2 - 1 \over 4} I_{N_a} \qquad K^i X^i = \oplus_a \alpha_a^3 {N_a^2 - 1 \over 4} I_{N_a} \qquad K^i = -{i \over 2} \epsilon_{ijk} [X^j,X^k] \; 
\end{equation}
so the number of zero eigenvalues for either of these gives the multiplicity of the trivial representation while for nonzero eigenvalues, $(\lambda^i_{XX})^3/(\lambda^i_{KX})^2$ gives $(N_a^2 - 1)/4$ with multiplicity $N_a$ times the number of irreducible representations of this size. 

\begin{figure}
    \centering
    \includegraphics[width=\linewidth]{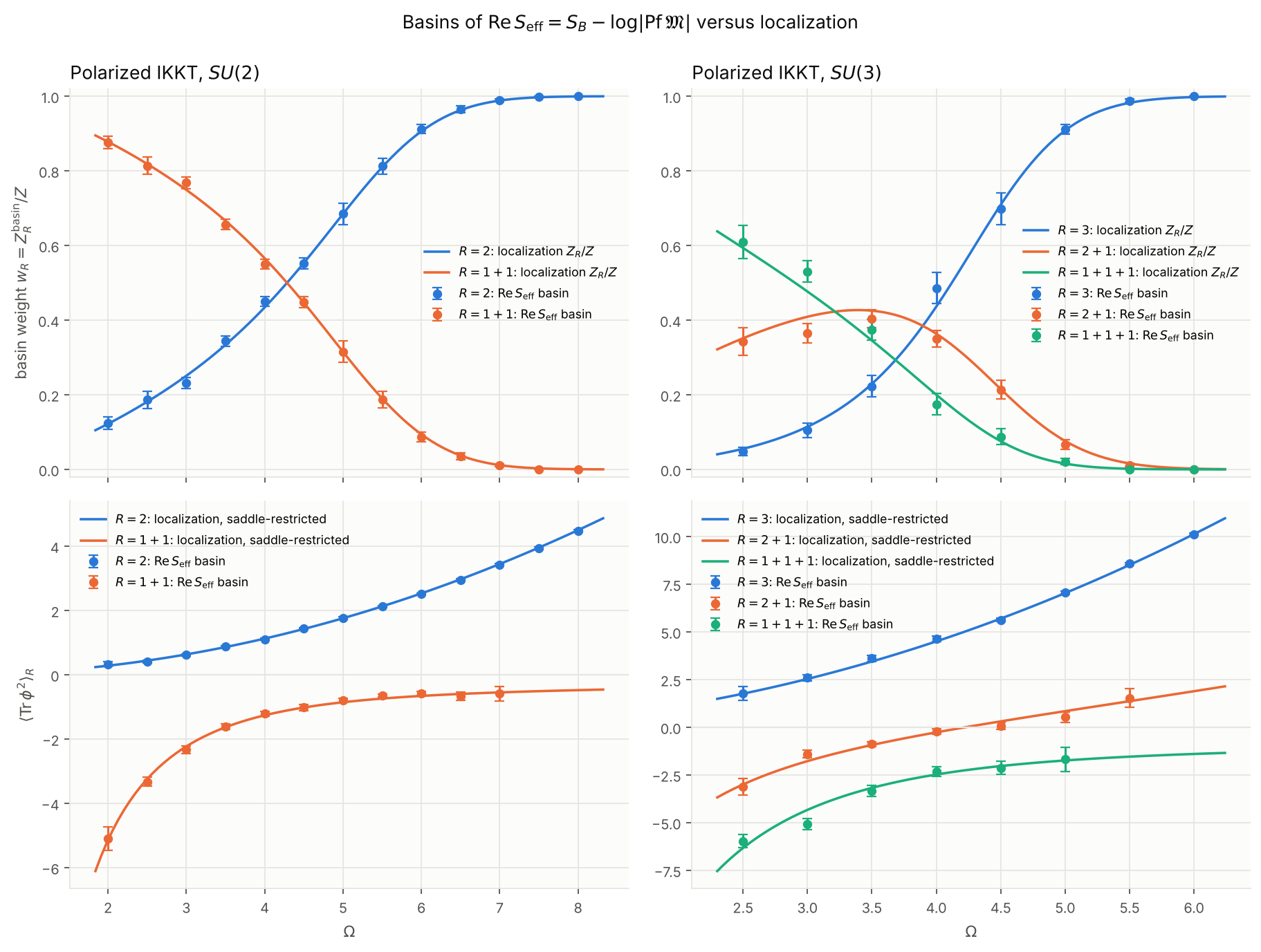}
    \caption{Basin-restricted Monte-Carlo evaluation of relative contributions to the partition function (top) and protected observable $\Tr(\phi^2)$, as a function of $\Omega = 8/3 \omega$. In each case, the results are compared with the corresponding exact representation-restricted contribution from localization.}
    \label{fig:basins}
\end{figure}

We have (using Claude) tested this procedure for $N=2$ and $N=3$, assigning a basin to each bosonic configuration by examining $X^i X^i$ and $K^i X^i$ for the endpoint of each gradient flow. The identification was successful except in $\approx 1\%$ of cases where the endpoint of the gradient flow had $X^i$ matrices without a clear representation structure. These configurations (which may represent saddles of a different type appearing at some finite $\omega$) were omitted from the calculation. We have calculated the basin-restricted partition functions and protected $\Tr(\phi^2)$ observables and compared the contributions from each basin to the exact contribution from the corresponding representation in the localization calculation. The results, plotted in Figure \ref{fig:basins}, shows good agreement over the full range of $\omega$ that we tested.

While these investigations were at very small $N$, they motivate the conjecture that supergravity observables calculated in a solution corresponding to some representation $\mathcal{R}$ are related to matrix model observables restricted to the corresponding basin. If this picture is correct, it may be that the precise definition of the basin boundaries affects observables only at a level that would correspond to non-perturbatively small corrections from the supergravity point of view. In this case, we could for example define the basin associated with some representation $\mathcal{R}$ to be the connected neighborhood of $M_\mathcal{R}$ for that representation defined by  $|V_{eff}(X) - V_{eff}^{min}(R)| < \Delta$ where $\Delta$ is chosen so that no extrema not associated with $\mathcal{R}$ lie in the domain.

\paragraph{Lefschetz thimbles} Our basin prescription may be related to and/or made more precise via the formalism of Lefschetz thimbles. Here, the basic idea is that a complex integral
\begin{equation}
\int \rd z_i e^{-S(\vec{z})}
\end{equation}
defined over a specific integration contour (in our case, the real slice) can be (under appropriate assumptions) rewritten as a sum over contributions associated with the various saddles:
\begin{equation}
\int \rd z_i e^{-S(\vec{z})} = \sum_a n_a \int_{T_a} \rd z_i e^{-S(\vec{z})} \, .
\end{equation}
Here, $a$ runs over complex saddles of the action, $n_a$ are integers, and $T_a$ is the ``Lefschetz thimble'' associated with the saddle $a$. With $d$ complex variables, this thimble is the real $d$-dimensional slice through saddle $a$ defined as the union of trajectories originating at the saddle and satisfying
\begin{equation}
{\rd z^i \over \rd \tau} = \overline{{\partial S \over \partial z^i}}
\end{equation}
This equation ensures that
\begin{equation}
{\partial \text{Re} S \over \partial \tau} = \sum_i \left|{\partial S \over \partial z^i}\right|^2 \ge 0 \qquad \qquad {\partial \text{Im} S \over \partial \tau} = 0
\end{equation}
so that saddle $a$ is a global minimum of the action on the thimble and the phase of $e^{-S}$ is constant on the slice.

In our situation, it is natural to take the action $S$ as the (complex) effective action obtained by integrating out the fermions. For the SU(2) case, it is clear that the thimble construction is similar to the basin construction since the effective action is real, so the thimble associated with any real minimum of the effective action will at least locally coincide with the basins we have defined. For higher $N$, the effective action is complex, so the thimbles will generally not lie in the real configuration space. It would be interesting to investigate further the relation between the basin approach and the thimble approach and their potential relation to the decomposition of gravity observables into contributions associated with the supergravity saddles. It seems plausible that different prescriptions for defining basins or decomposing the matrix integral into separate contributions associated with saddles may give results differing by amounts that would correspond only to non-perturbative corrections on the supergravity side.

\section{Discussion}
\label{sec:Discussion}

There are various interesting directions for future work.

\paragraph{Stringy physics from the matrix model}
It would be interesting to understand the emergence of stringy physics from the matrix model. While we have focused on the supergravity limit in the gravity picture, the matrix model should be dual to (or provide a definition of) the full type IIB string theory. 

There are non-supersymmetric saddles in the matrix model that can be thought of configurations intermediate between the supersymmetric saddles. In gravity, moving between saddles requires changing quantized fluxes in the geometry, so intermediate configurations should involve explicit branes carrying these fluxes. It would be interesting to find string/brane configurations in the various supergravity backgrounds and try to establish a correspondence with the matrix model, for example by matching actions. 

Another approach would be to consider matrix model expectation values that should directly probe string or brane worldsheets in the dual geometry. In standard AdS/CFT, inserting supersymmetric Wilson loop observables on a path $\gamma$ in the field theory path integral corresponds to adding a string or brane worldsheet in the geometry whose boundary traces a corresponding path $\hat{\gamma}$ at the conformal boundary.\footnote{The specific string theory object involved depends on the representation used in defining the Wilson line.} For the IKKT model, there are no obvious Wilson line observables, however in the D1-brane limit discussed in Section \ref{sec:Limits}, the matrix model becomes a continuum 2-dimensional gauge theory with standard Wilson line observables. These Wilson lines are limits of certain matrix model operators that can be understood using the standard map between matrix models and non-commutative field theories. These same matrix model operators can be studied away from the D1-brane decoupling limit and should presumably still correspond to adding string/brane worldsheets to the dual geometries.

Another approach would be to consider observables in the matrix model that are not $U(N)$-invariant (i.e. to treat $U(N)$ as a global symmetry of the model), following \cite{Maldacena_2018}. As in that paper, we might expect such observables to correspond to backgrounds where we have strings emerging from and returning to the asymptotic point in the geometry.

\paragraph{Other observables}
We have focused on one particular class of observables $\Tr(\phi^{2k})$, relating them to multipole moments in the supergravity description and to the asymptotic supergravity fields. However, using symmetries, we can relate this to many other matrix model observables. For example, consider the operators
\begin{equation}
\mathcal T^{(2k)}_{I_1\cdots I_{2k}}=\text{STr}(X_{I_1}\cdots X_{I_{2k}})_\text{traceless} \, ,
\end{equation}
which are symmetric traceless tensors of $SO(10)$. Their expectation values are captured by the coefficients of the homogeneous polynomial
\begin{equation}
F_k(y)=
\left\langle \mathcal T^{(2k)}_{I_1\cdots I_{2k}}\right\rangle
y^{I_1}\cdots y^{I_{2k}} \; .
\end{equation}
Via $SO(3)\times SO(7)$ symmetry, this polynomial must take the form 
\begin{equation}
F_k(y)=\sum_{r=0}^{k}c_r \left( \sum_{i=1}^{3}y_i^2 \right)^r \left( \sum_{p=4}^{10}y_p^2 \right)^{k-r} \, .
\end{equation}
The tracelessness of $\mathcal T^{(2k)}$ implies that $\partial_I^2 F = \partial_i^2 F + \partial_p^2 F = 0$, and this fixes all coefficients $c_r$ in terms of a single coefficient. Defining $v=e_3-i e_{10}$, we have that
\begin{equation}
F_k(v)=
\left\langle\operatorname{Tr}(X_3-iX_{10})^{2k}\right\rangle
=\langle \Tr(\phi^{2k}) \rangle \; ,
\end{equation}
which fixes the remaining undetermined coefficient. So far, we have only made use of $SO(3) \times SO(7)$ symmetry. Using supersymmetries, we can similarly relate expectation values for a whole multiplet of matrix model observables to the $\langle \Tr(\phi^{2k}) \rangle$ expectation values. 

For this whole family of observables, we have distinct contributions associated with each representation sector $\mathcal{R}$, so we can try to understand the relation with the corresponding supergravity saddle. As in more familiar examples of AdS/CFT, we should be able to establish a dictionary relating specific operators to specific asymptotic field fluctuations. 

\paragraph{Gravity interpretation of the dipole gas}

It would be interesting to understand whether the exact finite $N$ interpretation of the localization integral as a dipole gas can be given some gravity interpretation. An interesting aspect of this formulation is that for each configuration of dipoles, we already have a Coulomb field exactly satisfying the Laplace equation in the region above the dipole gas, so we can formally define the corresponding 4D Coulomb field and supergravity fields in the asymptotic regions. Thus, the integral over dipole locations becomes an integral over a specific class of gravitational configurations. This may be related to a localization deformation of the full gravity path integral.

\paragraph{Non-BPS observables}

It will be very interesting to better understand and test the matrix model - gravity correspondence beyond the protected sector. We have proposed a way to directly compute general observables in the matrix model through a basin-restricted Monte-Carlo integration. It would be interesting to explore this proposal (and its potential relation to the Lefshetz thimble formalism) in more detail and use it to calculate unprotected observables at somewhat larger values of $N$ where corrections to the supergravity approximation are small. It will be crucial to understand the appropriate gravity interpretation for unprotected observables such as $\Tr(X_i X_i)$ and $\Tr(X_a X_a)$ or to find other unprotected observables with a simple gravity interpretation so that a direct comparison can be made.

\section*{Acknowledgements}

This work is supported in part by the National Sciences and Engineering Research Council of Canada (NSERC) and the Simons Foundation via a Simons Investigator Award.
We have made significant use of AI during this project. ChatGPT6 and Claude Fable were used as a tool in various aspects of the research. In addition to performing various numerical calculations and checking our analytic calculations, AI tools (specifically ChatGPT Astra) made several significant novel suggestions including the approach to the action matching in section 4 and important observations leading to the Coulomb gas reformulation of section 5. The full text of the paper was written by the authors apart from the technical Appendix on Monte Carlo simulations. 

\appendix

\section{Type IIB conventions, field equations, and charges}
\label{app:TheoryAndEoM}

\subsection{Euclidean theory and supergravity action}

We work with the following form of the Euclidean type IIB supergravity action in string frame
\begin{equation}
\label{TheoryAndEoM:ActionStringFrame}
\begin{aligned}
\cI^{\rm bulk} = & \frac{1}{2 \kappa_0^2} \int \rd^{10} x \sqrt{g} e^{- 2 \Phi} \left( - R - 4 \left( \partial \Phi \right)^2 + \frac{1}{12} H_3^2 \right) \\
& + \frac{1}{2 \kappa_0^2} \int \rd^{10} x \sqrt{g} \left( - \frac{1}{2} \left( \partial C_0 \right)^2  - \frac{1}{12} \cF_3^2 \right) \, ,
\end{aligned}
\end{equation}
where $2 \kappa_0^2 = (2 \pi)^7 \alpha'^4$. Here, $C_0$ and $C_2$ are the RR potentials, $H_3 = \rd B_2$ is the NSNS three-form, and $\cF_3 = \rd C_2 - C_0 H_3$. We are also restricting to the sector with vanishing RR five-form field strength ($F_5 = 0$). The consistency of this restriction for the solutions we will study will be verified in appedix \ref{app:ElectrostaticSolutions}.

We are parametrizing the fields in a way adapted to the Euclidean signature solutions that are the main focus of this work, so that all variables appearing in the action are real in those solutions. In particular, \cite{Komatsu:2024bop} used purely imaginary axion and RR two-form, denoted $\chi_{\rm I}$ and $C_{2,{\rm I}}$ respectively. Our fields here are related to those as
\begin{equation}
\chi_{\rm I} = i g_s C_0 \, , \qquad C_{2,{\rm I}} = i g_s C_2 \, .
\end{equation}
This explains the ``wrong sign''  for the $(\partial C_0)^2$ and $\mathcal{F}_3^2$ terms in \eqref{TheoryAndEoM:ActionStringFrame}. Furthermore, we also absorb factors of the string coupling into the dilaton, $e^{\Phi} = g_s e^{\Phi_{\rm I}}$. Note that the string frame metric agrees in both conventions, $g_{\rm I} = g$, so there is a rescaling in Einstein frame of the form $g_{\rm I}^{(E)} = g_s^{1/2} g^{(E)}$.

The use of real Euclidean RR variables should not be interpreted as imposing an ordinary real supersymmetry structure on the full theory. The $S^6\times S^2$ backgrounds we study -- see appendix \ref{app:ElectrostaticSolutions} -- are naturally described within complexified type IIB supergravity. As explained in \cite{DHoker:2025nid,DHoker:2026eox}, the complexified Killing-spinor equations admit solutions associated with the required $F(4,\mathbb C)$ symmetry, but there is no corresponding real form with $SO(7)\times SO(3)$ bosonic symmetry. The analytically continued Killing spinors therefore need not satisfy Euclidean reality conditions. Our conventions concern the bosonic action and its classical equations, but they do not assume a real Euclidean superalgebra.

\subsection{Variational principle and equations of motion}

As usual, to have a stationary action with Dirichlet boundary conditions we will add a Gibbons-Hawking-York (GHY) boundary term. In string frame, this is
\begin{equation}
\mathcal{I}^{\rm GHY} = - \frac{1}{\kappa_0^2} \int_{\partial M} \rd^{9}x \sqrt{h} \, e^{-2 \Phi} K \, ,
\end{equation}
where $h_{ab}$ is the induced boundary string-frame metric, and $K_{ab}$ the extrinsic curvature measured with respect to the outward-pointing unit normal $n^{\mu}$. A first variation then gives
\begin{equation}
\delta \left( \mathcal{I}^{\rm bulk} + \mathcal{I}^{\rm GHY} \right) = \int_M \mathcal{E} \cdot \delta\Psi + \Theta_{\partial M} \, ,
\end{equation}
where the first term represents all the equations of motion $\mathcal{E}$, while the second collects the boundary contributions produced after integration by parts. 

The equations of motion are
\begin{subequations}
\label{TheoryAndEoM:EoMStringFrame}
\begin{align}
0 & = R + 4 \nabla^2\Phi - 4 (\partial\Phi)^2 -\frac{1}{12} H_3^2 =0 \, , \\
\nonumber 0 & = R_{\mu \nu} + 2 \nabla_{\mu} \nabla_{\nu} \Phi - \frac{1}{4} H_{\mu}{}^{\rho \sigma} H_{\nu \rho \sigma} \\
 & \quad + \frac{e^{2 \Phi}}{2} \left( \partial_{\mu} C_0 \partial_{\nu} C_0 + \frac{1}{2} \cF_{\mu}{}^{\rho \sigma} \cF_{\nu \rho \sigma} - \frac{1}{2} g_{\mu \nu} \left( (\partial C_0)^2 + \frac{1}{6} \cF_3^2 \right) \right) \, , \\
0 & = \rd \left( \star \, \cF_3 \right) \, , \\
0 & = \rd \left( \star \, \rd C_0 \right) + H_3 \wedge \star \, \cF_3 \, , \\
0 & = \rd \left( e^{- 2 \Phi} \star H_3 + C_0 \star \cF_3 \right)  \, .
\end{align}
\end{subequations}
It is sometimes useful to combine the trace of the metric equation with the dilaton equation, obtaining
\begin{equation}
(\partial \Phi)^2 - \frac{1}{2} \nabla^2 \Phi - \frac{1}{24} H_3^2 - e^{2 \Phi} \left( \frac{1}{2} (\partial C_0)^2 + \frac{1}{24} \cF_3^2 \right) = 0 \, .
\end{equation}

We will also need the form of the boundary term for some of the arguments in the main text. For convenience, we split it into contributions from variations of the metric and dilaton and contributions from the $p$-form potentials,
\begin{equation}
\Theta_{\partial M} = \Theta_{g,\Phi} + \Theta_{\rm forms} \, .
\end{equation}
The two contributions can be shown to be
\begin{subequations}
\label{TheoryAndEoM:VariationBoundaryTerms}
\begin{align}
\nonumber \Theta_{g,\Phi} & = \frac{1}{2 \kappa_0^2} \int_{\partial M} \rd^9 x \sqrt{h} \, e^{-2 \Phi} \left[ \left( K^{ab} - K h^{ab} + 2 h^{ab} n^{\mu} \partial_{\mu} \Phi \right) \delta h_{ab} \right. \\
& \qquad \qquad \qquad \qquad \qquad \qquad \qquad \qquad \quad \left. + \left( 4 K - 8 n^{\mu} \partial_{\mu} \Phi \right) \delta \Phi \right] \, , \\
\Theta_{\rm forms} & = \frac{1}{2 \kappa_0^2} \int_{\partial M} \left[ - \delta C_0 \wedge \star \rd C_0 - \delta C_2 \wedge \star \mathcal{F}_3 + \delta B_2 \wedge \left( e^{-2 \Phi} \star H_3 + C_0 \star \mathcal{F}_3 \right) \right] \, .
\end{align}
\end{subequations}
%

Finally, we derive an expression for the on-shell bulk action as a total derivative. The dilaton equation implies, using differential form notation
\begin{equation}
e^{-2 \Phi} \left[ -R - 4 \left( \partial \Phi \right)^2 + \frac{1}{12} H_3^2 \right] \star 1 = 4 \, \rd \left( e^{-2 \Phi} \star \rd \Phi \right)
\end{equation}
Furthermore, the trace relation above gives
\begin{equation}
\rd \left( e^{-2 \Phi} \star \rd \Phi \right) = - \frac{1}{2} e^{-2 \Phi} H_3 \wedge \star H_3 - \rd C_0 \wedge \star \rd C_0 - \frac{1}{2} \mathcal{F}_3 \wedge \star \mathcal{F}_3 \, .
\end{equation}
Combining these identities, the form-field equations of motion, and the definitions of $H_3 = \rd B_2$ and $\mathcal{F}_3 = \rd C_2 - C_0 H_3$, the on-shell bulk action becomes
\begin{equation}
\label{TheoryAndEoM:OnShellAction}
\hspace{-0.1cm} \mathcal{I}^{\rm bulk}_{\rm on-shell} = \frac{1}{2 \kappa_0^2} \int_M \rd \left[ \frac{9}{2} e^{-2 \Phi} \star \rd \Phi + \frac{1}{4} B_2 \wedge \left( e^{-2 \Phi} \star H_3 + C_0 \star \mathcal{F}_3 \right) - \frac{1}{4} C_2 \wedge \star \mathcal{F}_3 \right] \, .
\end{equation}
%

\subsection{Conserved charges}

Based on the form-field equations of motion and the Bianchi identities, we obtain electric and magnetic charges by integrating conserved forms along appropriate cycles:
\begin{subequations}
\label{TheoryAndEoM:Charges}
\begin{align}
N_{D(-1)} & = \frac{1}{(2 \pi)^8 \alpha'^4} \int_{\Sigma_9} \left( \star\, \rd C_0 + B_2 \wedge \star\, \cF_3 \right) \equiv \frac{1}{(2 \pi)^8 \alpha'^4} \int_{\Sigma_9} J_{9,D(-1)} \, , \\
N_{D1} & = \frac{1}{(2 \pi)^6 \alpha'^3} \int_{\Sigma_7} \star\, \cF_3 \equiv \frac{1}{(2 \pi)^6 \alpha'^3} \int_{\Sigma_7} J_{7,D1} \, , \\
N_{F1} & = \frac{1}{(2 \pi)^6 \alpha'^3} \int_{\Sigma_7} \left( e^{-2 \Phi} \star H_3 + C_0 \star \cF_3 \right) \equiv \frac{1}{(2 \pi)^6 \alpha'^3} \int_{\Sigma_7} J_{7,F1} \, , \\
N_{D5} & = \frac{1}{(2 \pi)^2 \alpha'} \int_{\Sigma_3} \rd C_2 \equiv \frac{1}{(2 \pi)^2 \alpha'} \int_{\Sigma_3} J_{3, D5} \, , \\
N_{NS5} & = \frac{1}{(2 \pi)^2 \alpha'} \int_{\Sigma_3} H_3 \equiv \frac{1}{(2 \pi)^2 \alpha'} \int_{\Sigma_3} J_{3,NS5} \, , \\
N_{D7} & = \int_{\Sigma_1} \rd C_0 \equiv \int_{\Sigma_1} J_{1,D7} \, .
\end{align}
\end{subequations}
For the D7 charge, the corresponding classical flux is the axion monodromy, which vanishes around every closed one-cycle for any globally single-valued $C_0$. Note that the D1 and NS5 currents are both gauge invariant and closed, but other currents contain the gauge potentials explicitly and are thus interpreted as Page-type expressions \cite{Marolf:2000cb}. 

A related feature is the transformation of the conserved charges under constant axion shifts. Consider
\begin{equation}
C_0 \rightarrow C_0 + c \, , \qquad C_2 \rightarrow C_2 + c B_2 \, ,
\end{equation}
with constant $c$. The gauge-invariant field strengths are unchanged, but the conserved currents transform as
\begin{equation}
J_{7, F1} \rightarrow J_{7,F1} + c \, J_{7,D1} \, , \qquad J_{3, D5} \rightarrow J_{3,D5} + c \, J_{3, NS5} \, ,
\end{equation}
so that the charges also shift accordingly,
\begin{equation}
\label{TheoryAndEoM:ChargesTransformationAxionShift}
N_{F1} \rightarrow N_{F1} + c \, N_{D1} \, , \qquad N_{D5} \rightarrow N_{D5} + c \, N_{NS5} \, .
\end{equation}
This is the classical axion-translation symmetry of the bosonic action, part of the $SL(2,\mathbb{R})$ duality group. 

Finally, the conserved seven-forms are precisely the
electric fluxes appearing in the two-form boundary
variation derived in \eqref{TheoryAndEoM:VariationBoundaryTerms}:
\begin{equation}
\Theta_{\rm 2-forms} = \frac{1}{2 \kappa_0^2} \int_{\partial M} \left( - \delta C_2 \wedge J_{7,D1} + \delta B_2 \wedge J_{7,F1} \right) \, .
\end{equation}
Thus the original action naturally defines a variational
problem with fixed two-form potentials. Changing to fixed electric fluxes requires modifying the corresponding boundary terms

\section{Supergravity solutions from the electrostatic problem}
\label{app:ElectrostaticSolutions}

In this appendix, we record the explicit expressions for the supergravity fields in terms of the electrostatics potential $V = V(\rho, z)$. We will also calculate some charges using these solutions.

\subsection{Electrostatic supergravity fields}

Consider the potential $V(\rho, z)$ satisfying the four-dimensional Laplace equation with the boundary conditions specified in section \ref{sec:electrostatics}. The supergravity fields are obtained from this potential as \cite{Komatsu:2024bop}
\begin{subequations}
\label{ElectrostaticSolutions:FieldsFromV}
\begin{align}
\frac{\rd s^2}{\alpha'} & = A(\rho, z) \left( F(\rho, z) \left( \rd \rho^2 + \rd z^2 \right) + G_6(\rho, z) \, \rd \Omega_6^2 + G_2(\rho,z) \, \rd\Omega_2^2 \right) \, , \\
e^{\Phi} & = - 54\pi \frac{3 \dot{V} + \rho V''}{\rho \sqrt{3 \Delta \dot{V} (-V'')}} \, , \\
C_0 & = - \frac{1}{54\pi}\frac{3 \dot{V} (V' + \rho \dot{V}') + \rho V' V''}{3 \dot{V} + \rho V''} \, , \\
B_2 & = - 2\pi\alpha' \left( z - \frac{\rho \dot{V} \dot{V}'}{\Delta} \right) \omega_2 \, , \\
C_2 & = \frac{\alpha'}{27} \left( V - \rho \frac{\dot{V}}{\Delta} \left( V' \dot{V}' + 3 \dot{V} (- V'') \right) \right) \omega_2 \, ,
\end{align}
\end{subequations}
where $\omega_2$ is the volume form of the round $S^2$, and we have introduced the functions
\begin{equation}
\begin{gathered}
A(\rho, z) = 2\pi \sqrt{\frac{( - 3 \dot{V} - \rho V'' )}{\rho (-V'')}} \, , \qquad F(\rho, z) = - \frac{V''}{\dot{V}} \, , \\
G_2(\rho, z) = - \frac{\rho V'' \dot{V}}{\Delta} \, , \qquad  G_6(\rho, z)  = 3 \rho \, , \qquad \Delta = 3 \dot{V} V'' + \rho ( \dot{V}'^2 + V''^2 ) \, ,
\end{gathered}
\end{equation}
The notation for derivatives is $\dot{V} \equiv \partial_{\rho} V$ and $V' \equiv \partial_z V$. 

We note that we may consistently set $F_5=0$ for this family. Indeed, $B_2 = b \, \omega_2$ and $C_2 = c \, \omega_2$ are both proportional to $\omega_2$, so that
\begin{equation}
H_3 = \rd b \wedge \omega_2 \, , \qquad \mathcal{F}_3 = \left( \rd c - C_0 \, \rd b \right) \wedge \omega_2 \, , \qquad H_3 \wedge F_3 = 0 \, .
\end{equation}
This implies that the five-form Bianchi identity and equation of motion admit $F_5=0$: $\rd F_5 = \rd \star F_5 = H_3 \wedge \mathcal{F}_3$. The Chern--Simons terms defining $F_5$ also vanish on this ansatz, allowing $C_4=0$.

\paragraph{Normalization conventions.}
The original solution of \cite{Komatsu:2024bop} is written in terms of dimensionful coordinates, a mass parameter $\mu$, and a reference string coupling $g_s$. Denoting its electrostatic variables by a superscript ${\rm I}$, our dimensionless variables are related to them by
\begin{equation}
\begin{aligned}
\rho^{\rm I} & = \frac{3\pi}{4}\mu \alpha' \rho \, , & z^{\rm I} & = \frac{3\pi}{4}\mu \alpha' z \, , & V^{\rm I} & = \frac{g_s \mu^4 \alpha'}{72} V \, , \\
Q^{\rm I}_s & = \frac{\pi^4g_s \mu^6 \alpha'^3}{32} Q_s \, , & P^{\rm I} & = \frac{3\pi^5g_s \mu^7 \alpha'^4}{128} P \, , & \eta^{\rm I} & = \frac{g_s \eta}{54\pi} \, .
\end{aligned}
\end{equation}
The parameter $\eta^{\rm I}$ is defined as in \cite{Komatsu:2024bop}, where the linear term in the background potential is $-\eta^{\rm I}\mu^3 z^{\rm I}$. Furthermore, we redefine $\mu$ into a dimensionless quantity
\begin{equation}
\frac{\mu^4 \alpha'^2}{g_s} = \frac{512}{81\pi^3}\omega^4 \, .
\end{equation}
Note that $g_s$ is the constant reference coupling appearing in the normalization of the fields and parameters of \cite{Komatsu:2024bop}. As described in
appendix \ref{app:TheoryAndEoM}, the rescaling $e^\Phi=g_s e^{\Phi_{\rm I}}$, together with the corresponding rescaling of the RR potentials, absorbs this reference coupling into our fields. After the coordinate and potential rescalings above, neither $g_s$ nor $\mu$ appears separately in the dimensionless supergravity solution; their relevant combination is encoded in $\omega$. In our conventions, the background potential becomes
\begin{equation}
\label{ElectrostaticSolutions:BackgroundPotential}
V_{\rm bg}(\rho, z) = - \eta z + \frac{3}{2}\omega^4 \left( z \rho^2 - z^3 \right) \, ,
\end{equation}
and the relation between the data fixed in the electrostatic problem and the matrix-model representation is
\begin{equation}
z_s = \frac{N_s}{2} \, , \qquad Q_s = n_s \, .
\end{equation}

\paragraph{Axion shifts.}
The linear term in the background potential \eqref{ElectrostaticSolutions:BackgroundPotential} implements a constant axion shift. Indeed, if we transform
\begin{equation}
V \rightarrow V - \delta \eta \, z \, ,
\end{equation}
the only derivative affected is $V' \rightarrow V' - \delta \eta$ (note that $\Delta$ stays invariant). It follows from \eqref{ElectrostaticSolutions:FieldsFromV} that the only fields that transform non-trivially are
\begin{equation}
C_0 \rightarrow C_0 + \frac{\delta \eta}{54\pi} \, , \qquad C_2 \rightarrow C_2 + \frac{\delta \eta}{54\pi} \, B_2 \, .
\end{equation}
This is just an axion shift transformation as discussed in appendix \ref{app:TheoryAndEoM}.

\subsection{Fluxes and charges}

We now evaluate the conserved fluxes obtained in \eqref{TheoryAndEoM:Charges} on the electrostatic solutions. Most of these calculations were performed in \cite{Komatsu:2024bop}, we will review them here briefly using our conventions for completeness. However, we find an important discrepancy regarding the F1 charge which we will explain in greater detail.

All charges are computed using the ten-dimensional orientation
\begin{equation}
{\rm vol}_{10} = \alpha'^{5} A^{5} F G_6^3 G_2 \, \rd \rho \wedge \rd z \wedge \omega_6 \wedge \omega_2 \, ,
\end{equation}
with $\omega_6$ and $\omega_2$ volume forms for the round unit spheres, with total volumes ${\rm vol}(S^6) = 16 \pi^3 / 15$ and ${\rm vol}(S^2) = 4 \pi$. We will integrate over three-cycles $\Sigma_{3,s}$ obtained by fibering the $S^2$ over a path joining the $s$-th and $(s-1)$-th conductors, oriented from $z_s$ to $z_{s-1}$ (we include the grounded plane at $z_0 = 0$). Similarly, the seven-cycle $\Sigma_{7,s}$ is obtained by fibering the $S^6$ over a curve in the $(\rho, z)$ plane enclosing the $s$-th conductor. This contour is oriented so that its upper face runs from $\rho = 0$ to $\rho = \rho_s$, while its lower face runs in the opposite direction. Finally, the asymptotic nine-cycle $\Sigma_{9,\infty}$ has the outward orientation.

\paragraph{Magnetic charges.} 

In the interior of the conducting balls, the potential is fixed to $V(\rho, z_s) = V_s$ and the two-form potentials approach
\begin{equation}
B_2 \rightarrow - 2\pi\alpha' z_s \, \omega_2 \, , \qquad C_2 \rightarrow \frac{\alpha'}{27} V_s \, \omega_2 \, .
\end{equation}
Thus, integrating $H_3 = \rd B_2$ and $\rd C_2$ over $\Sigma_{3,s}$ gives
\begin{subequations}
\begin{align}
N_{NS5,s} & = 2 \left(z_s - z_{s-1} \right) \, , \\
N_{D5,s} & = - \frac{1}{27\pi} \left(V_s - V_{s-1} \right) \, .
\end{align}
\end{subequations}
Up to a sign, fixed here by keeping the same orientation of $\Sigma_{3,s}$ for both charges and our choice for the potentials in \eqref{ElectrostaticSolutions:FieldsFromV}, these results agree with \cite{Komatsu:2024bop}. Note that $V_s$ has a contribution of the form $- \eta z_s$ from the background potential, consistent with the transformation under axion shifts $N_{D5,s} \rightarrow N_{D5,s} + \frac{\delta \eta}{54\pi} \, N_{NS5,s}$.

\paragraph{D-instanton charge.}
At the asymptotic nine-sphere $\Sigma_{9, \infty}$, the D-instanton charge is found to be
\begin{equation}
N_{D(-1)} = P \, ,
\end{equation}
with $P = 2 \sum_s z_s Q_s$. This result is obtained by using the multipole expansion \eqref{multipole}. Since the axion has no monodromy in the electrostatic solutions, the D7 flux vanishes.

\paragraph{Electric charges.}
The calculations of the D1 and F1 charges are similar, so we will analyze them together. The main tool is the expansion of the potential close to one of the conducting balls. Using Laplace's equation, this is
\begin{equation}
V(\rho,z) = V_s + g_{\pm}(\rho) (z-z_s) - \frac{\rho \ddot{g}_{\pm}(\rho) + 2 \dot{g}_{\pm}(\rho)}{6 \rho} (z-z_s)^3 + \mathcal{O}((z-z_s)^5) \, ,
\end{equation}
where $\pm$ indicates whether we are approaching the ball at $z = z_s$ from above or below. Some lengthy algebra produces then the currents $J_{7,D1}$ and $J_{7,F1}$. Since we are going to integrate them over a constant-$z$ surface, the relevant Hodge dual is
\begin{equation}
\star \left(\rd z \wedge \omega_2 \right) = - \alpha'^2 \frac{A^2 G_6^3}{G_2} \, \rd \rho \wedge \omega_6 \, .
\end{equation}
Using this, the above expansions close to $z_s$, and the forms of the fields in \eqref{ElectrostaticSolutions:FieldsFromV}, we obtain
\begin{align}
\label{ElectrostaticSolutions:D1Current}
\frac{j_{D1}[g]}{\alpha'^3} & = 4\pi^2 \left[ - 3 \partial_{\rho} \left( \frac{\rho^4 \dot{g}^2}{\rho \ddot{g} + 2 \dot{g}} \right) + 5 \rho^3 \dot{g} \right]  \, , \\
\label{ElectrostaticSolutions:F1Current}
\frac{j_{F1}[g]}{\alpha'^3} & = \frac{2\pi}{27} \left[ 3 \partial_{\rho} \left( \frac{\rho^4 \dot{g}^2 (g + \rho \dot{g})}{\rho \ddot{g} + 2 \dot{g}} \right) - \frac{5}{2} \rho^3 \partial_{\rho} (g^2) \right] \, .
\end{align}
where $J_{7,D1} = j_{D1} \, \rd \rho \wedge \omega_6$, $J_{7,F1} = j_{F1} \, \rd \rho \wedge \omega_6$, and $g = g_+$ or $g = g_-$ depending on which side we are approaching the conductor from.

The contribution at $\rho = 0$ is fixed by the expansion of the potential
\begin{equation}
V(\rho, z) = f(z) - \frac{\rho^2}{6} f''(z) + \mathcal{O}(\rho^4) \, .
\end{equation}
This implies that $\partial_{\rho} g_{\pm} = \mathcal{O}(\rho)$, and in particular $g_{\pm}$ is sufficiently regular as $\rho \to 0$ for the term inside the derivative in the first piece of \eqref{ElectrostaticSolutions:D1Current} and \eqref{ElectrostaticSolutions:F1Current} to vanish. Therefore, these terms will not produce any boundary contribution at $\rho = 0$.

For the other endpoint at $\rho = \rho_s$, more care is needed. The behavior near the conductor tip is established in section \ref{subsec:HalfPlaneElectrostatics} -- concretely, in \eqref{HalfPlaneElectrostatics:Tip} and the paragraph below -- using a holomorphic reformulation of the electrostatic problem. For our purposes here, the important conclusion is that $g_{\pm}$ admit expansions in $\sqrt{\rho_s - \rho}$, and that their leading nonanalytic terms proportional to $\sqrt{\rho_s - \rho}$ have opposite coefficients on the two faces. Thus
\begin{equation}
\lim_{\rho \to \rho_s^-} \frac{\rho^4 \dot{g}_{\pm}^2}{\rho \ddot{g}_{\pm} + 2 \dot{g}_{\pm}} = 0 \, , \qquad \lim_{\rho \to \rho_s^-} \frac{\rho^4 \dot{g}_{+}^2 (g_+ + \rho \dot{g}_+)}{\rho \ddot{g}_{+} + 2 \dot{g}_{+}} = \lim_{\rho \to \rho_s^-} \frac{\rho^4 \dot{g}_{-}^2 (g_- + \rho \dot{g}_-)}{\rho \ddot{g}_{-} + 2 \dot{g}_{-}} \, .
\end{equation}
Again, this implies that total derivatives in the first terms of \eqref{ElectrostaticSolutions:D1Current} and \eqref{ElectrostaticSolutions:F1Current} do not give any contribution at $\rho = \rho_s$.

With these results, we can now evaluate the charges. For the D1,
\begin{equation}
\begin{aligned}
N_{D1,s} & = \frac{{\rm vol}(S^6)}{(2 \pi)^6} \int_0^{\rho_s} \rd \rho \left( j_{D1}[g_+] - j_{D1}[g_-] \right) = \frac{1}{3 \pi} \int_0^{\rho_s} \rd \rho \, \rho^3 \left(\dot{g}_+ - \dot{g}_- \right) \\
& = \frac{1}{\pi} \int_0^{\rho_s} \rd \rho \, \rho^2 \left( g_- - g_+ \right) = 4 \pi \int_0^{\rho_s} \rd \rho \, \rho^2 \sigma_s(\rho) = Q_s = n_s \, .
\end{aligned}
\end{equation}
This agrees with \cite{Komatsu:2024bop}, once we translate to our conventions. The F1 charge follows similarly
\begin{equation}
\begin{aligned}
N_{F1,s} & = \frac{{\rm vol}(S^6)}{(2 \pi)^6} \int_0^{\rho_s} \rd \rho \left( j_{F1}[g_+] - j_{F1}[g_-] \right) = - \frac{1}{324 \pi^2} \int_0^{\rho_s} \rd \rho \, \rho^3 \partial_{\rho} \left( g_+^2 - g_-^2 \right) \\
& = \frac{1}{108 \pi^2} \int_0^{\rho_s} \rd \rho \, \rho^2 \left( g_+^2 - g_-^2 \right) \, .
\end{aligned}
\end{equation}
In the last step, we have used that $g_+^2 - g_-^2 \to 0$ as $\rho \to \rho_s^-$. This expression has a simple electrostatic interpretation. At a point above or below the charged ball, the vertical electric field is $- g_{\pm} = E_{\rm reg} \pm 2\pi^2\sigma_s$, where $\pm 2\pi^2\sigma_s$ is the contribution from the infinitesimal charged element below or above the point, and $E_{\rm reg}$ is a regular contribution from all the other charges (and background field) in the system. This regular contribution does not jump. The vertical force experienced by the infinitesimal charged element is $E_{\rm reg} \sigma_s$, which can be written as
\begin{equation}
E_{\rm reg} \sigma_s = - \frac{ g_+ + g_-}{2} \frac{g_- - g_+}{4\pi^2} = \frac{g_+^2 - g_-^2}{8\pi^2} \, .
\end{equation}
Thus, we conclude that the F1 charge is directly related to the total vertical force on the conductor,
\begin{equation}
\label{ElectrostaticSolutions:F1Charge}
N_{F1,s} = \frac{2}{27} \int_0^{\rho_s} \rd \rho \, \rho^2 \sigma_s E_{\rm reg} = \frac{F_s}{54 \pi} \, .
\end{equation}
Unlike the conclusion originally drawn in \cite{Komatsu:2024bop}, we find that the F1 charge need not vanish. The fact that the electrostatic solutions have $N_{F1, s} \neq 0$ was also noted in \cite{DHoker:2026eox}. However, our expression above follows directly from the ten-dimensional conserved current and differs from the intermediate electrostatic integral used in those references.

As a final consistency check, under the axion shift $V\to V - \delta\eta \, z$, the functions $g_\pm$ transform as $g_\pm \to g_\pm - \delta\eta$. It follows that
\begin{equation}
N_{F1,s} \rightarrow N_{F1,s} + \frac{\delta \eta}{54 \pi} N_{D1,s} \, ,
\end{equation}
in agreement with the transformation \eqref{TheoryAndEoM:ChargesTransformationAxionShift} under the axion shift $C_0 \to C_0 + \delta \eta / (54 \pi)$.

\section{Fixing the finite counterterm by the variational principle.}
\label{app:counterterms}

In this appendix, we show that the coefficient of the finite counterterm $\mathcal{I}_{\text{ct},3}$ (third term in \eqref{eq:local-ct-action}) in the gravitational action is fixed directly by the asymptotic
variational principle in the electrostatic sector. The idea is to keep $\omega$, $P$ and the axion offset $\eta = 0$ fixed, allowing variations of subleading multipoles (in particular, $\delta q_3 \neq 0$). We then require to have a well-defined variational principle after adding the counterterms, i.e., a stationary action under the allowed variations. Note that a variation of the type we are considering can be extended off-shell into the interior and chosen to vanish near the compact cycles, so that the RR Legendre term \eqref{OnShellMatching:LegendreTerm} does not contribute to it.

The relevant condition will come from the variation at the asymptotic boundary, where we will get a contribution proportional to $\delta q_3$ that does not vanish as we remove the regulator, $R \to \infty$. To show this, we plug the electrostatic solution into the variation in \eqref{TheoryAndEoM:VariationBoundaryTerms}, expanding the fields and varying with $\delta \omega = \delta P = \delta \eta = 0$ but nonzero variations of the subleading multipoles. In particular, the potential varies as
\begin{equation}
\delta V =\frac{4 z(z^2 - \rho^2)}{r^8} \delta q_{3} + \mathcal{O}(r^{-7}) \, .
\end{equation}
To make the cancellation of potentially divergent terms explicit, it is useful to separate the part of the string-frame metric variation proportional to the dilaton variation. Define for this
\begin{equation}
\widehat{\delta h}_{ab} \equiv \delta h_{ab} - \frac{1}{2} h_{ab} \delta\Phi \, ,
\end{equation}
where $h_{ab}$ is the induced boundary metric in string frame. Substituting this decomposition into \eqref{TheoryAndEoM:VariationBoundaryTerms}, the metric and dilaton contribution can be written as
\begin{equation}
\Theta_{g,\Phi} = \mathcal{D}_R+\Theta_{\widehat{h}} \, , \qquad \mathcal{D}_R = \frac{1}{2\kappa_0^2} \int_{\Sigma_R} \rd^9 x \, \sqrt{h}\,e^{-2\Phi} (n^\mu\partial_\mu\Phi) \delta\Phi \, ,
\end{equation}
where $\Theta_{\widehat h}$ contains the terms proportional to $\widehat{\delta h}_{ab}$. On the other hand, using $C_0=e^{-\Phi}-W$, the axion contribution to $\Theta_{\rm forms}$ contains precisely $-\mathcal{D}_R$. These terms therefore cancel identically, before taking the asymptotic limit.

The introduction of $W$ is useful because, while $C_0$ and $e^{-\Phi}$ grow as $R^8$, their difference is much smaller. Indeed, from \eqref{SUGRAsolutions:AsymptoticRelationPhiC0} we have
\begin{equation}
W = \frac{\omega^4 r^2}{18\pi}(1+2\cos^2\theta) + \mathcal{O}(r^{-4}) \, , \qquad \delta W = \mathcal{O}(r^{-4}) \, ,
\end{equation}
for the variations considered here. After the cancellation of $\mathcal{D}_R$, the remaining axion contribution is
\begin{equation}
\Theta_{C_0}+\mathcal{D}_R = -\frac{1}{2\kappa_0^2} \int_{\Sigma_R} \rd^9 x \, \sqrt{h} \left( e^{-\Phi} \left(n^\mu\partial_\mu\Phi\,\delta W + n^\mu\partial_\mu W \, \delta\Phi \right) + n^\mu \partial_\mu W\,\delta W \right) \, .
\end{equation}
Using $\sqrt{h}=\mathcal{O}(R^{-9})$, $e^{-\Phi} = \mathcal{O}(R^8)$, $n^r=\mathcal{O}(R^2)$, and $\delta\Phi=\mathcal{O}(R^{-2})$, the three terms in this integral scale as $\mathcal{O}(R^{-4})$, $\mathcal{O}(1)$ and $\mathcal{O}(R^{-10})$, respectively, including the boundary measure. Thus, only the term proportional to $n^\mu\partial_\mu W\,\delta\Phi$ gives a finite contribution. The remaining metric contribution $\Theta_{\widehat h}$ and the two-form contributions are also finite.

Evaluating all the remaining terms using the large-distance expansions of the supergravity fields, we obtain
\begin{equation}
\Theta_{g,\Phi} = \mathcal{D}_R -\frac{31\omega^4}{180} \delta q_3 + o(1) \, ,\qquad \Theta_{\rm forms} = -\mathcal{D}_R + \frac{5\omega^4}{54}\delta q_3+o(1) \, .
\end{equation}
Their sum finally gives
\begin{equation}
\label{counterterms:AsymptoticVariation}
\Theta_{\infty} \equiv \Theta_{g,\Phi} + \Theta_{\rm forms} = - \frac{43 \omega^4}{540} \delta q_3 + o(1) \, .
\end{equation}

We see that the action is not stationary under a variation with free $\delta q_3$ unless the counterterms cancel the $\delta q_3$ piece. The contributions from the first two counterterms in \eqref{eq:local-ct-action} are computed in \eqref{eq:local-ct-expansion-1} and \eqref{eq:local-ct-expansion-2}, so taking a variation we get
\begin{equation}
\delta \mathcal{I}_{\rm ct,1} = \frac{31 \omega^4}{180} \delta q_3 + o(1) \, , \qquad \delta \mathcal{I}_{\rm ct,2} = - \frac{7 \omega^4}{135} \delta q_3 + o(1) \, .
\end{equation}
Combining these with the boundary term from the original action we get
\begin{equation}
\Theta_{\infty} + \delta \mathcal{I}_{\rm ct,1} + \delta \mathcal{I}_{\rm ct,2} = \frac{11 \omega^4}{270} \delta q_3 + o(1) \, .
\end{equation}
The coefficient of the remaining finite counterterm is set to cancel this.

To show that this can be done, consider
\begin{equation}
\mathcal{J} = \frac{1}{\kappa_0^2} \int_{\Sigma_R} \rd^9 x \sqrt{h^{(E)}} \frac{e^{\Phi}}{\sqrt{\chi}} D_a W D^a \Phi \, ,
\end{equation}
where we are exceptionally using Einstein frame quantities, because the counterterms are more naturally written using this form of the metric -- in particular, boundary covariant derivatives $D_a$ in the previous expression are defined with respect to the Einstein frame metric, and contracted with it. We recall from the main text that $h^{(E)}_{ab} = e^{-\Phi/2} h^{(s)}_{ab}$, $W = e^{-\Phi} - C_0$ and $\chi = \mathcal{R}[h^{(E)}]/72$. Evaluating this quantity asymptotically on the electrostatic solution and taking a variation, we obtain
\begin{equation}
\mathcal{J} = \frac{14 \omega^4 q_3}{27} + o(1) \, , \qquad \delta \mathcal{J} = \frac{14 \omega^4}{27} \delta q_3 + o(1) \, .
\end{equation}
We thus see that we can have a stationary action without fixing $q_3$ if we take
\begin{equation}
\mathcal{I}_{\rm ct,3} = - \frac{11}{140} \mathcal{J} \, ,
\end{equation}
as finite counterterm. This is precisely the term used in \eqref{eq:local-ct-action}.

We close this appendix with a comment about other possible equivalent forms of the finite counterterm. Using $W = e^{-\Phi} - C_0$, we can show that
\begin{equation}
\mathcal{J} = - \frac{1}{2} \tilde{\mathcal{J}} - \frac{1}{2} \mathcal{B} \, , 
\end{equation}
where
\begin{align}
\tilde{\mathcal{J}} & = \frac{1}{\kappa_0^2} \int_{\Sigma_R} \rd^9 x \sqrt{h^{(E)}} \frac{(D\Phi)^2 - e^{2 \Phi} (DC_0)^2}{\sqrt{\chi}} \, , \\
\mathcal{B} & = \frac{1}{\kappa_0^2} \int_{\Sigma_R} \rd^9 x \sqrt{h^{(E)}} \frac{e^{2\Phi}}{\sqrt{\chi}} (D W)^2 \, .
\end{align}
Using the asymptotic behavior of the fields,
\begin{equation}
\sqrt{h^{(E)}} = \mathcal{O}(R^9) \, , \qquad e^{\Phi} = \mathcal{O}(R^{-8}) \, , \qquad \frac{1}{\sqrt{\chi}} = \mathcal{O}(R) \, , \qquad (DW)^2 = \mathcal{O}(R^2) \, ,
\end{equation}
we can check that
\begin{equation}
\mathcal{B} = \mathcal{O}(R^{-4}) \, ,
\end{equation} 
with $\delta \mathcal{B}$ falling at least as fast for the variations considered. It follows that both $\mathcal{J}$ and $- \tilde{\mathcal{J}}/2$ have the same finite asymptotic value and variation, so we can use either to build the finite counterterm. 

\section{Conformal mapping solution for one-ball electrostatics}
\label{sec:OneDiskConformal}

In this section, we provide details of the solution for the single-ball electrostatics problem using the conformal mapping method.

For one ball set $a=1$ and take $c>1$. The remaining undetermined map parameters are $(b,c,C)$; the conductor potential $V_{\star}$ entering $F$ is also initially unknown. Explicitly, we have
\begin{equation}
Z'(u)=C\frac{u-b}{\sqrt u\sqrt{u-1}\sqrt{u-c}},
\qquad 1<b<c.
\end{equation}
The conditions \eqref{HalfPlaneElectrostatics:Prevertices} give that 
\begin{eqnarray*}
    C \int_0^1 dt \frac{b-t}{\sqrt t\sqrt{1-t}\sqrt{c-t}} &=& z_{\star} \cr
    C \int_1^c dt \frac{b-t}{\sqrt t\sqrt{t-1}\sqrt{c-t}} &=& 0 \cr
    C \int_1^b dt \frac{b-t}{\sqrt t\sqrt{t-1}\sqrt{c-t}} &=& \rho_{\star} 
\end{eqnarray*}
The second equation fixes $b$ in terms of $c$:
\begin{equation}
b=
\frac{\displaystyle\int_1^c\frac{t\,dt}
{\sqrt{t(t-1)(c-t)}}}
{\displaystyle\int_1^c\frac{dt}
{\sqrt{t(t-1)(c-t)}}}
=c\,\frac{E(m)}{K(m)},\qquad m=1-\frac1c.
\label{ConformalAppendix:SingleBallClosing}
\end{equation}
where
\begin{equation}
K(m)=\int_0^{\pi/2}\frac{d\theta}{\sqrt{1-m\sin^2\theta}},
\qquad
E(m)=\int_0^{\pi/2}\sqrt{1-m\sin^2\theta}\,d\theta.
\end{equation}
Here $K$ and $E$ use the parameter convention $m=k^2$, whereas $E$ in \eqref{K2} uses the modulus $k$; the definitions above fix the convention throughout this appendix.
The first equation above then determines $C$ in terms of $c$. Explicitly,
\begin{equation}
\displaystyle C=\frac{z_{\star}K(m)}{\pi\sqrt c}.
\label{ConformalAppendix:SingleBallScale}
\end{equation}
This follows by substituting $b=cE(m)/K(m)$ into that integral and using Legendre's relation
\begin{equation}
K(m)E(1-m)+E(m)K(1-m)-K(m)K(1-m)=\frac\pi2,
\end{equation}
which gives
\begin{equation}
\int_0^1\frac{b-t}{\sqrt{t(1-t)(c-t)}}\,dt
=\frac{\pi\sqrt c}{K(m)}.
\end{equation}

Along the slit, the radial coordinate is
\begin{equation}
\rho(t)=C\int_1^t\frac{b-v}{\sqrt{v(v-1)(c-v)}}\,dv,
\qquad 1\leq t\leq c,
\qquad \rho_{\star}=\rho(b).
\label{ConformalAppendix:SingleBallRadius}
\end{equation}
It increases from zero to $\rho_{\star}$ and then returns to zero.  Thus $c$
is the only remaining shape parameter.

Now we write the tip equation $F'(b) = 0$ more explicitly. We have
\begin{equation}
g(t)=A \rho(t)-B \rho(t)^3,
\qquad
A=V_{\star}+\eta z_{\star}+\frac{3\omega^4}{2} z_{\star}^3,
\qquad B=\frac{3\omega^4}{2} z_{\star}.
\label{ConformalAppendix:BoundaryPolynomial}
\end{equation}
Integrating by parts in the derivative of the Schwarz integral gives
\begin{equation}
F'(u)=\mathcal W'_{\rm bg}(Z(u))Z'(u)
+\frac1\pi \int_{1}^{c}\frac{g'(t)}{t-u}\,dt.
\end{equation}
The boundary terms in the integration by parts vanish because $g(1)=g(c)=0$. Taking the limit $u\to b$ from the upper half-plane as in \eqref{HalfPlaneElectrostatics:BoundaryDerivative}, the background term vanishes and $g'(b)=0$, yielding the real equation
\begin{equation}
0=\frac{1}{\pi} \int_{1}^{c}
\frac{\bigl[A-3B\rho(t)^2\bigr]\rho'(t)}{t-b}\,dt.
\label{ConformalAppendix:RegularTipEquations}
\end{equation}
The apparent singularity at $b$ is removable, since
\begin{equation}
\frac{g'(t)}{t-b}=-\frac{C\,[A-3B\rho(t)^2]}{\sqrt{t(t-1)(c-t)}}.
\end{equation}
Thus the tip equation is an ordinary improper integral, with only integrable endpoint singularities.

Defining
\begin{equation}
J_0=\int_1^c\frac{dt}{\sqrt{t(t-1)(c-t)}}=\frac{2K(m)}{\sqrt c},\qquad
J_2=\int_1^c\frac{\rho(t)^2\,dt}{\sqrt{t(t-1)(c-t)}} \; ,
\end{equation}
equation~\eqref{ConformalAppendix:RegularTipEquations} reduces to
$AJ_0=3BJ_2$. Hence the potential on the ball is determined by 
\begin{equation}
\displaystyle
V_{\star}=-\eta z_{\star}-\frac{3\omega^4}{2} z_{\star}^3
+\frac{9\omega^4}{2} z_{\star}\frac{J_2}{J_0}.
\label{ConformalAppendix:SingleBallPotential}
\end{equation}
In the normalization of appendix~\ref{app:ElectrostaticProblem}, we have
$V_0=V_{\star}+\eta z_{\star}+\frac{3\omega^4}{2} z_{\star}^3$ and
$\nu=\frac{3\omega^4}{2} z_{\star}\rho_{\star}^2/V_0$ and the same result reads
\begin{equation}
\displaystyle \nu=\frac{\rho_{\star}^2J_0}{3J_2}.
\label{ConformalAppendix:SingleBallNu}
\end{equation}

For each trial $c$, equations
\eqref{ConformalAppendix:SingleBallClosing}--\eqref{ConformalAppendix:SingleBallPotential}
determine the map and conductor potential.  The Schwarz integral then
determines $F$, and \eqref{HalfPlaneElectrostatics:Charge} gives a
calculated charge $Q_{\rm calc}(c)$. For one ball, this charge can also be obtained directly from the dipole coefficient. Since
\begin{equation}
Z(u)\sim2C\sqrt u,\qquad
F(u)-\mathcal W_{\rm bg}(Z(u))
=-\frac{1}{\pi u}\int_1^c g(t)\,dt+\mathcal O(u^{-2}),
\end{equation}
comparison with $\mathcal W-\mathcal W_{\rm bg}\sim-P/\zeta^2$ and $P=2z_{\star}Q_{\rm calc}(c)$ for the trial configuration gives
\begin{equation}
\displaystyle
Q_{\rm calc}(c)=\frac{2C^2}{\pi z_{\star}}\int_1^c g(t)\,dt.
\label{ConformalAppendix:SingleBallCharge}
\end{equation}
The remaining equation is
$Q_{\rm calc}(c)=Q_{\star}$, with $Q_{\star}$ fixed by the matrix model data. This needs to be solved numerically by varying $c$.

\paragraph{Numerical procedure for several balls.}
The following gives a numerical procedure for $q$ balls with prescribed heights $z_s$ and charges $Q_s$.  
\begin{enumerate}
\item Choose trial radii, $\rho_s$, set $a_1=1$, and use
\begin{equation}
Z(a_s)=Z(c_s)=z_s,
\qquad Z(b_s)=z_s+i\rho_s.
\end{equation}
to determine $(a_s,b_s,c_s)$ and $C$.

\item On $I_s=(a_s,c_s)$, compute
$r_s(t)=\operatorname{Im}Z(t)$ and write
\begin{equation}
g_s(t)=A_s r_s(t)-\frac{3\omega^4}{2} z_s r_s(t)^3,
\qquad A_s=V_s+\eta z_s+\frac{3\omega^4}{2} z_s^3.
\end{equation}
Determine the $A_s$ from the $q$ linear equations
\begin{equation}
\sum_{r=1}^q\int_{I_r}
\frac{[A_r-\frac{9\omega^4}{2} z_r r_r(t)^2]r_r'(t)}
     {t-b_s}\,dt=0,
\qquad s=1,\ldots,q.
\end{equation}
\item Construct
\begin{equation}
F(u)=\mathcal W_{\rm bg}(Z(u))
+\frac1\pi\sum_{r=1}^q\int_{I_r}\frac{g_r(t)}{t-u}\,dt
\end{equation}
and evaluate the individual charges
\begin{eqnarray*}
Q_s^{\rm calc}
&=&-\frac1\pi\int_{I_s}r_s(t)\operatorname{Re}F'(t+i0)\,dt. \cr
&=&-\frac1{\pi^2}\sum_{r=1}^q\int_{I_s}dx\int_{I_r}dt\,
r_s'(x)g_r'(t)\log|t-x|.
\end{eqnarray*}
\item Adjust the radii until $Q_s^{\rm calc}=Q_s$ for every ball,
repeating the preceding steps at each iteration.
\end{enumerate}

\section{The electrostatic problem with a single charged ball}
\label{app:ElectrostaticProblem}

In this appendix, we will explain how to find the potential $V(\rho, z)$ for a 4d axisymmetric configuration of charges that includes a grounded plane at $z = 0$ and a ball with charge $\Qs$ at position $\zs > 0$. Like in the main text, the radius of the ball, $\rs$, will be fixed to have vanishing charge density at the tip, assuming we are in the presence of a background potential of the form
\begin{equation}
V_{\rm bg}(\rho, z) = - \eta z + \frac{3\omega^4}{2}(z \rho^2 - z^3) \, .
\end{equation}
If we let $\psi(\rho,z)$ be the potential generated by the ball, the problem reduces to solving the 4d Laplace equation with axial symmetry and the following boundary conditions:
\begin{equation}
\psi(\rho,0) = 0 \quad  \forall \rho > 0 \, , \qquad \psi(\rho, \zs) = V_{\star} + \eta \zs + \frac{3\omega^4 \zs^3}{2} - \frac{3\omega^4 \zs}{2} \rho^2 \quad {\rm if}\; 0 < \rho < \rs \, .
\end{equation}
$\psi(\rho, z)$ is also regular at the axis $\rho = 0$, it vanishes at infinity, and it is continuous everywhere and smooth outside of the conducting ball. The value of the potential at $\zs$, $V_{\star}$, is fixed by the condition to have total charge $\Qs$. This electrostatic problem was also studied in \cite{Komatsu:2024ydh}, using an integral equation formulation for the charge density. Here we develop an alternative approach based on integral transforms, which is particularly convenient for numerical implementation.

It is convenient to use the scaling symmetry of the Laplace equation to write the problem in a normalized form. Define
\begin{equation}
\hV(\hr, \hz) \equiv \frac{1}{V_0} V(\rs \hr, \rs \hz) = \hat{V}_{\rm bg}(\hr, \hz) + \hpsi(\hr, \hz) \, , \qquad V_0 \equiv V_{\star} + \eta \zs + \frac{3\omega^4 \zs^3}{2} \, .
\end{equation}
where hats will always refer to normalized quantities. The potential $\hV$ (or $\hpsi$) also satisfies the 4d Laplace equation, but now with boundary conditions
\begin{equation}
\hpsi(\hr,0) = 0 \quad  \forall \hr > 0 \, , \qquad \hpsi(\hr, \hzs) = 1 - \nu \hr^2 \quad {\rm if}\; 0 < \hr < 1 \, ,
\end{equation}
where
\begin{equation}
\nu = \frac{3\omega^4 \zs \rs^2}{2 V_0} \, .
\end{equation}
This normalized problem is thus fully characterized by $\hzs$ and the combination $\nu$. The condition that the charge density vanishes at the tip of the ball will impose a relation between these two parameters, as we will see shortly. For future reference, it is convenient to compute the normalized charge density in terms of the one of the original problem
\begin{equation}
\hat{\sigma} (\hr) = - \frac{1}{4\pi^2} \left( \left. \partial_{\hz} \right|_{\hzs^+} \hpsi - \left. \partial_{\hz} \right|_{\hzs^-} \hpsi \right) = \frac{\rs}{V_0} \sigma(\rs \hr) \, ,
\end{equation}
so the condition $\sigma(\rs) = 0$ sets $\hat{\sigma}(1) = 0$. The total charge is
\begin{equation}
\hat{Q}_{\star} = \int_0^1 \rd \hr \, 4 \pi \hr^2 \hat{\sigma}(\hr) = \frac{1}{V_0 \rs^2} \Qs \, .
\end{equation}

In order to impose $\hat{\sigma}(1) = 0$, we have to solve the electrostatic problem. Using axisymmetry, continuity at $\hzs$, and the boundary conditions at infinity, $\hr = 0$, and $\hz = 0$, the potential $\hpsi$ can be expanded in modes as
\begin{equation}
\hpsi(\hr,\hz) = \begin{cases}
    \displaystyle 2\int_0^{\infty} \rd k \, c(k) \sinh(k \hzs) \frac{\sin(k \hr)}{k \hr} e^{-k \hz} & {\rm if} \; \hzs \leq \hz \\
    \displaystyle 2\int_0^{\infty} \rd k \, c(k) e^{-k \hzs} \frac{\sin(k \hr)}{k \hr} \sinh(k \hz) & {\rm if} \; 0 \leq \hz \leq \hzs
\end{cases} \, .
\end{equation}
We now impose the Dirichlet condition on the ball, as well as continuity of $\partial_{\hz} \hpsi$ for $\hz = \hzs$ and $\hr > 1$. This translates into two integral equations for the coefficient function $c(k)$:
\begin{equation}
\label{ElectrostaticProblem:DualEquations}
\begin{aligned}
\int_0^{\infty} \rd k \, c(k) \left( 1 - e^{-2 k \hzs} \right) \frac{\sin(k \hr)}{k \hr} & = 1 - \nu \hr^2 & & {\rm for}\; 0 < \hr < 1 \, , \\
\int_0^{\infty} \rd k \, c(k) \sin(k \hr) & = 0 & & {\rm for}\; 1 < \hr \, . \\
\end{aligned}
\end{equation}

The second equation can be directly solved by viewing it as a restriction on the sine transform of $c(k)$. Indeed, we can define
\begin{equation}
g(\hr) \equiv \int_0^{\infty} \rd k \, c(k) \sin(k \hr) \, ,
\end{equation}
then the inverse transform gives
\begin{equation}
c(k) = \frac{2}{\pi} \int_0^{\infty} \rd \hr \, g(\hr) \sin(k \hr) = \frac{2}{\pi} \int_0^{1} \rd \hr \, g(\hr) \sin(k \hr) \, ,
\end{equation}
where in the second step we have imposed that $g(\hr) = 0$ for $\hr > 1$, automatically solving the second integral equation. The first equation in terms of $g(\hr)$ adopts the form
\begin{equation}
\label{ElectrostaticProblem:IntegralEquationFirstKind}
\int_0^1 \rd u \, K_{\hzs} (\hr, u) g(u) = \frac{\pi}{2} \hr \left(1 - \nu \hr^2 \right) \, , \qquad {\rm for}\;0<\hr < 1 \, , 
\end{equation}
with the kernel given by
\begin{equation}
\begin{aligned}
K_{\hzs} (\hr, u) & \equiv \int_0^{\infty} \frac{\rd k}{k} \left( 1 - e^{-2 \hzs k} \right) \sin(k \hr) \sin(k u) \\
& = \frac{1}{4} \log \left( \frac{(\hr + u)^2 \left( 4 \hzs^2 +(\hr -u)^2 \right)}{(\hr - u)^2 \left( 4 \hzs^2 + (\hr + u)^2 \right)} \right) \, .
\end{aligned}
\end{equation}

The charge density and potential can be rewritten in terms of the function $g(\hr)$ for $0 < \hr < 1$ determined by this integral equation. Indeed, the charge density is simply
\begin{equation}
\hat{\sigma}(\hr) = \frac{1}{2\pi^2\hr} \int_0^{\infty} \rd k \, c(k) \sin(k \hr) = \frac{g(\hr)}{2\pi^2\hr} \, ,
\end{equation}
so regularity at the axis requires $g(\hr) = \cO(\hr)$ as $\hr \to 0$. For the potential, using
\begin{equation}
\int_0^{\infty} \frac{\rd k}{k} e^{- b k} \sin(k \hr) \sin(k u) = \frac{1}{4} \log \left( \frac{(\hr + u)^2 + b^2}{(\hr -u)^2 + b^2} \right) \, ,
\end{equation}
for $b > 0$, we obtain
\begin{equation}
\hpsi(\hr, \hz) = \frac{1}{2 \pi \hr} \int_0^{1} \rd u \,g(u) \log \left( \frac{\left[ (\hz - \hzs)^2 + (\hr + u)^2 \right] \left[ (\hz + \hzs)^2+(\hr-u)^2 \right]}{\left[ (\hz - \hzs)^2+(\hr-u)^2 \right] \left[ (\hz + \hzs)^2 + (\hr + u)^2 \right]} \right) \, .
\end{equation}

\subsection*{Large $\hzs$ limit}

In the limit in which $\hzs = \zs / \rs \gg 1$, we can solve the integral equation \eqref{ElectrostaticProblem:IntegralEquationFirstKind} explicitly. Indeed, in that case the kernel simplifies and we get, after taking a derivative $\partial_{\hr}$:
\begin{equation}
\int_0^1 \rd u \frac{u}{u^2 - \hr^2} g(u) = \frac{\pi}{2} (1 - 3 \nu \hr^2) \equiv F(\hr) \, . 
\end{equation}
with the left-hand side understood in the Cauchy principal value sense. This integral equation (which has the form of a finite Hilbert transform by changing the integration variable to $v = u^2$) has a well-known solution \cite{}:
\begin{equation}
g(\hr) = \frac{1}{\hr \sqrt{1 - \hr^2}} \left( C - \frac{4}{\pi^2} \int_0^1 \rd x \frac{x^2\sqrt{1-x^2}}{x^2 - \hr^2} F(x) \right) \, ,
\end{equation}
where $C$ is a constant to be fixed by the condition $g(0) = 0$, and we are again using an implicit principal value. Using the form of our function $F(x)$, we obtain $C = (4 + 3 \nu)/8$ and then
\begin{equation}
g(\hr) = \frac{\hr}{\sqrt{1 - \hr^2}} \left( 1 + \frac{3 \nu}{2} - 3 \nu \hr^2 \right) \, .
\end{equation}

With this explicit solution for $g(\hr)$ we can then obtain the constraint that having a vanishing charge density at the edge of the ball imposes. From $g(1) = 0$, we conclude
\begin{equation}
\nu = \frac{2}{3} \, , \qquad g(\hr) = 2 \hr \sqrt{1 - \hr^2} \, ,
\end{equation}
so the charge density and total charge are
\begin{equation}
\hat{\sigma}(\hr) = \frac{1}{\pi^2} \sqrt{1 - \hr^2} \, , \qquad \hat{Q}_{\star} = \frac14 \, .
\end{equation}

\subsection*{Small $\hzs$ limit}

The analysis in the limit $\hzs = \zs / \rs \ll 1$ is a bit more involved, because the kernel of the integral equation \eqref{ElectrostaticProblem:IntegralEquationFirstKind} vanishes essentially everywhere except in the region $u \approx \hr$, where it is large. The physical situation is clear though: the charged ball at $\hzs$ is becoming very close to the grounded plane at $\hz = 0$, so away from the edge the problem is well approximated by an infinite parallel plate capacitor (in 4d). If we ignore for a moment the edge effects (so we fix $0 < \hr < 1$), we can recover the results for such an ideal capacitor as follows. Split the kernel as
\begin{equation}
K_{\hzs}(\hr, u) = \frac{1}{4} \log \left( 1 + \frac{4 \hzs^2}{(\hr - u)^2}\right) - \frac{1}{4} \log \left( 1 + \frac{4 \hzs^2}{(\hr + u)^2}\right) \, .
\end{equation}
For small $\hzs$, the first term is a very narrow peak in the region $x = |\hr - u| \sim \cO(\hzs)$, with an area
\begin{equation}
 \frac{1}{4} \int_{-\infty}^{\infty} \rd x \, \log \left( 1 + \frac{4 \hzs^2}{x^2} \right) = \pi \hzs \, .
\end{equation}
The second term is small (of order $\cO(\hzs^2)$) for fixed $0 < \hr < 1$, because $\hr + u$ is bounded away from zero. Thus, to leading order under integration against a smooth bulk profile, we can take
\begin{equation}
K_{\hzs \to 0} (\hr, u) \sim \pi \hzs \delta(\hr - u) \, .
\end{equation}
This gives the leading bulk solution of the integral equation
\begin{equation}
g(\hr) = \frac{\hr}{2 \hzs} \left( 1 - \nu \hr^2 \right) + o(\hzs^{-1}) \, , \qquad \hat{\sigma}(\hr) = \frac{1 - \nu \hr^2}{4\pi^2\hzs} + o(\hzs^{-1}) \, .
\end{equation}
The previous analysis ignores edge effects, because we cannot approximate the kernel by a $\delta$-function if we take $1 - \hr = \cO(\hzs)$ as $\hzs \to 0$.
The regularity condition at the edge must therefore be imposed using the
full solution. The exact conformal solution, expanded in
appendix \ref{app:SmallZsLimit}, gives $\nu\to1$ and the leading total
charge. Combining these results with the bulk solution, we conclude
\begin{equation}
g(\hr) \sim \frac{\hr}{2 \hzs} \left( 1 - \hr^2 \right) \, , \qquad \hat{\sigma}(\hr) \sim \frac{1 - \hr^2}{4\pi^2\hzs} \, , \qquad \hat{Q}_{\star} \sim \frac{2}{15\pi\hzs} \, ,
\end{equation}
where the density expressions hold at fixed $0<\hr<1$.

\subsection*{Numerical solution at general $\hzs$}

For general values of $\hzs$, we can try to find a numerical solution to the integral equation \eqref{ElectrostaticProblem:IntegralEquationFirstKind}. However, this is somewhat challenging because it is an inhomogeneous Fredholm equation of the first kind, and inverting the kernel for such equations is typically unstable due to small singular values of $K_{\hzs}$. We can obtain a related second kind equation, much nicer for numerical implementation, as follows. The method is a generalization to a four-dimensional setup of standard techniques in three-dimensional electrostatic problems \cite{}.

The second integral equation for $c(k)$ in \eqref{ElectrostaticProblem:DualEquations} can be solved not only via the sine transform, but also using well-known identities involving Bessel functions. In particular, if we write
\begin{equation}
c(k) = k \int_0^1 \rd t\, t h(t) J_0(k t)  \; \Longleftrightarrow \; h(t) = \int_0^{\infty} \rd k \, c(k) J_0(k t) \, ,
\end{equation}
then the identity
\begin{equation}
\int_0^{\infty} \rd k \, k J_0(kt) \sin(k \hr) = \begin{cases}
    \displaystyle - \frac{\hr}{(t^2 - \hr^2)^{3/2}} & {\rm if}\;0 < \hr < t \\
    \displaystyle 0 & {\rm if}\;t < \hr \\
\end{cases} \, ,
\end{equation}
guarantees that the second equation in \eqref{ElectrostaticProblem:DualEquations} holds (note that $0 < t < 1$ in the parametrization of $c(k)$). The new $h(t)$ is related to our previous $g(u)$ as
\begin{equation}
h(t) = \frac{2}{\pi} \int_t^1 \rd u \frac{g(u)}{\sqrt{u^2 - t^2}} = 4\pi\int_t^1\rd u\,\frac{u\hat{\sigma}(u)}{\sqrt{u^2-t^2}} \, ,
\end{equation}
where we have used the relation
\begin{equation}
\label{ElectrostaticProblem:BesselIdentity}
\int_0^{\infty} \rd k J_0(kt) \sin(ku) = \Theta(u-t) \frac{1}{\sqrt{u^2- t^2}} \, .
\end{equation}
Looking at $t \to 1^{-}$, the condition $g(1) = 0$ then implies that $h(1) = 0$ (vanishing faster than $\sqrt{1-t}$). This is the regularity condition at the edge in terms of $h(t)$, and it will be used to fix $\nu$ for a given $\hzs$.

The first integral equation in \eqref{ElectrostaticProblem:DualEquations} in terms of $h(t)$ becomes, using again \eqref{ElectrostaticProblem:BesselIdentity} for the piece without the exponential
\begin{equation}
\int_0^{\hr} \rd t \frac{t h(t)}{\sqrt{\hr^2 - t^2}} - \int_0^1 \rd t \, t h(t) \int_0^{\infty} \rd k \, e^{-2 \hzs k} J_0(kt) \sin (k \hr) = \hr ( 1 - \nu \hr^2) \, .
\end{equation}
The first transform can be inverted using the relation
\begin{equation}
B(\hr) = \int_0^{\hr} \rd t \frac{A(t)}{\sqrt{\hr^2 - t^2}} \; \Longleftrightarrow \; A(t) = \frac{2}{\pi} \frac{\rd}{\rd t} \int_0^t  \rd \hr \frac{\hr B(\hr)}{\sqrt{t^2 - \hr^2}} \, .
\end{equation}
Applying then the operator on the right to the whole equation, we get
\begin{equation}
\label{ElectrostaticProblem:IntegralEquationSecondKind}
h(x) - \int_0^1 \rd t \, t \, \cK_{\hzs}(x,t) h(t)  = 1 - \frac{3 \nu}{2} x^2 \, ,
\end{equation}
with the kernel given by
\begin{equation}
\begin{aligned}
\cK_{\hzs}(x,t) & = \int_0^{\infty} \rd k \, k e^{- 2 \hzs k} J_0(kx) J_0(kt) \\
& = \frac{4 \hzs}{\pi \sqrt{4 \hzs^2 +(x+t)^2} \left( 4 \hzs^2 +(x-t)^2 \right)} E \left(\sqrt{\frac{4 x t}{4 \hzs^2 + (x+t)^2}}\right) \, .
\end{aligned}
\label{K2}
\end{equation}
Here, $E$ is the complete elliptic integral of the second kind.\footnote{In Mathematica, $E(m)$ must be input as $\texttt{EllipticE[m}^2\texttt{]}$.}

\begin{figure}[t!]
    \centering
    \includegraphics[width = 1.0 \linewidth]{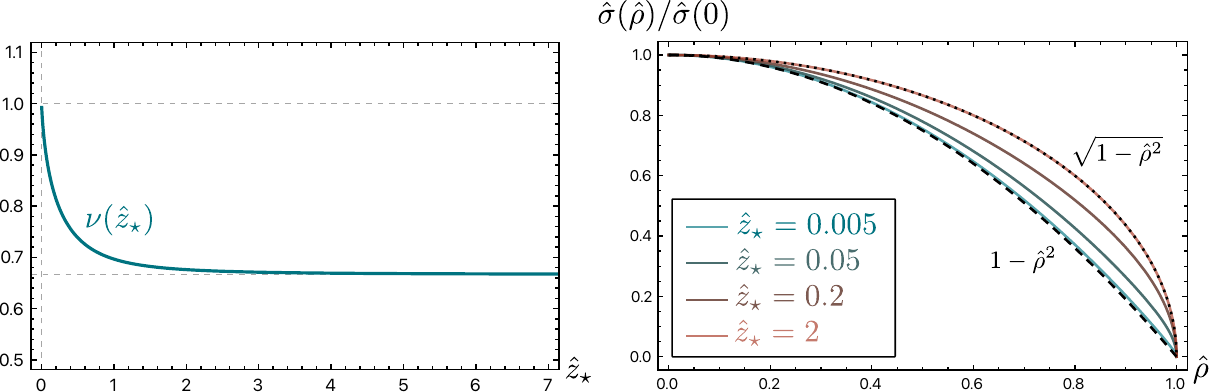}
    \caption{Left: Value $\nu(\hat{z}_{\star})$ for which the charge density vanishes at the edge of the ball. The two limiting values as $\hat{z}_{\star} \to 0$ and $\hat{z}_{\star} \to \infty$ are shown with the horizontal dashed lines. Right: Normalized charge density $\hat{\sigma}(\hat{\rho})/\hat{\sigma}(0)$ for four different values of $\hat{z}_{\star}$. Dashed and dotted lines show the limiting values as $\hat{z}_{\star} \to 0$ ($1 - \hat{\rho}^2$, valid only away from the edge) and $\hat{z}_{\star} \to \infty$ ($\sqrt{1 - \hat{\rho}^2}$).}
    \label{fig:PlotNuAndSigma}
\end{figure}

The equation \eqref{ElectrostaticProblem:IntegralEquationSecondKind} is much nicer to implement numerically (it is an inhomogeneous Fredholm equation of the second kind). We can discretize the kernel on $[0,1] \times [0,1]$ and use linearity to solve separately for
\begin{equation}
h_0(x) - \int_0^1 \rd t \, t \, \cK_{\hzs}(x,t) h_0(t)  = 1 \, , \qquad h_2(x) - \int_0^1 \rd t \, t \, \cK_{\hzs}(x,t) h_2(t)  = x^2 \, .
\end{equation}
Once these are found, the solution is $h(x) = h_0(x) - (3 \nu / 2) h_2(x)$, so the value of $\nu$ for which $h(1) = 0$ is
\begin{equation}
\nu (\hzs) = \frac{2 h_0(1)}{3h_2(1)} \, ,
\end{equation}
where we have emphasized that this depends on $\hzs$. Sampling different values of $\hzs$ we can reconstruct this function, the results are shown in Figure \ref{fig:PlotNuAndSigma} -- we solve the integral equations using a Nyström discretization with Gauss–Legendre quadrature in an auxiliary variable $y\in[0,1]$, mapped to $x=1-(1-y)^2$ to cluster the nodes near $x=1$; the number of quadrature points is increased until the results converge to the desired numerical precision. For each value of $\hzs$, from the solution $h(t)$ we can also reconstruct $g(\hr)$ and thus the charge density and total charge,
\begin{equation}
\hat{\sigma}(\hr) = \frac{g(\hr)}{2\pi^2\hr} = - \frac{1}{2\pi^2\hr} \frac{\rd}{\rd \hr} \int_{\hr}^1 \rd t \frac{t h(t)}{\sqrt{t^2 - \hr^2}} \, , \qquad \hat{Q}_{\star} = \int_0^1 \rd t \, t h(t) \, .
\end{equation}
We show some results for the charge density in Figure \ref{fig:PlotNuAndSigma}.

Finally, the potential can also be written directly in terms of $h(t)$, by direct substitution of the expression for $c(k)$:
\begin{equation}
\begin{aligned}
\hpsi(\hr, \hz) & = \frac{1}{\hr} \int_0^1 \rd t \, t h(t) \int_0^{\infty} \rd k \, J_0(kt) \sin(k \hr) \left[ e^{-k |\hz - \hzs|} - e^{-k(\hz +\hzs)} \right] \\
& = \frac{1}{\hr} \int_0^1 \rd t \, t h(t) \; {\rm Im} \left[ \frac{1}{\sqrt{t^2 + (|\hz - \hzs| - i \hr)^2}} - \frac{1}{\sqrt{t^2 + (\hz + \hzs - i \hr)^2}} \right] \, ,
\end{aligned}
\end{equation}
where the second line was obtained using the Laplace transform of $J_0$, and the first term is understood as the limit $|\hz - \hzs| \to 0^+$ when evaluating the potential on the plane of the ball.

\subsection*{Small $\hzs$ limit II}
\label{app:SmallZsLimit}

Next, we further analyze the small $\hat{z}_\star$ limit to derive the scaling \eqref{Limits:LambdaScalingSmallZstar} and the
charge-radius relation needed to express it in terms of $\Qs$ and
$\omega$. Set $d=\zs$. We use the one-ball
conformal solution of appendix~\ref{sec:OneDiskConformal}, with
\begin{equation}
\begin{gathered}
\zeta=Z(u),\qquad
Z'(u)=C\frac{u-b}{\sqrt u\sqrt{u-1}\sqrt{u-c}},\qquad
Z(0)=0,\quad Z(1)=Z(c)=d,\quad Z(b)=d+i\rs,\\
m=1-\frac1c,\qquad K=K(m),\quad E=E(m),\qquad
b=\frac{cE}{K},\qquad C=\frac{dK}{\pi\sqrt c}.
\end{gathered}
\label{SmallZsLimit:ExactMap}
\end{equation}
The limit $\rs/d\to\infty$ corresponds to $c\to\infty$, for which
$K\to\infty$, $E=1+\cO(K/c)$ and
$c^{-1}=16e^{-2K}[1+\cO(K/c)]$.

\paragraph{The subtracted potential and first mode.}
Let $F(u)=\mathcal W(Z(u))$, where
$\operatorname{Im}\mathcal W(\zeta)=\rho V(\rho,z)$, and subtract the
parallel-plate solution:
\begin{equation}
\Phi(u)=F(u)-\frac{V_\star}{2d}Z(u)^2,
\qquad
\operatorname{Im}\Phi(u)
=\rho\left(V(\rho,z)-\frac{V_\star}{d}z\right).
\label{SmallZsLimit:SubtractedPotential}
\end{equation}
The boundary values of $\Phi$ are real on the entire real $u$ axis. Indeed, $Z^2$ is real on either coordinate axis, while on the slit $Z=d+i\rho$ and $\operatorname{Im}(V_\star Z^2/(2d))=V_\star \rho$, matching the behavior of $\text{Im}(F(u))$. The extension
\begin{equation}
\Phi_{ext}(u) = \left\{ \begin{array}{cc} \Phi(u) &\qquad \text{Im}(u) \ge 0 \cr & \cr
\overline{\Phi(\bar{u})} & \qquad \text{Im}(u) < 0 \end{array} \right.
\end{equation}
is therefore an entire function. Since $Z(u)\sim2C\sqrt u$ and
$\mathcal W_{\rm bg}(\zeta)=-\eta\zeta^2/2-3\omega^4\zeta^4/8$,
this function has at most quadratic growth and leading coefficient
$-6\omega^4 C^4$. The result $Z'(b)=0$ and the regular-tip condition $F'(b)=0$  together require $\Phi'(b)=0$ so we must have
\begin{equation}
\Phi(u)=-6\omega^4 C^4(u-b)^2+D,\qquad D\in\mathbb R.
\label{SmallZsLimit:ExactPolynomial}
\end{equation}
The constant $D$ does not affect the electrostatic potential.

On the $\rho=0$ axis, $\mathcal W'(z)=V(0,z)$ (see Eq. \ref{HolomorphicElectrostatics:Density}) and $0<u<1$ maps to $0<z<d$.
Differentiating \eqref{SmallZsLimit:ExactPolynomial} gives
\begin{equation}
V(0,Z(u))-\frac{V_\star}{d}Z(u)
=\frac{\Phi'(u)}{Z'(u)}
=12\omega^4 C^3\sqrt{u(1-u)(c-u)}.
\label{SmallZsLimit:AxisPotential}
\end{equation}
On this interval, the map derivative is
$Z'(u)=d[1+\cO(K/c)]/(\pi\sqrt{u(1-u)})$ uniformly, where we have used $E = 1 + {\cal O}(K/c)$. Thus,
\begin{equation}
\frac{Z(u)}d=\frac2\pi\arcsin\sqrt u+\cO(K/c).
\label{SmallZsLimit:AxisMap}
\end{equation}
It follows that
\begin{equation}
V(0,z)-\frac{V_\star}{d}z
=6\omega^4 C^3\sqrt c\left[\sin\frac{\pi z}{d}+\cO(K/c)\right],
\end{equation}
uniformly for $0\leq z\leq d$. Projecting onto the first sine mode
of \eqref{Limits:PotentialSineExpansion} at $\rho=0$ therefore gives
\begin{equation}
c_1=6\omega^4 C^3\sqrt c\,[1+\cO(K/c)]
=\frac{96\omega^4d^3K^3}{\pi^3}e^{-2K}
\left[1+\cO(K/c)\right].
\label{SmallZsLimit:FirstModeK}
\end{equation}
The bound in section~\ref{subsec:NS5} already implies that keeping
$c_1>0$ finite suppresses all higher modes on fixed compact regions
of the gap and gives \eqref{Limits:PotentialLimitC1}.

\paragraph{The physical radius.}
Using $\rho(1)=\rho(c)=0$, we can evaluate the tip radius along the
outer face of the conductor:
\begin{equation}
\rs=C\int_b^c\frac{t-b}{\sqrt{t(t-1)(c-t)}}\,\rd t.
\label{SmallZsLimit:OuterFaceRadius}
\end{equation}
Since $t\geq b\sim c/K$, replacing $\sqrt{t(t-1)}$ by $t$ has
uniform relative error $\cO(K/c)$. The remaining integral is elementary,
\begin{equation}
\int_b^c\frac{1-b/t}{\sqrt{c-t}}\,\rd t
=2\sqrt{c-b}-\frac{2b}{\sqrt c}
\operatorname{arcosh}\sqrt{\frac cb},
\end{equation}
and hence
\begin{equation}
\begin{aligned}
\frac{\pi\rs}{d}
&=2\left[K\sqrt{1-\frac EK}
-E\operatorname{arcosh}\sqrt{\frac KE}\right]+\cO(K^2/c)\\
&=2K-\log(4K)-1+\cO(K^{-1}).
\end{aligned}
\label{SmallZsLimit:RadiusK}
\end{equation}
In particular, $e^{-\pi\rs/d}=4eK e^{-2K}[1+\cO(K^{-1})]$.
Combining this with \eqref{SmallZsLimit:FirstModeK}, and writing
$h=d/\rs=\hzs$, gives
\begin{equation}
c_1=\frac{6\omega^4d\rs^2}{\pi e}e^{-\pi\rs/d}
\left[1+\cO\left(h\log\frac1h\right)\right].
\label{SmallZsLimit:FirstModePhysicalRadius}
\end{equation}
This is precisely \eqref{Limits:LambdaScalingSmallZstar}.

\paragraph{Charge and the charge-based radius.}
The charge formula \eqref{HalfPlaneElectrostatics:Charge} reads
\begin{equation}
\Qs=-\frac1\pi\int_1^c\rho(t)\operatorname{Re}F'(t)\,\rd t.
\end{equation}
On the slit, $Z=d+i\rho$ and
$\operatorname{Re}F'=\Phi'-(V_\star/d)\rho\rho'$.
The parallel-plate term integrates to zero, and an integration by parts
using \eqref{SmallZsLimit:ExactPolynomial} gives the exact expression
\begin{equation}
\Qs=\frac{6\omega^4}{\pi}C^5\int_1^c
\frac{(t-b)^3}{\sqrt{t(t-1)(c-t)}}\,\rd t.
\label{SmallZsLimit:ExactCharge}
\end{equation}
Set $\epsilon=1/c$ and introduce the moments
\begin{equation}
M_j=\int_\epsilon^1
\frac{x^j\,\rd x}{\sqrt{x(x-\epsilon)(1-x)}}.
\end{equation}
Then $M_0=2K$, $M_1=2E$, while the two remaining moments needed here are
$M_2=4/3+\cO(\epsilon)$ and $M_3=16/15+\cO(\epsilon)$.
Changing variables to $t=cx$ in \eqref{SmallZsLimit:ExactCharge} therefore yields
\begin{equation}
\begin{aligned}
\Qs&=\frac{6\omega^4d^5K^5}{\pi^6}
\left[M_3-\frac{3E}{K}M_2+\frac{4E^3}{K^2}\right]\\
&=\frac{32\omega^4d^5K^5}{5\pi^6}
\left[1-\frac{15}{4K}+\cO(K^{-2})\right].
\end{aligned}
\label{SmallZsLimit:ChargeK}
\end{equation}
Define
\begin{equation}
R=\left(\frac{5\pi\Qs}{\omega^4}\right)^{1/5},
\qquad X=\frac{\pi R}{d}.
\label{SmallZsLimit:ChargeRadius}
\end{equation}
Equation~\eqref{SmallZsLimit:ChargeK} gives
\begin{equation}
X=2K-\frac32+\cO(K^{-1}).
\label{SmallZsLimit:XK}
\end{equation}
Retaining this constant shift in \eqref{SmallZsLimit:FirstModeK} gives
\begin{equation}
c_1=\frac{12\omega^4R^3}{e^{3/2}}e^{-\pi R/d}
\left[1+\cO(d/R)\right],
\label{SmallZsLimit:FirstModeR}
\end{equation}
while comparison with \eqref{SmallZsLimit:RadiusK} gives
\begin{equation}
\rs=R-\frac d\pi\left(\log\frac{2\pi R}{d}-\frac12\right)
+\cO\left(\frac{d^2}{R}\right).
\label{SmallZsLimit:RadiusShift}
\end{equation}
These are \eqref{Limits:C1ChargeScaling} and
\eqref{Limits:RadiusCorrectionSmallZstar}; substituting the definition
of $R$ into \eqref{SmallZsLimit:FirstModeR} gives the fixed charge and
matrix-model combinations stated in section~\ref{subsec:NS5}.
Although $\rs/R\to1$, the logarithmic shift cannot be dropped inside
the exponential: it changes the algebraic prefactor from $\rs^2$ to
$R^3$. All relative errors above are independent of $\omega$, since
both the charge and the subtracted potential are linear in $\omega^4$.

\paragraph{Normalized quantities.}
For completeness, the same formulas give the leading normalized quantities
quoted in the earlier small-$\hzs$ analysis.
To determine $\nu$, use the large-$u$ expansion
$Z^2=4C^2[u-(1+c-2b)+\cO(u^{-1})]$; the additive constant in $Z$
vanishes because $Z$ is real on $u>c$ and purely imaginary on $u<0$.
Matching the coefficient of $u$ in \eqref{SmallZsLimit:ExactPolynomial}
to the background asymptotics gives
\begin{equation}
\nu=\frac{\rs^2}{d^2+4C^2(1+c-3b)}.
\label{SmallZsLimit:ExactNu}
\end{equation}
Using \eqref{SmallZsLimit:ExactMap} and \eqref{SmallZsLimit:RadiusK},
both the numerator and denominator are asymptotic to $4d^2K^2/\pi^2$,
so $\nu\to1$. Equations~\eqref{SmallZsLimit:ChargeK} and
\eqref{SmallZsLimit:RadiusK} also give
$\Qs\sim\omega^4\rs^5/(5\pi)$. Using
$V_0=3\omega^4d\rs^2/(2\nu)$ and $\hat Q_\star=\Qs/(V_0\rs^2)$, we obtain
\begin{equation}
\hat Q_\star=\frac{2\nu\Qs}{3\omega^4d\rs^4}
\sim\frac{2}{15\pi h},\qquad h=\frac d\rs=\hzs.
\end{equation}

\section{Notes on the Monte Carlo evaluation (Claude)}

The expectation values of $\mathrm{Tr}\,X_3^2$ and $\mathrm{Tr}\,X_{10}^2$ shown in
Figs.~\ref{fig:su2_traces}--\ref{fig:su4_traces} were obtained by direct Monte Carlo evaluation of
the matrix integral for $SU(N)$, $N=2,3,4$, with the fermions integrated out exactly.
Writing the fermionic action in the 16-component form
$-\tfrac12\psi^{T}\bar\gamma^{I}[X_I,\psi]+\tfrac{i}{3}\omega\,\psi^{T}\bar\gamma^{123}\psi$
and expanding $X_I=X_I^A T^A$ in generators with $\mathrm{Tr}\,T^AT^B=\delta^{AB}$, the
Grassmann integral gives $\mathrm{Pf}\,\mathfrak{M}(X)$ for the complex antisymmetric
matrix $\mathfrak{M}=-i\sum_{I\le 9}\gamma^{I}\otimes\mathcal{A}_I+\mathbbm{1}\otimes\mathcal{A}_{10}
+\tfrac{2i\omega}{3}\gamma^{123}\otimes\mathbbm{1}$, $\mathcal{A}_I^{AB}=-X_I^{C}f^{CAB}$,
of dimension $16(N^2-1)$ (the normalization of the measure was fixed against the exactly
solvable free limit). Since the bosonic action is real, the only complex factor is the
Pfaffian phase $e^{i\theta}$; we sampled the phase-quenched weight
$e^{-S_B}\,|\mathrm{Pf}\,\mathfrak{M}|$ by hybrid Monte Carlo and reweighted,
$\langle\mathcal{O}\rangle=\langle\mathcal{O}e^{i\theta}\rangle_{\rm pq}/\langle e^{i\theta}\rangle_{\rm pq}$.
The sign problem is absent for $N=2$ ($\mathrm{Pf}\,\mathfrak{M}>0$) and mild for
$N=3,4$, being worst in the crossover region, $\langle e^{i\theta}\rangle_{\rm pq}\ge 0.86$
for all points shown. The molecular-dynamics force uses the exact derivative
$\partial\log\mathrm{Pf}\,\mathfrak{M}=\tfrac12\mathrm{Tr}(\mathfrak{M}^{-1}\partial\mathfrak{M})$,
which requires one dense inverse per step; because the bosonic force is far cheaper but
stiffer, the two were integrated on separate time scales with a nested
Omelyan (Sexton--Weingarten) scheme. Trajectory lengths were tied to the softest modes,
the seven $SO(7)$ matrices with frequency $\sqrt2\omega/3$, and set to a quarter of
their period, $3\sqrt2\pi/(4\omega)$, with randomized lengths; each trajectory was
supplemented by a collective dilatation update $X_i\to e^{s_3}X_i$, $X_p\to e^{s_7}X_p$
accepted with the Jacobian $e^{(N^2-1)(3s_3+7s_7)}$, which moves the chain efficiently along
the nearly flat commuting directions at small $\omega$. Ergodicity across the
$P(N)$ fuzzy-sphere saddles was ensured in two ways: where one saddle dominates
($\omega\gtrsim 1.875$) the basin of each saddle was sampled separately and the averages were
combined with the exact relative weights $Z_{\mathcal R}/Z$ from localization; through the
multi-saddle crossover and down to the smallest $\omega$ we used replica exchange along
ladders in $\omega$ (spacing $\Delta\omega=0.075$--$0.1875$, swap acceptance $0.3$--$0.7$),
with every rung a measurement point and several independent ladders merged. At
$\omega\lesssim 0.5625$ a chain started near the origin never reaches the commuting valley
that carries the weight while appearing converged; starting configurations were therefore
drawn from the one-loop effective eigenvalue distribution of the commuting valley
(\cite{Komatsu:2024ydh} Sec.~4.2). Each point rests on $[1$--$1.2]\times10^{4}$ measurements with
integrated autocorrelation times of $1$--$3$ trajectories; errors are blocked-jackknife
estimates of the reweighted ratios, enlarged to the scatter between independent
replicas or ladders whenever that is larger. As an end-to-end check, the supersymmetry-protected
combination $\langle\mathrm{Tr}\,\phi^2\rangle=\langle\mathrm{Tr}\,X_3^2\rangle-\langle\mathrm{Tr}\,X_{10}^2\rangle$
(the cross term vanishes by $SO(3)\times SO(7)$) was compared at every point with its exact
localization value; all points agree within about two standard
deviations, as shown in the lower panels.

\section{Extrema of the effective action}
\label{sec:EffectiveActionSaddles}

In this appendix, we discuss the extrema of the effective action $V_{eff}$ obtained by integrating out the fermions and taking the real part. For any representation $\mathcal{R}$ built from $m$ irreducible representations of sizes $\{N_a\}$ and any $\omega$, we must have associated extrema of $V_{eff}$ of the form 
\begin{equation}
X^i = \oplus_a \alpha_a L_{(N_a)}^i \qquad X^{p} = \oplus_a m^p_a I_{N_a} \; .
\end{equation}
To show this, note that configurations of this form are exactly the subset of all configurations invariant under a subgroup $H = SU(2)_\mathcal{R} \times T^m$ of $SO(3) \times SO(7) \times U(N)$ where an element of $SU(2)_\mathcal{R}$ is a rotation $M^i {}_j$ in $SO(3)$ combined with a compensating $U(N)$ transformation $u^\mathcal{R} = \oplus_a u_{N_a}$ where
\begin{equation}
M^i {}_j L^j_{(N_a)} = u_{N_a} L^i_{(N_a)} (u_{N_a})^\dagger \; ,
\end{equation}
and an element of the torus group $T^m$ is a unitary $\oplus_a e^{i \theta_a} I_{N_a}$.

Extrema of the effective potential restricted to the subset invariant under $H$ must be extrema of the full effective potential, according to the principle of symmetric criticality \cite{palais1979principle}. 

The effective potential restricted to the set of matrices invariant under $H$ is a function $F(\{\alpha_a\}, \{m^p_a\})$ that is real, bounded below, and growing to infinity for large values of the parameters, so it must have a minimum. We can understand this physically. The parameters $\alpha_a$ determine the fuzzy sphere radii, while the $m^p_a$ are coordinates of the fuzzy-sphere centers in the transverse seven-dimensional space. Before taking into account the Pfaffian term, the potential for $\alpha_a$ is minimized by setting them all equal and the quadratic potential for $m^p_a$ is minimized by setting them all to zero. The Pfaffian modifies the potential for the $\alpha_a$ so that they may become unequal; the Pfaffian also gives a repulsive pairwise interaction between the coordinates $m^p_a$; for $\omega$ below some critical value, this causes the $m^p_a$ to spread out.\footnote{The existence of a critical value can be seen from the fact that the pairwise potential is quadratic at short distances with a strength that increases with decreasing $\omega$, while the overall quadratic potential has a strength that increases with increasing $\omega$.} 

As an example, for the representation with $n \le 8$ copies of the $M$-dimensional irreducible representation, we can take all the $\alpha_a$s equal and take the coordinates $m^p_a$ to be equidistant from one another. There is a family of such configurations related by $SO(7)$ rotations. For $n > 8$ it is not possible to choose all the $\vec{m}_a$ equidistant from one another, so we will end up with some type of frustrated crystal of fuzzy sphere centers.

For this $(n,M)$ example with $n\geq2$, it is possible to calculate explicitly the critical value of $\omega$ below which $\vec{m}_a$ become nonzero for the minima. The effective potential for $F(\alpha, \{\vec{m}_a\})$ can be computed explicitly in this case, to yield
\begin{align}
  F(\alpha,\{\vec m_a\})
  &= \frac{nM(M^2-1)}{4}\Big(\frac{\alpha^4}{2}-\frac{8\omega\,\alpha^3}{9}
     +\frac{\omega^2\alpha^2}{3}\Big)
   +\frac{\omega^2 M}{9}\sum_{a=1}^n|\vec m_a|^2
  \nonumber\\
  &\quad-4n\sum_{l=0}^{M-1}\Big[(l+1)\log\big(l\alpha+\tfrac{2\omega}{3}\big)^2
     + l\log\big((l+1)\alpha-\tfrac{2\omega}{3}\big)^2\Big]
  \nonumber\\
  &\quad-8\sum_{a<b}\sum_{l=0}^{M-1}\Big[(l+1)\log\Big(\big(l\alpha+\tfrac{2\omega}{3}\big)^2+\delta_{ab}^2\Big)
     + l\log\Big(\big((l+1)\alpha-\tfrac{2\omega}{3}\big)^2+\delta_{ab}^2\Big)\Big]\\
  & \quad + 8\log\frac{2\omega}{3},
  \label{eq:F-nM}
\end{align}
with $\delta_{ab}=|\vec m_a-\vec m_b|$. The first line is the bosonic action; the second line is the Pfaffian for the $n$
diagonal blocks; the third line is the contribution of the off-diagonal block
pairs, where the relative modulus $\delta_{ab}$ enters as a mass. The final constant comes
from removing the $\mathfrak u(1)$ trace mode and plays no role in what follows. Since $F$ depends on the centers only through
$|\vec m_a|^2$ and $|\vec m_a-\vec m_b|^2$, the configuration $\vec m_a=0$ is always
stationary with respect to variations in $\vec m_a$; demanding that such a configuration is also stationary with respect to variations in  $\alpha$, i.e. taking 
$\partial_\alpha F|_{\vec m=0}=0$, determines a radius $\alpha^*$. Writing $\hat\alpha=\alpha/\omega$ (and noting that
the $l=0$ terms do not depend on $\alpha$), the condition determining $\alpha^*$ is
\begin{equation}
  \frac{\omega^4 M(M^2-1)}{16\,n}\;
  \hat\alpha\Big(\hat\alpha-\frac13\Big)\Big(\hat\alpha-1\Big)
  \;=\;\sum_{l=1}^{M-1} l(l+1)\left[\frac{1}{l\hat\alpha+\frac23}
      +\frac{1}{(l+1)\hat\alpha-\frac23}\right]\;
  \label{eq:alpha-star}
\end{equation}
For $M>1$ and $1/3<\hat\alpha\le 1$ the left-hand side is non-positive while the right-hand side
is positive, so there is no solution; for $\hat\alpha>1$ the left-hand side is positive
and increasing while the right-hand side is positive and decreasing, so
\eqref{eq:alpha-star} has exactly one solution $\hat\alpha^*>1$ on the branch
$\hat\alpha>1/3$. 

For large enough $\omega$, this saddle is a stable minimum. However, as $\omega$ decreases, the repulsive pairwise interactions coming from the Pfaffian can lead to unstable directions and vevs for $\vec{m}_a$. A direct analysis of the quadratic action for $\vec{m}_a$ including the background term from the bosonic potential and the pairwise repulsion shows that the $\vec{m}_a = 0$ saddle is a stable minimum for $\omega > \omega_c$ where
\begin{equation}
  \omega_c^4 = 36\,\frac{n}{M}\sum_{l=0}^{M-1}\left[\frac{2l+2}{(l\hat\alpha^*+\tfrac23)^2}
      +\frac{2l}{((l+1)\hat\alpha^*-\tfrac23)^2}\right] .
  \label{eq:Omega-c}
\end{equation}
The equations \eqref{eq:alpha-star} and \eqref{eq:Omega-c} must be solved
together, and give
\begin{equation}
\omega_c=(n K_M)^{1/4},\qquad
K_M=\frac{36}{M}\sum_{l=0}^{M-1}
\left[\frac{2l+2}{(l\hat\alpha_M+\tfrac23)^2}
+\frac{2l}{((l+1)\hat\alpha_M-\tfrac23)^2}\right],
\end{equation}
where (for $M>1$) $\hat\alpha_M>1$ is the solution to
\begin{equation}
\begin{split}
  &\frac94 (M^2-1)\;
  \hat\alpha_M\Big(\hat\alpha_M-\frac13\Big)\Big(\hat\alpha_M-1\Big)
  \sum_{l=0}^{M-1}\left[\frac{2l+2}{(l\hat\alpha_M+\tfrac23)^2}
      +\frac{2l}{((l+1)\hat\alpha_M-\tfrac23)^2}\right]
  \\&= \sum_{l=1}^{M-1} l(l+1)\left[\frac{1}{l\hat\alpha_M+\frac23}
      +\frac{1}{(l+1)\hat\alpha_M-\frac23}\right].
\end{split}
  \label{eq:critical-alpha-M}
\end{equation}
For $M=1$ the radius parameter is absent, and the critical condition directly gives $K_1=162$. We find $K_2 \approx 121.46$, $K_4 \approx 83.08$, $K_8 \approx 53.24$ and $K_M \sim 144\,\frac{\ln M+0.78}{M}$ for large $M$.


\bibliography{refs.bib}

@article{Ishibashi:1996xs,
    author = "Ishibashi, N. and Kawai, H. and Kitazawa, Y. and Tsuchiya, A.",
    title = "{A Large N reduced model as superstring}",
    eprint = "hep-th/9612115",
    archivePrefix = "arXiv",
    reportNumber = "KEK-TH-503",
    doi = "10.1016/S0550-3213(97)00290-3",
    journal = "Nucl. Phys. B",
    volume = "498",
    pages = "467--491",
    year = "1997"
}

@article{palais1979principle,
  title={The principle of symmetric criticality},
  author={Palais, Richard S},
  journal={Communications in Mathematical Physics},
  volume={69},
  number={1},
  pages={19--30},
  year={1979},
  publisher={Springer}
}

@article{Lau:1999dp,
    author = "Lau, Stephen R.",
    title = "{Lightcone reference for total gravitational energy}",
    eprint = "gr-qc/9903038",
    archivePrefix = "arXiv",
    doi = "10.1103/PhysRevD.60.104034",
    journal = "Phys. Rev. D",
    volume = "60",
    pages = "104034",
    year = "1999"
}

@article{Mann:1999pc,
    author = "Mann, Robert B.",
    title = "{Misner string entropy}",
    eprint = "hep-th/9903229",
    archivePrefix = "arXiv",
    doi = "10.1103/PhysRevD.60.104047",
    journal = "Phys. Rev. D",
    volume = "60",
    pages = "104047",
    year = "1999"
}

@article{Kraus:1999di,
    author = "Kraus, Per and Larsen, Finn and Siebelink, Ruud",
    title = "{The gravitational action in asymptotically AdS and flat space-times}",
    eprint = "hep-th/9906127",
    archivePrefix = "arXiv",
    doi = "10.1016/S0550-3213(99)00549-0",
    journal = "Nucl. Phys. B",
    volume = "563",
    pages = "259--278",
    year = "1999"
}

@article{OoguriSkenderis1998DInstantons,
  author        = {Ooguri, Hirosi and Skenderis, Kostas},
  title         = {On the field theory limit of {D-instantons}},
  journal       = {JHEP},
  volume        = {11},
  pages         = {013},
  year          = {1998},
  doi           = {10.1088/1126-6708/1998/11/013},
  eprint        = {hep-th/9810128},
  archivePrefix = {arXiv},
  primaryClass  = {hep-th},
  url           = {https://arxiv.org/abs/hep-th/9810128}
}

@article{Ling:2006up,
    author = "Ling, Henry and Mohazab, Ali Reza and Shieh, Hsien-Hang and van Anders, Greg and Van Raamsdonk, Mark",
    title = "{Little string theory from a double-scaled matrix model}",
    eprint = "hep-th/0606014",
    archivePrefix = "arXiv",
    doi = "10.1088/1126-6708/2006/10/018",
    journal = "JHEP",
    volume = "10",
    pages = "018",
    year = "2006"
}

@article{Kazakov:1998ji,
    author = "Kazakov, Vladimir A. and Kostov, Ivan K. and Nekrasov, Nikita A.",
    title = "{D particles, matrix integrals and KP hierarchy}",
    eprint = "hep-th/9810035",
    archivePrefix = "arXiv",
    reportNumber = "HUTP-98-A051, ITEP-TH-35-98, CERN-TH-98-302, SACLAY-SPH-T-98-102, LPTENS-98-40",
    doi = "10.1016/S0550-3213(99)00393-4",
    journal = "Nucl. Phys. B",
    volume = "557",
    pages = "413--442",
    year = "1999"
}

@article{GibbonsGreenPerry1996Instantons,
  author        = {Gibbons, Gary W. and Green, Michael B. and Perry, Malcolm J.},
  title         = {Instantons and seven-branes in type {IIB} superstring theory},
  journal       = {Phys. Lett. B},
  volume        = {370},
  pages         = {37--44},
  year          = {1996},
  doi           = {10.1016/0370-2693(95)01565-5},
  eprint        = {hep-th/9511080},
  archivePrefix = {arXiv},
  primaryClass  = {hep-th},
  url           = {https://arxiv.org/abs/hep-th/9511080}
}

@article{Berenstein_2002,
   title={Strings in flat space and pp waves from Script N = 4 Super Yang Mills},
   volume={2002},
   ISSN={1029-8479},
   url={http://dx.doi.org/10.1088/1126-6708/2002/04/013},
   DOI={10.1088/1126-6708/2002/04/013},
   number={04},
   journal={Journal of High Energy Physics},
   publisher={Springer Science and Business Media LLC},
   author={Berenstein, David and Maldacena, Juan and Nastase, Horatiu},
   year={2002},
   month=Apr, pages={013–013} }

@article{Maldacena_2018,
   title={To gauge or not to gauge?},
   volume={2018},
   ISSN={1029-8479},
   url={http://dx.doi.org/10.1007/JHEP04(2018)084},
   DOI={10.1007/jhep04(2018)084},
   number={4},
   journal={Journal of High Energy Physics},
   publisher={Springer Science and Business Media LLC},
   author={Maldacena, Juan and Milekhin, Alexey},
   year={2018},
   month=Apr }

@article{Martina2026MassDeformations,
  author        = {Martina, Adrien},
  title         = {Massive deformations of supersymmetric {Yang-Mills} matrix models},
  journal       = {JHEP},
  volume        = {03},
  pages         = {250},
  year          = {2026},
  doi           = {10.1007/JHEP03(2026)250},
  eprint        = {2507.17813},
  archivePrefix = {arXiv},
  primaryClass  = {hep-th},
  url           = {https://arxiv.org/abs/2507.17813}
}

@article{LinMaldacena2006Fivebranes,
  author        = {Lin, Hai and Maldacena, Juan},
  title         = {Fivebranes from gauge theory},
  journal       = {Phys. Rev. D},
  volume        = {74},
  pages         = {084014},
  year          = {2006},
  doi           = {10.1103/PhysRevD.74.084014},
  eprint        = {hep-th/0509235},
  archivePrefix = {arXiv},
  primaryClass  = {hep-th},
  url           = {https://arxiv.org/abs/hep-th/0509235}
}

@article{AsanoEtAl2014Bubbling,
  author        = {Asano, Yuhma and Ishiki, Goro and Okada, Takashi and Shimasaki, Shinji},
  title         = {Emergent bubbling geometries in the plane wave matrix model},
  journal       = {JHEP},
  volume        = {05},
  pages         = {075},
  year          = {2014},
  doi           = {10.1007/JHEP05(2014)075},
  eprint        = {1401.5079},
  archivePrefix = {arXiv},
  primaryClass  = {hep-th},
  url           = {https://arxiv.org/abs/1401.5079}
}

@article{BobevEtAl2018SphericalBranes,
  author        = {Bobev, Nikolay and Bomans, Pieter and Gautason, Fridrik Freyr},
  title         = {{Spherical Branes}},
  journal       = {JHEP},
  volume        = {08},
  pages         = {029},
  year          = {2018},
  doi           = {10.1007/JHEP08(2018)029},
  eprint        = {1805.05338},
  archivePrefix = {arXiv},
  primaryClass  = {hep-th},
  url           = {https://arxiv.org/abs/1805.05338}
}

@article{Mann:2005yr,
    author = "Mann, Robert B. and Marolf, Donald",
    title = "{Holographic renormalization of asymptotically flat spacetimes}",
    eprint = "hep-th/0511096",
    archivePrefix = "arXiv",
    doi = "10.1088/0264-9381/23/9/010",
    journal = "Class. Quant. Grav.",
    volume = "23",
    pages = "2927--2950",
    year = "2006"
}

@article{Maldacena:2002rb,
    author = "Maldacena, Juan Martin and Sheikh-Jabbari, Mohammad M. and Van Raamsdonk, Mark",
    title = "{Transverse five-branes in matrix theory}",
    eprint = "hep-th/0211139",
    archivePrefix = "arXiv",
    reportNumber = "SU-ITP-02-33",
    doi = "10.1088/1126-6708/2003/01/038",
    journal = "JHEP",
    volume = "01",
    pages = "038",
    year = "2003"
}

@article{Troyer_2005,
   title={Computational Complexity and Fundamental Limitations to Fermionic Quantum Monte Carlo Simulations},
   volume={94},
   ISSN={1079-7114},
   url={http://dx.doi.org/10.1103/PhysRevLett.94.170201},
   DOI={10.1103/physrevlett.94.170201},
   number={17},
   journal={Physical Review Letters},
   publisher={American Physical Society (APS)},
   author={Troyer, Matthias and Wiese, Uwe-Jens},
   year={2005},
   month=May }

@article{Bonelli:2002mb,
    author = "Bonelli, Giulio",
    title = "{Matrix strings in pp wave backgrounds from deformed superYang-Mills theory}",
    eprint = "hep-th/0205213",
    archivePrefix = "arXiv",
    reportNumber = "ULB-TH-02-15",
    doi = "10.1088/1126-6708/2002/08/022",
    journal = "JHEP",
    volume = "08",
    pages = "022",
    year = "2002"
}

@article{Hartnoll:2024csr,
    author = "Hartnoll, Sean A. and Liu, Jun",
    title = "{The polarised IKKT matrix model}",
    eprint = "2409.18706",
    archivePrefix = "arXiv",
    primaryClass = "hep-th",
    doi = "10.1007/JHEP03(2025)060",
    journal = "JHEP",
    volume = "03",
    pages = "060",
    year = "2025"
}

@article{Komatsu:2024bop,
    author = "Komatsu, Shota and Martina, Adrien and Penedones, Jo{\~a}o and Vuignier, Antoine and Zhao, Xiang",
    title = "{Einstein gravity from a matrix integral. Part I}",
    eprint = "2410.18173",
    archivePrefix = "arXiv",
    primaryClass = "hep-th",
    doi = "10.1007/JHEP12(2025)029",
    journal = "JHEP",
    volume = "12",
    pages = "029",
    year = "2025"
}

@article{Komatsu:2024ydh,
    author = "Komatsu, Shota and Martina, Adrien and Penedones, Joao and Vuignier, Antoine and Zhao, Xiang",
    title = "{Einstein gravity from a matrix integral. Part II}",
    eprint = "2411.18678",
    archivePrefix = "arXiv",
    primaryClass = "hep-th",
    doi = "10.1007/JHEP12(2025)030",
    journal = "JHEP",
    volume = "12",
    pages = "030",
    year = "2025"
}

@article{Hartnoll:2025ecj,
    author = "Hartnoll, Sean A. and Liu, Jun",
    title = "{Statistical physics of the polarised IKKT matrix model}",
    eprint = "2504.06481",
    archivePrefix = "arXiv",
    primaryClass = "hep-th",
    doi = "10.21468/SciPostPhys.19.4.099",
    journal = "SciPost Phys.",
    volume = "19",
    number = "4",
    pages = "099",
    year = "2025"
}

@article{Chou:2025rwy,
    author = "Chou, Chien-Yu and Nishimura, Jun and Wang, Cheng-Tsung",
    title = "{Monte~Carlo Studies of the Emergent Spacetime in the Polarized Type IIB Matrix Model}",
    eprint = "2507.18472",
    archivePrefix = "arXiv",
    primaryClass = "hep-th",
    reportNumber = "KEK-TH-2740",
    doi = "10.1103/y1rm-n85b",
    journal = "Phys. Rev. Lett.",
    volume = "135",
    number = "22",
    pages = "221601",
    year = "2025"
}

@article{DHoker:2025nid,
    author = "D'Hoker, Eric and Gutperle, Michael and Uhlemann, Christoph F.",
    title = "{Exact solutions to complex Type IIB supergravity for complex superalgebra F (4) and its real forms}",
    eprint = "2508.14959",
    archivePrefix = "arXiv",
    primaryClass = "hep-th",
    doi = "10.1007/JHEP07(2026)148",
    journal = "JHEP",
    volume = "07",
    pages = "148",
    year = "2026"
}

@article{DHoker:2026eox,
    author = "D'Hoker, Eric and Gutperle, Michael and Uhlemann, Christoph F.",
    title = "{Type IIB supergravity solutions for polarized IKKT}",
    eprint = "2609.35963",
    archivePrefix = "arXiv",
    primaryClass = "hep-th",
    month = "9",
    year = "2026"
}

@inproceedings{Marolf:2000cb,
    author = "Marolf, Donald",
    title = "{Chern-Simons terms and the three notions of charge}",
    booktitle = "{International Conference on Quantization, Gauge Theory, and Strings: Conference Dedicated to the Memory of Professor Efim Fradkin}",
    eprint = "hep-th/0006117",
    archivePrefix = "arXiv",
    reportNumber = "SUGP-00-6-1",
    pages = "312--320",
    month = "6",
    year = "2000"
}

@article{Kanitscheider:2008kd,
    author = "Kanitscheider, Ingmar and Skenderis, Kostas and Taylor, Marika",
    title = "{Precision holography for non-conformal branes}",
    eprint = "0807.3324",
    archivePrefix = "arXiv",
    primaryClass = "hep-th",
    reportNumber = "ITFA-2008-25",
    doi = "10.1088/1126-6708/2008/09/094",
    journal = "JHEP",
    volume = "09",
    pages = "094",
    year = "2008"
}

@article{Wiseman:2008qa,
    author = "Wiseman, Toby and Withers, Benjamin",
    title = "{Holographic renormalization for coincident Dp-branes}",
    eprint = "0807.0755",
    archivePrefix = "arXiv",
    primaryClass = "hep-th",
    doi = "10.1088/1126-6708/2008/10/037",
    journal = "JHEP",
    volume = "10",
    pages = "037",
    year = "2008"
}

\end{document}